\documentclass[useAMS,usenatbib]{mn2e}

\usepackage{graphicx}
\newcommand{\eq}[1]{\begin{equation}  #1 \end{equation}}

\newcommand{\eqa}[1]{\begin{eqnarray}   #1 \end{eqnarray}}

\newcommand{\br}[1]{\left( #1 \right)}
\newcommand{\bc}[1]{\left\{ #1 \right\}}
\newcommand{\bb}[1]{\left[ #1 \right]}
\newcommand{\ba}[1]{\left\langle #1 \right\rangle}

\newcommand{\nn}{\nonumber}

\newcommand{\dd}{{\rm d}}
\newcommand{\expo}[1]{~{\rm e}^{ #1 }}
\newcommand{\vek}[1]{\mbox{\boldmath $#1$}}
\newcommand{\svek}[1]{\mbox{\boldmath \scriptsize $#1$}}  
\newcommand{\ic}{{\rm i}}

\title{Cosmological information in Gaussianised weak lensing signals}
\author[B. Joachimi, A.N. Taylor \& A. Kiessling]
  {B.~Joachimi,$^1$\thanks{E-mail: bj@roe.ac.uk}
  A.N.~Taylor$^1$ and A. Kiessling$^1$\\
  $^1$Institute for Astronomy, University of Edinburgh, Royal Observatory, Blackford Hill, Edinburgh, EH9 3HJ, U.K.}
\date{Accepted . Received ; in original form }

\pagerange{\pageref{firstpage}--\pageref{lastpage}} \pubyear{2011}

\begin{document}
\label{firstpage}

\maketitle

\begin{abstract}
Gaussianising the one-point distribution of the weak gravitational lensing convergence has recently been shown to increase the signal-to-noise contained in two-point statistics. We investigate the information on cosmology that can be extracted from the transformed convergence fields. Employing Box-Cox transformations to determine optimal transformations to Gaussianity, we develop analytical models for the transformed power spectrum, including effects of noise and smoothing. We find that optimised Box-Cox transformations perform substantially better than an offset logarithmic transformation in Gaussianising the convergence, but both yield very similar results for the signal-to-noise. None of the transformations is capable of eliminating correlations of the power spectra between different angular frequencies, which we demonstrate to have a significant impact on the errors on cosmology. Analytic models of the Gaussianised power spectrum yield good fits to the simulations and produce unbiased parameter estimates in the majority of cases, where the exceptions can be traced back to the limitations in modelling the higher-order correlations of the original convergence. In the idealistic case, without galaxy shape noise, we find an increase in cumulative signal-to-noise by a factor of 2.6 for angular frequencies up to $\ell=1500$, and a decrease in the area of the confidence region in the $\Omega_{\rm m}-\sigma_8$ plane, measured in terms of $q$-values, by a factor of 4.4 for the best-performing transformation. When adding a realistic level of shape noise, all transformations perform poorly with little decorrelation of angular frequencies, a maximum increase in signal-to-noise of $34\,\%$, and even slightly degraded errors on cosmological parameters. We argue that, to find Gaussianising transformations of practical use, it will be necessary to go beyond transformations of the one-point distribution of the convergence, extend the analysis deeper into the non-linear regime, and resort to an exploration of parameter space via simulations.
\end{abstract}


\section{Introduction}

Weak gravitational lensing of distant galaxies by the large-scale structure is considered as one of the most powerful probes of cosmological physics (\citealp{albrecht06,peacock06}; see \citealp{munshi08} for a recent review). Planned surveys from the ground (e.g. LSST\footnote{\texttt{http://www.lsst.org}}) and from space (e.g. Euclid\footnote{\texttt{http://sci.esa.int/euclid}}) will measure the dark energy equation of state, properties of dark matter, and possible deviations from general relativity with unprecedented precision, reaching percentage accuracy on some parameters \citep[e.g.][]{refregier10}.

The standard analysis employs two-point statistics of the gravitational shear, which would fully specify the properties of the underlying matter distribution if it were distributed according to a Gaussian random field. However, non-linear structure formation induces correlations between different angular scales in Fourier space and hence reduces the cosmological information contained in weak lensing two-point statistics. At the same time extra information is generated in higher-order statistics of the shear, which, if it can be extracted, improves parameter constraints. The most widespread approaches to exploit higher-order correlations make use of shear three-point statistics \citep[e.g.][]{semboloni11} and peak statistics \citep[e.g.][]{berge08}.

The goal of highly precise inference on cosmological parameters entails the need for access to the constraining power created by non-linear effects on the shear fields in an effective way, and for the guarantee that none of the steps in the observations and analysis introduce uncertainty or systematics at a level which significantly affects errors on cosmology. These requirements have raised a range of issues driving current research, among them an efficient choice of higher-order statistic \citep{berge10}, the determination of accurate covariances \citep{takada09,pielorz10,kiessling11}, and the derivation of the functional form of the likelihood \citep{hartlap09,Schneider09}.

The problems listed above can, at least in principle, all be solved if one could find a bijective mapping of the observed gravitational shear field, or equivalently the weak lensing convergence field, such that the transformed field is described by a Gaussian random field. This field is completely determined by its power spectrum which consequently contains all cosmological information present in the original field. Therefore only two-point statistics have to be considered in the likelihood analysis whose covariance can also be expressed in terms that are second-order in the shear or convergence. In addition, the common assumption of a simple Gaussian likelihood becomes exact when formulating it for the transformed convergence (for a similar ansatz in the context of cosmic microwave background temperature fluctuations see \citealp{bond00}).

The recent work by \citet{seo11} suggests that such a beneficial transformation is approximately realised by taking the logarithm of the positively offset convergence, decorrelating angular frequencies and boosting the signal-to-noise in the transformed power spectrum. Logarithmic transformations are widely used in statistics to reduce the skewness in distributions which in the context of large-scale structure is caused by the excess of high-density regions due to non-linear evolution. \citet{coles91} provided a heuristic physical justification by demonstrating that the one-point distribution of the matter density contrast, $\delta$, is lognormal in Lagrangian coordinates if one assumes the Zel'dovich approximation and Gaussian initial conditions for the matter density and velocity fields \citep[for an exact calculation of the one-point distribution under these assumptions see][]{kofman94}.

\citet{kayo01} showed by means of N-body simulations that a lognormal model accurately describes the one-point distribution of $\delta$ well into the non-linear regime. This fact has fostered the use of the lognormal distribution, or equivalently $\ln (1+\delta)$ as a \lq natural\rq\ variable, in the modelling of large-scale structure \citep[e.g.][]{szapudi03,kitaura10}. The weak lensing convergence $\kappa$ is a weighted projection of the matter density contrast and hence should inherit a skewed shape of the one-point distribution, although its minimum varies as $\kappa$ does in practice not reach its theoretical lower limit, as opposed to $\delta \geq -1$ \citep[see the discussion in][]{taruya02}. Indeed \citet{taruya02} found that $\kappa$ is well described empirically as lognormal distributed, with some deviations reported for high source galaxy redshifts and the tails of the convergence distribution.

As presented in \citet{neyrinck09,neyrinck11}, Gaussianising the one-point distribution of the matter density contrast via logarithmic transformation, or by matching the cumulative distribution function to a Gaussian one (referred to as rank-order Gaussianisation henceforth), increases the signal-to-noise in the transformed matter power spectrum. This result has triggered the analogous studies on the weak lensing convergence by \citet{seo11} considering logarithmic transformations and by \citet{yu11} who investigate the bispectrum and higher moments of the rank-order Gaussianised convergence. Note also the approach of \citet{zhang11} and \citet{yu11b} who apply a non-linear Wiener filter to the matter density contrast and the convergence, respectively, instead of transforming these quantities. Recently, \citet{neyrinck11b} presented a simulation-based study of the effect on cosmological parameters due to the Gaussianisation of the matter density contrast, finding significantly improved constraints in some cases.

This work is aimed at elucidating the cosmological information content of transformed convergence fields and their ability to constrain cosmological parameters using analytical models. To this end, we employ Box-Cox transformations which encompass a range of transformations frequently applied in statistics, including the logarithm, and which provide us with an efficient maximum likelihood formalism to estimate their free parameters. This allows us to quantify how well logarithmic transformations fare in Gaussianising the one-point distribution of $\kappa$, and derive optimal transformations.

Contrary to rank-order Gaussianisation, it is conceptually easy to determine the statistics of the transformed convergence analytically for a parametrised form as given by the Box-Cox transformations. We assess the accuracy and limitations of our models in fitting the power spectra obtained from a large suite of simulations of transformed convergence fields, and investigate the constraints on cosmological parameters that can be achieved with these models for different transformations as well as convergence maps with and without galaxy shape noise.

The article is structured as follows: In Section \ref{sec:simulations} we summarise the simulations underlying our analysis. Section \ref{sec:theory} describes the transformations we apply to the weak lensing convergence and their optimisation, as well as the modelling of the statistics of the transformed fields. Our results for optimal transformations, the analytical models of the power spectrum of the transformed fields, the noise properties, and the constraints on cosmology are presented in Section \ref{sec:results}. In Section \ref{sec:discussion} we discuss our findings and their implications, before summarising and concluding in Section \ref{sec:conclusions}.

\section{Weak lensing simulations}
\label{sec:simulations}

To test the performance of our transformations and models, and perform a mock likelihood analysis, we use 100 independent realisations of weak lensing convergence fields generated by the {\small SUNGLASS} pipeline \citep{kiessling11}. They are based on medium-resolution dark matter-only N-body simulations with a box size of $512 h^{-1}\,{\rm Mpc}$ and periodic boundary conditions. The simulations are populated by $512^3$ particles with mass $7.5 \times 10^{10}\,M_\odot$, using a force softening length of $33h^{-1}\,{\rm kpc}$. A flat $\Lambda$CDM cosmology with WMAP7 parameters \citep{jarosik11}, particularly $\Omega_{\rm m}=0.27$ and $\sigma_8=0.81$, is adopted.

The weak lensing convergence is given by a weighted integral over the matter density contrast, $\delta$ \citep[see e.g.][]{bartelmann01},
\eq{
\label{eq:kappaprojection}
\kappa(\vek{\theta},z_{\rm s}) = \frac{3 H_0^2 \Omega_{\rm m}}{2 c^2} \int_0^{\chi(z_{\rm s})} \!\!\!\!\!\!\! \dd \chi'\; \frac{\chi'\, \bb{\chi(z_{\rm s})-\chi'}}{\chi(z_{\rm s})\, a(\chi')}\; \delta\br{\chi' \vek{\theta},\chi'}\;,
}
where $\chi$ denotes comoving distance and $a$ the cosmic scale factor. The convergence depends on the redshift $z_{\rm s}$ of the galaxies which serve as sources for the weak lensing signal. Usually the source galaxies follow a distribution in redshift, but for simplicity we will assume a single source redshift $z_{\rm s}=1$, which is close to the median redshift of upcoming surveys. 

{\small SUNGLASS} assumes the Born approximation and computes the convergence directly via a discretised version of equation (\ref{eq:kappaprojection}), using the three-dimensional particle positions. The light cone out to $z_{\rm s}=1$ is constructed from 19 snapshots with a separation of $128h^{-1}\,{\rm Mpc}$ each. To avoid the repetition of structure after spanning distances exceeding the box size of the simulation, boxes are randomly translated and rotated. The convergence maps are computed on a grid with $N^2$ points with $N=2048$, covering an area of $A_{\rm field}=10 \times 10\,{\rm deg}^2$. Since the realisations are fully independent, we obtain a total survey size of $10,000\,{\rm deg}^2$ by jointly analysing all convergence maps.

We are also interested in considering convergence maps with a realistic level of noise which in weak lensing is governed by the random distribution of intrinsic galaxy shapes. To add shape noise, the convergence fields are Fourier-transformed and converted to shear fields via
\eq{
\label{eq:gamkap}
\gamma(\vek{\ell}) = {\rm e}^{2 \ic \varphi_\ell}\; \kappa(\vek{\ell})\;,
}
where $\gamma$ denotes the complex gravitational shear (quantifying the ellipticity and the position angle of the galaxy shape), and $\varphi_\ell$ the polar angle of the angular frequency vector $\vek{\ell}$. Note that, to ease the notation, we will throughout use the same symbol to designate quantities and their respective Fourier transforms.

After Fourier-transforming back to real space, an intrinsic ellipticity is added to each shear component at every grid point of the shear map by randomly drawing values from a Gaussian distribution with dispersion $\sigma_\epsilon/\sqrt{2 n_{\rm gal} A_{\rm field} / N^2}$, where $\sigma_\epsilon=0.35$ is a typical intrinsic ellipticity dispersion, and $n_{\rm gal}=30\,{\rm arcmin}^{-2}$ the assumed number density of galaxies. Inverting equation (\ref{eq:gamkap}), one readily calculates the convergence from the shear maps with shape noise included by again transforming to Fourier space and back.

Since we work with the convergence maps, the convergence power spectrum is the two-point statistic of choice for the subsequent likelihood analysis. We employ the estimator
\eq{
\label{eq:psestimator}
\hat{P}_\kappa(\ell) = \frac{2\pi}{\ell\, \Delta \ell} \sum_{\svek{\ell}_j \in\, {\rm shell}(\ell)} |\kappa_{\svek{\ell}_j}|^2\;,
}
where $\Delta \ell$ is the width of the angular frequency bin, which we choose to be constant in log-space with $\ln \Delta \ell \approx 0.23$. We use the notation $\kappa_{\svek{\ell}}$ for the convergence values on a discretely Fourier-transformed grid; see also Appendix \ref{app:likekappa}. The sum runs over all angular frequency vectors which lie in a shell with central radius $\ell$ and width $\Delta \ell$. To avoid aliasing in the power spectrum due to the edges of the convergence fields, a Hann window is applied to the convergence values in the margins covering the outmost $12.5\,\%$ of the maps. For further details on the simulations and power spectrum estimation we refer the reader to \citet{kiessling11}.

\section{Transformations of convergence and power spectrum}
\label{sec:theory}

In the following we will detail the transformations that we apply to Gaussianise the convergence maps, including the procedure to estimate the free parameters in the transformation equations. We then proceed to express the power spectrum of the transformed convergence in terms of the statistics of the original convergence, the central prerequisite that will allow us compute the analytical models required for the likelihood analysis.

\subsection{Box-Cox transformations}
\label{sec:boxcox}

\begin{figure}
\centering
\includegraphics[scale=.34,angle=270]{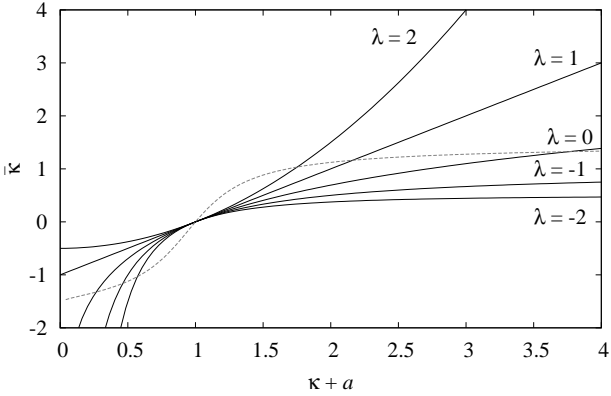}
\caption{Illustration of Box-Cox transformations. Shown is the mapping of the shifted original convergence $\kappa + a$ to the transformed convergence $\bar{\kappa}$ for several values of $\lambda$. Note that $\lambda=0$ corresponds to the logarithmic transformation, and that $\lambda=1$ leaves the convergence unchanged (except for an offset by $a$). In addition we have plotted a log-arctan transformation with $s=3$ as grey dashed curve; see Section \ref{sec:shapenoise} for details.}
\label{fig:boxcoxillustrate}
\end{figure}

\citet{box64} introduced a parametrised set of power transformations that are widely used in statistical data analysis, encompassing the logarithmic transformation which has recently gained attention in attempts to boosting information in cosmological density fields. For a given random sample of data, in our case the $n=N^2$ grid point values of the convergence in one map, the Box-Cox transformation reads
\eq{
\label{eq:boxcox}
\bar{\kappa}_i(\lambda,a) = \left\{ \begin{array}{ll} \bb{\br{\kappa_i + a}^{\lambda}-1}/\lambda & \lambda \neq 0\\ \ln (\kappa_i + a) & \lambda = 0 \end{array} \right.\;,
}
for all $i=1,\,..\,,n$. We will consider both the power $\lambda$ and the shift $a$ as free parameters of the transformation. Note that the transformed convergence is denoted by a bar, and that the dependence on $\lambda$ and $a$ will mostly be suppressed henceforth. The normalisation of equation (\ref{eq:boxcox}) is chosen such that the transformation is continuous in $\lambda$ at $\lambda=0$. We have illustrated the mapping according to equation (\ref{eq:boxcox}) for a few exemplary cases in Fig.$\,$\ref{fig:boxcoxillustrate}.

The Box-Cox parameters $(\lambda,a)$ shall be determined from the sample $\bc{\kappa_i}$ such that the one-point distribution of transformed convergence values, $\bar{{\cal P}}_{\rm 1pt}(\bar{\kappa})$, is Gaussian\footnote{Note that Box-Cox transformations are not limited to one-dimensional distributions. We follow earlier work by concentrating on transforming the one-point distribution only, but discuss possible ways beyond this ansatz in Section \ref{sec:discussion}.}. The relation to the distribution of the original convergence is given by
\eq{
\label{eq:distributionchangeboxcox}
{\cal P}_{\rm 1pt}(\kappa) = \bar{{\cal P}}_{\rm 1pt}(\bar{\kappa})\; \prod_{i=1}^n (\kappa_i+a)^{\lambda-1}\;,
}
where the last term is the Jacobian of the Box-Cox transformation. Equation (\ref{eq:distributionchangeboxcox}) provides a model for the distribution ${\cal P}_{\rm 1pt}(\kappa)$ from which the data sample $\bc{\kappa_i}$ is drawn, featuring only the two Box-Cox parameters and the mean and variance of the assumed Gaussian $\bar{{\cal P}}_{\rm 1pt}(\bar{\kappa})$ as undetermined parameters. Employing the maximum likelihood estimators for mean and variance, one can derive the concentrated log-likelihood for $\lambda$ and $a$ (\citealp{box64}; see also \citealp{joachimi11}), resulting in
\eqa{
\label{eq:lmaxboxcox}
{\cal L}_{\rm max}(\lambda,a) &=& - \frac{n}{2}\; \ln \bc{ \frac{1}{n} \sum_{i=1}^n \bigl[ \bar{\kappa}_i(\lambda,a) -\ba{\bar{\kappa}(\lambda,a)} \bigr]^2}\\ \nn
&& +\; (\lambda -1)\; \sum_{i=1}^n \ln (\kappa_i+a)\;.
}
Here, the term in curly brackets is the maximum likelihood estimate for the variance of the transformed convergence field, the angular brackets denoting the mean. Note that the exponential of the Gaussian likelihood is unity if the maximum likelihood estimate of the mean is unbiased. Equation (\ref{eq:lmaxboxcox}) constitutes a model for the distribution of the original convergence with only $\lambda$ and $a$ as free parameters. Maximising this equation with respect to the Box-Cox parameters provides us with maximum-likelihood estimates for $\lambda$ and $a$ and thus with a method to determine an optimal transformation to Gaussianity which is entirely driven by the data itself.

Note that the Box-Cox transformation changes the dimension of the data set under consideration which could be corrected for e.g. by dividing by a function of the geometric mean of the original data set \citep{box64}. However, since $\kappa$ is dimensionless, we prefer to keep the simplest possible form of the transformation as given by equation (\ref{eq:boxcox}). Unlike the original convergence, $\bar{\kappa}$ has a non-vanishing mean on average which generally is a function of all moments of the original field. In principle this is irrelevant as there is no cosmological information in the mean, but due to the finite size of the convergence maps large-scale modes might become biased. Hence we correct all transformed fields to a mean of zero.

\subsection{Transformed power spectrum}
\label{sec:transformedps}

One of the major advantages of Box-Cox transformations (including the logarithmic transformation) over rank-order Gaussianisation techniques as e.g. applied by \citet{neyrinck09} and \citet{yu11} is the analytical relation between original and transformed values in each grid point of the field. This permits us to calculate the power spectrum of the Box-Cox transformed convergence in terms of the statistics of the original convergence, which in turn are computed from the cosmological model. We begin by Taylor-expanding the term in parentheses appearing in equation (\ref{eq:boxcox}) for $\lambda \neq 0$,
\eqa{
\label{eq:kappataylor}
\br{\kappa+a}^\lambda &=& a^\lambda + \lambda a^{\lambda-1} \kappa + \frac{\lambda (\lambda-1)}{2} a^{\lambda-2} \kappa^2\\ \nn
&& + \frac{\lambda (\lambda-1) (\lambda-2)}{6} a^{\lambda-3} \kappa^3 + {\cal O}(\kappa^4)\;.
}
With this result the Fourier transform of the transformed convergence reads, to the same order in $\kappa$,
\eqa{
\label{eq:kappatrafo}
\bar{\kappa}(\vek{\ell}) &=& \int \dd^2 \theta\; \bar{\kappa}(\vek{\theta})\; \expo{\ic \svek{l} \cdot \svek{\theta}}\\ \nn
&=& \frac{a^\lambda}{\lambda}\; \sum_{j=0}^\infty \frac{\prod_{k=0}^{j-1} (\lambda-k)}{j!\, a^j} \int \dd^2 \theta\; \bb{\kappa(\vek{\theta})}^j \expo{\ic \svek{l} \cdot \svek{\theta}}\\ \nn
&& -\; \frac{(2\pi)^2}{\lambda}\; \delta^{(2)}(\vek{\ell})\\ \nn
&=& \frac{(2\pi)^2}{\lambda} \br{a^\lambda-1} \delta^{(2)}(\vek{\ell}) + a^{\lambda-1} \kappa(\vek{\ell})\\ \nn
&& +\; \frac{\lambda-1}{2}\; a^{\lambda-2} \int \frac{\dd^2 \ell_1}{(2 \pi)^2}\; \kappa(\vek{\ell}_1)\; \kappa(\vek{\ell}-\vek{\ell}_1)\\ \nn
&& +\; \frac{(\lambda-1) (\lambda-2)}{6}\; a^{\lambda-3} \int \frac{\dd^2 \ell_1}{(2 \pi)^2} \int \frac{\dd^2 \ell_2}{(2 \pi)^2}\\ \nn
&& \;\; \times\; \kappa(\vek{\ell}_1)\; \kappa(\vek{\ell}_2)\; \kappa(\vek{\ell}-\vek{\ell}_1-\vek{\ell}_2)\; +\; {\cal O}(\kappa^4)\;,
}
where $\delta^{(2)}(\vek{\ell})$ denotes the Dirac-delta distribution. To arrive at the last equality, we applied the convolution theorem several times. Note that the corresponding calculation for the case $\lambda=0$, i.e. based on the expansion of $\ln (\kappa + a)$, yields the same result except for the zeroth-order term. This term is generated by the non-vanishing mean of the transformed convergence, but contributes due to the Dirac-delta function only to the DC ($\ell=0$) component of the power spectrum and is hence irrelevant. As an aside, since we correct $\bar{\kappa}$ to zero mean, there are in principle more terms contributing to the zeroth order in equation (\ref{eq:kappatrafo}), but we have omitted this step to keep the formalism simple.

It is evident from equation (\ref{eq:kappatrafo}) that the two-point correlation of $\bar{\kappa}$ is a function of all n-point correlations of the original convergence. The $d$-th spectrum of the convergence is defined via
\eq{
\label{eq:defspectra}
\ba{ \prod_{i=1}^d \kappa(\vek{\ell}_i) }_{\rm c} = (2\pi)^2\; \delta^{(2)} \br{ \sum_{i=1}^d \vek{\ell}_i }\; S_\kappa^d(\vek{\ell}_1,\,...\,,\vek{\ell}_d)\;,
}
where the subscript c denotes the connected moments. We identify the second-order spectrum $S_\kappa^2 \equiv P_\kappa$ with the power spectrum, $S_\kappa^3 \equiv B_\kappa$ with the convergence bispectrum, and $S_\kappa^4 \equiv T_\kappa$ with the connected convergence trispectrum. Making use of these definitions, one arrives at the following expression, containing terms up to fourth order in $\kappa$,
\eqa{
\label{eq:pstrafo_basic_approx}
P_{\bar{\kappa}}(\ell) &=& a^{2\lambda-2} \biggl\{ P_{\kappa}(\ell) \\ \nn
&& \hspace*{-1.6cm} +\; (\lambda-1)\; a^{-1} \int \frac{\dd^2 \ell_1}{(2 \pi)^2}\; B_\kappa(\ell,\ell_1,|\vek{\ell}-\vek{\ell}_1|)\\ \nn
&& \hspace*{-1.6cm}  +\; \frac{(\lambda-1) (\lambda-2)}{3}\; a^{-2} \Bigl[ 3 P_\kappa(\ell) \int \frac{\dd^2 \ell_1}{(2 \pi)^2}\; P_\kappa(\ell_1)\\ \nn
&& \hspace*{-1.5cm} + \int \frac{\dd^2 \ell_1}{(2 \pi)^2} \int \frac{\dd^2 \ell_2}{(2 \pi)^2}\; T_\kappa(\vek{\ell},-\vek{\ell}_1,-\vek{\ell}_2,\vek{\ell}_1+\vek{\ell}_2-\vek{\ell}) \Bigr]\\ \nn
&& \hspace*{-1.6cm} +\; \frac{(\lambda-1)^2}{4}\; a^{-2} \Bigl[ 2 \int \frac{\dd^2 \ell_1}{(2 \pi)^2}\; P_\kappa(\ell_1) P_\kappa(|\vek{\ell}-\vek{\ell}_1|)\\ \nn
&& \hspace*{-1.5cm} + \int \frac{\dd^2 \ell_1}{(2 \pi)^2} \int \frac{\dd^2 \ell_2}{(2 \pi)^2}\; T_\kappa(\vek{\ell}_1,\vek{\ell}-\vek{\ell}_1,-\vek{\ell}_2,\vek{\ell}_2-\vek{\ell}) \Bigr] + {\cal O}(\kappa^5) \biggr\}.
}
We provide the technical details of this and the following computations in Appendix \ref{app:transformedps}. Third- or higher-order contributions consist of integrals that run up to infinitely high angular frequencies. The finite resolution of simulations or observational data and our ignorance of e.g. the bispectrum in the highly non-linear regime necessitate a truncation of these integrals, which is achieved by Gaussian smoothing of the convergence maps before the transformation. The Fourier transform of the Gaussian kernel reads
\eq{
\label{eq:gausskernel}
W(\ell) = \expo{- \ell^2 \sigma_c^2/2}\;,
}
where $\sigma_c$ denotes the dispersion of the Gaussian. We will specify $\sigma_c$ in terms of the number of pixels in the convergence map that it covers, where the pixel size is approximately $0.3\,{\rm arcmin}$.

Moreover, equation (\ref{eq:pstrafo_basic_approx}) illustrates that any non-linear transformation entails a more complex dependence on noise. Although we have made the simple assumption that the shape noise is Gaussian distributed and hence fully described by its power spectrum, the transformed power spectrum still receives extra contributions, e.g. from the Gaussian four-point terms. If the distribution of intrinsic ellipticities is not Gaussian \citep[see e.g.][]{vwaer00}, $P_{\bar{\kappa}}(\ell)$ will also contain its higher moments. In the cases without shape noise that we will consider, shot noise caused by the discrete summation over particle positions in the simulation to obtain the convergence might become relevant. This particle shot noise should be approximately Gaussian, and we will thus represent both shape and shot noise by a scale-independent power spectrum $P_{\rm noise}$.

Taking the effects of smoothing and noise into account, one arrives at the following model for the transformed convergence power spectrum,
\eqa{
\label{eq:pstrafo_full_approx}
P_{\bar{\kappa}}(\ell) &\approx& a^{2\lambda-2} \bc{ P_\kappa(\ell) + P_{\rm noise}}\; W^2(\ell)\\ \nn
&& +\; (\lambda-1)\; a^{2\lambda-3}\; W(\ell) \int_0^\infty \frac{\dd \ell_1\; \ell_1}{2\pi}\; W(\ell_1)\\ \nn
&& \; \times \int_0^\pi \frac{\dd \varphi}{\pi} B_\kappa\bb{\ell,\ell_1,\ell_\Delta(\ell,\ell_1,\varphi)} W\bb{\ell_\Delta(\ell,\ell_1,\varphi)}\\ \nn
&& +\; (\lambda-1) (\lambda-2) a^{2\lambda-4} \bc{ P_\kappa(\ell) + P_{\rm noise}} W^2(\ell)\\ \nn
&& \; \times\; \int_0^\infty \frac{\dd \ell_1\; \ell_1}{2 \pi}\; \bc{P_\kappa(\ell_1) + P_{\rm noise}}\; W^2(\ell_1)\\ \nn
&& +\; \frac{(\lambda-1)^2}{2}\; a^{2\lambda-4} \int_0^\infty \frac{\dd \ell_1\; \ell_1}{2\pi}\; \bc{P_\kappa(\ell_1) + P_{\rm noise}}\\ \nn
&& \; \times\; W^2(\ell_1) \int_0^\pi \frac{\dd \varphi}{\pi}\; \bc{P_\kappa\bb{\ell_\Delta(\ell,\ell_1,\varphi)} + P_{\rm noise}}\\ \nn
&& \; \times\; W^2\bb{\ell_\Delta(\ell,\ell_1,\varphi)} + {\cal H}(\ell)\;,
}
where we defined
\eq{
\label{eq:thirdell}
\ell_\Delta(\ell,\ell_1,\varphi) = \sqrt{\ell^2 + \ell_1^2 - 2 \ell \ell_1 \cos \varphi}\;
}
for convenience, and introduced a term ${\cal H}(\ell)$ which includes a number of higher-order contributions as detailed below. The smoothing kernel, $W$, now suppresses the remaining integrands exponentially for large values of angular frequencies. If $P_{\rm noise}$ attains values of the same order of magnitude as the original convergence power spectrum over a range of angular frequencies which are well outside the smoothing regime, it yields significant contributions in particular to the four-point term.

A priori it is not guaranteed that the connected trispectrum terms or higher-order correlations are small; indeed we find that including terms only up to the Gaussian four-point level results in substantially biased parameter constraints. Thus we incorporate a number of higher-order contributions into the model, subsumed into the term ${\cal H}(\ell)$. Here, we only provide a brief synopsis of the calculation of ${\cal H}$, deferring the technical details to Appendix \ref{app:higherorder}.

While results for the trispectrum from tree-level perturbation exist in the literature \citep{fry84} and higher orders could be derived analogously, their non-linear evolution is very likely to be important in our modelling, yet unknown to date. As will be discussed further in Section \ref{sec:meanps}, even the modelling of the convergence bispectrum in the mildly non-linear regime introduces already a significant amount of uncertainty. Hence we take a different approach and assume that the original convergence follows a lognormal distribution, which should be reasonably accurate given the good performance of the logarithmic transformations to Gaussianity at the one-point level (see below), and which allows us to proceed analytically.

Equation (\ref{eq:pstrafo_full_approx}) contains all terms up to second order in $P_\kappa$, so that we include all terms proportional to $P_\kappa^3$ under the assumption of multivariate lognormality into ${\cal H}$. This includes the lowest-order contribution to the lognormal trispectrum, the unconnected part of the convergence five-point correlation, and the Gaussian six-point correlation. We find that the connected moments calculated from the lognormal model significantly underestimate the moments of the original convergence fields as measured from the simulations. Therefore we re-calibrate the amplitudes of the different contributions to the model to match the respective simulation signal, thereby implicitly assuming that the angular dependence of the lognormal model is accurate.

\section{Results}
\label{sec:results}

The main goal of this work is to assess the cosmological information contained in the Box-Cox transformed convergence fields, which we will quantify in terms of a figure of merit for the two best-constrained cosmological parameters in weak lensing surveys, $\Omega_{\rm m}$ and $\sigma_8$. In a first step we determine transformations that optimally Gaussianise the one-point distribution of the convergence before testing how well our analytic models recover the simulation results. The performance of the transformed convergence is further investigated via the signal-to-noise, power spectrum covariances, and a likelihood analysis in the $\Omega_{\rm m}-\sigma_8$ plane. We will first work with the idealistic case of convergence maps that only contain low levels of discreteness noise from the underlying N-body simulations, but repeat the analysis adding a realistic amount of shape noise in Section \ref{sec:shapenoise}.

\subsection{Optimal transformations}
\label{sec:trafos}

\begin{figure*}
\begin{minipage}[c]{0.5\textwidth}
\centering
\includegraphics[scale=.33,angle=270]{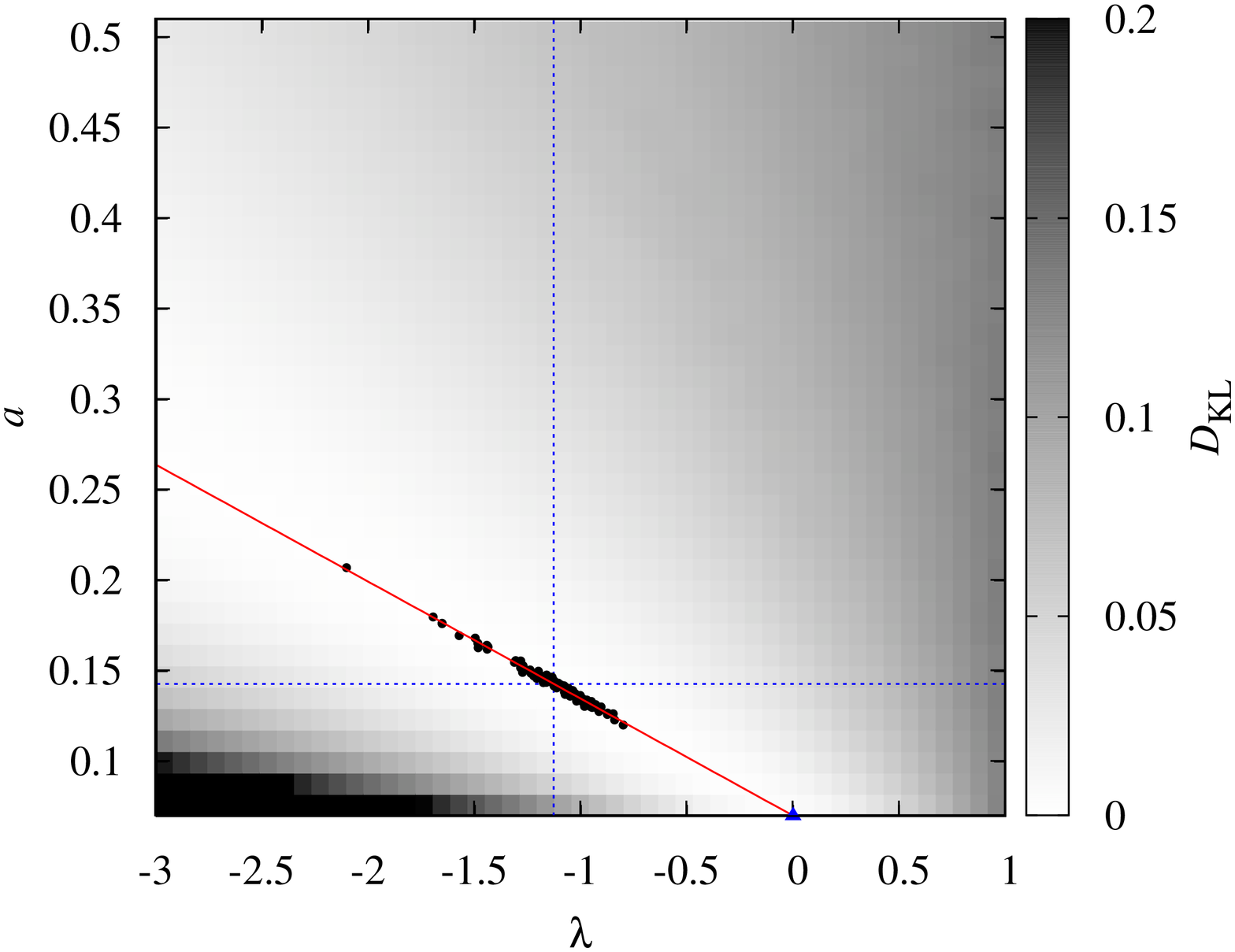}
\end{minipage}%
\begin{minipage}[c]{0.5\textwidth}
\centering
\includegraphics[scale=.33,angle=270]{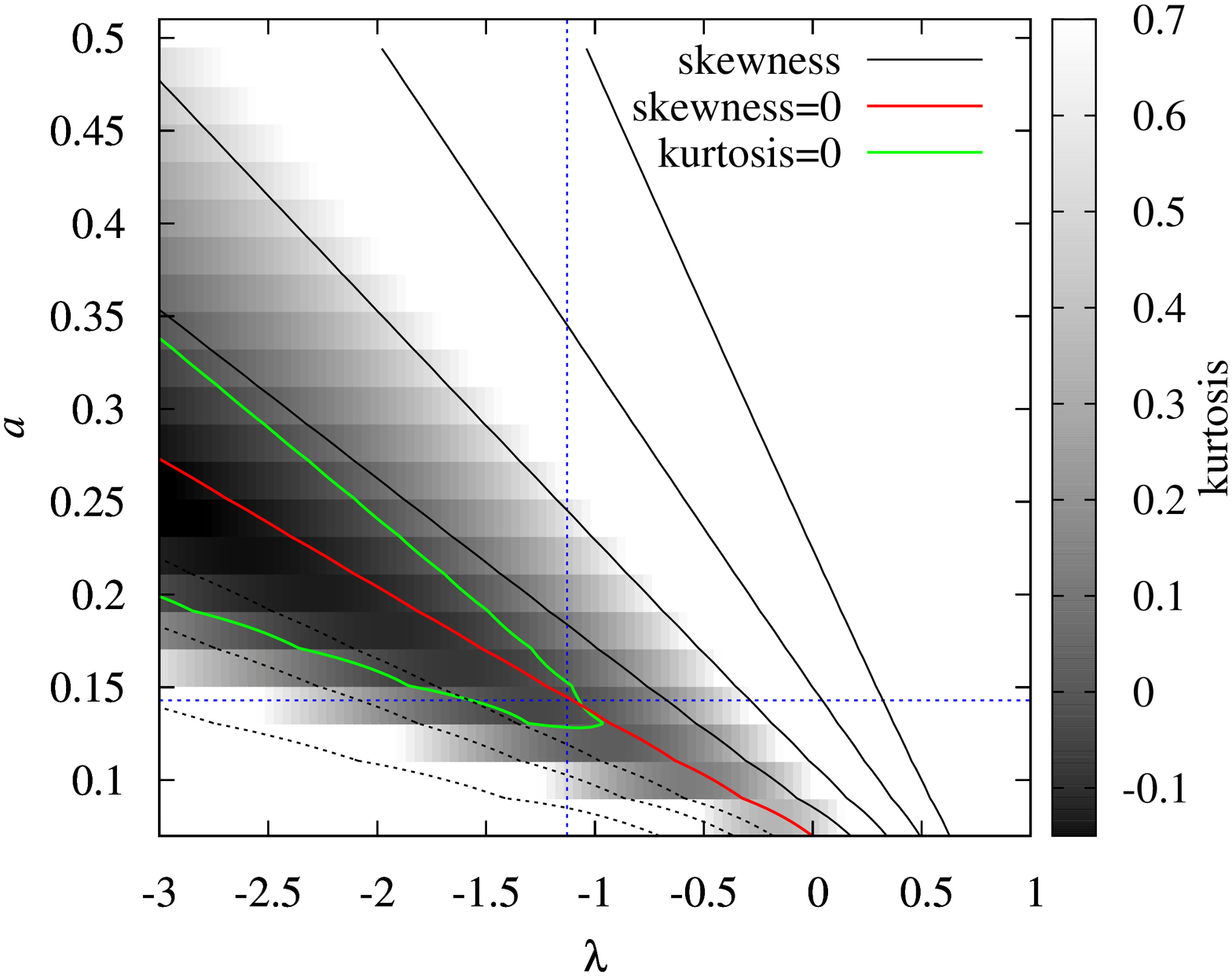}
\end{minipage}
\caption{\textit{Left panel}: Kullback-Leibler divergence $D_{\rm KL}$ as a function of the Box-Cox parameters $\lambda$ and $a$ for one randomly chosen realisation of an unsmoothed convergence field. The grey shading corresponds to $D_{\rm KL}$, as indicated by the colour bar, with smallest values shown in white. We also plot the distribution of optimal Box-Cox parameters $\lambda$ and $a$ determined from each of the 100 realisations of convergence fields as black points. The red line indicates the linear fit (\ref{eq:lambdaafit}). The blue dotted lines indicate the values of $\lambda$ and $a$ used for the Box-Cox transformation BC1. The values of $\lambda$ and $a$ that correspond to the logarithmic transformation are marked by the blue triangle. \textit{Right panel}: Same as above, but for skewness and kurtosis of the Box-Cox transformed convergence field. Black contours correspond to the skewness and are linearly spaced with steps of 0.25, with values of 0 given by the red line and negative values shown as dotted lines. The grey shading corresponds to the kurtosis, as indicated by the colour bar, with values above 0.7 shown in white. The green curve indicates vanishing kurtosis.}
\label{fig:bcplane}
\end{figure*} 

Each convergence map contains $2048^2$ pixels whose convergence values we use as the input data-vector for equation (\ref{eq:lmaxboxcox}) to determine the optimal Box-Cox parameters. To gain further insight and keep the numerics tractable, we compute optimal values for $\lambda$ and $a$ for each realisation individually and obtain the final pair of Box-Cox parameters by taking the mean over the 100 realisations.

\begin{table}
\centering
\caption{Overview on the transformations used in this work. In each case we consider a Box-Cox transformation with optimised parameters and a logarithmic transformation (which corresponds to $\lambda=0$. Parameters are determined from the unsmoothed convergence fields, from fields smoothed by a Gaussian of width 5 pixels, and from fields with shape noise added. The identifiers used in the remainder of this paper, and the values of the Box-Cox parameters $\lambda$ and $a$ are also listed.}
\begin{tabular}[t]{lllrr}
\hline\hline
transformation & convergence map & identifier & $\lambda$ & $a$ \\
\hline
Box-Cox        & unsmoothed      & BC1        & -1.13     & 0.14\\
logarithmic    & unsmoothed      & LOG1       & 0         & 0.07\\
Box-Cox        & smoothed        & BC2        & -2.20     & 0.08\\
logarithmic    & smoothed        & LOG2       & 0         & 0.03\\
Box-Cox        & shape noise     & BCs        & -7.47     & 0.62\\
logarithmic    & shape noise     & LOGs       & 0         & 0.07\\
\hline
\end{tabular}
\label{tab:trafos}
\end{table}

For comparison we also investigate logarithmic transformations, i.e. $\lambda=0$ in the parametrisation given by equation (\ref{eq:boxcox}), where we determine the shift $a$ to be slightly larger than the absolute value of the minimum convergence in all 100 realisations. Since it is clear that the convergence maps need to be smoothed for the likelihood analysis, the question arises whether the transformation parameters shall be estimated from the untreated or the smoothed fields. We will investigate both cases, where throughout a smoothing kernel with a width of 5 pixels is used which, as will be demonstrated below, is suited to suppress noise and modelling uncertainty on small scales.

\begin{figure*}
\begin{minipage}[c]{0.5\textwidth}
\centering
\includegraphics[scale=.33,angle=270]{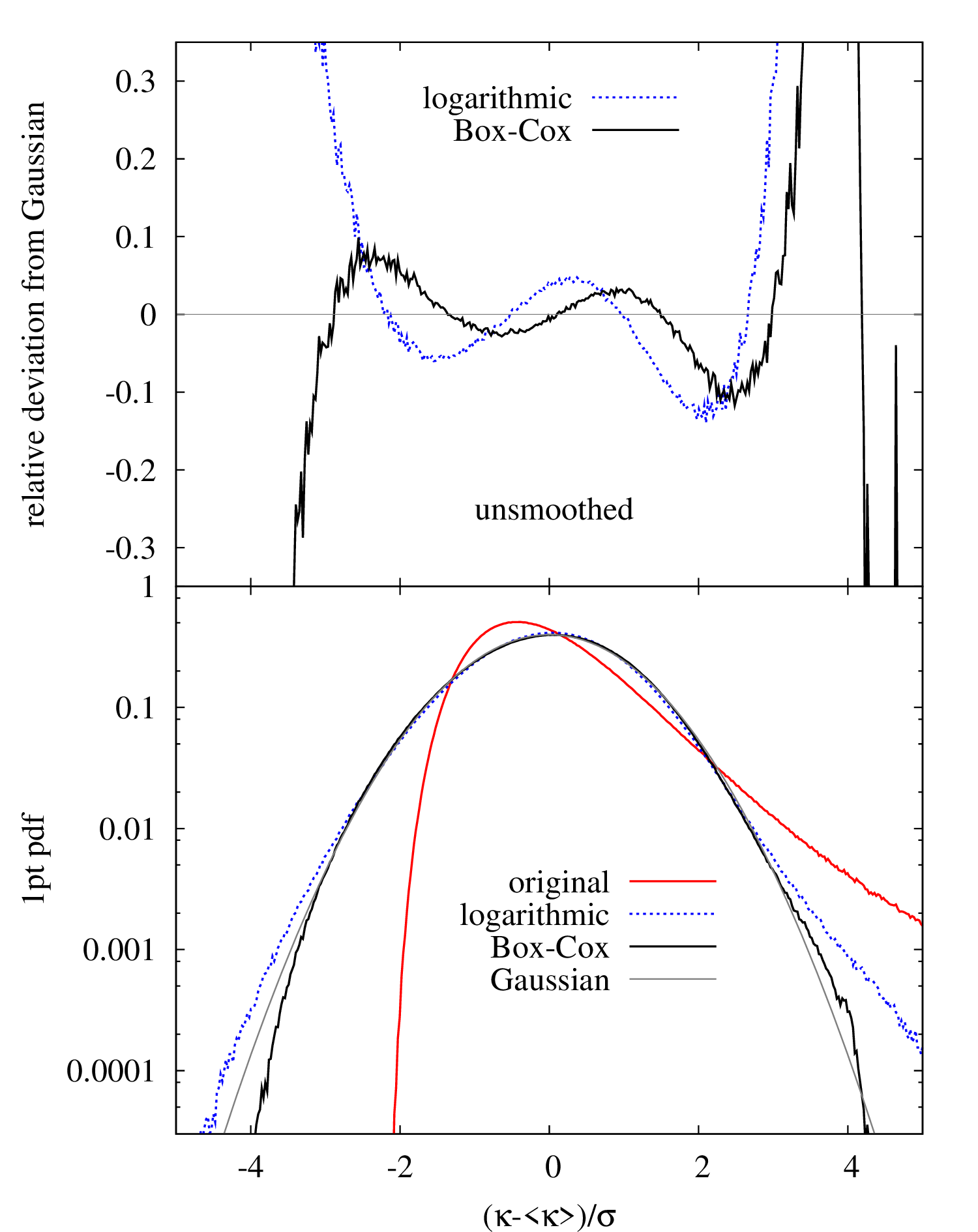}
\end{minipage}%
\begin{minipage}[c]{0.5\textwidth}
\centering
\includegraphics[scale=.33,angle=270]{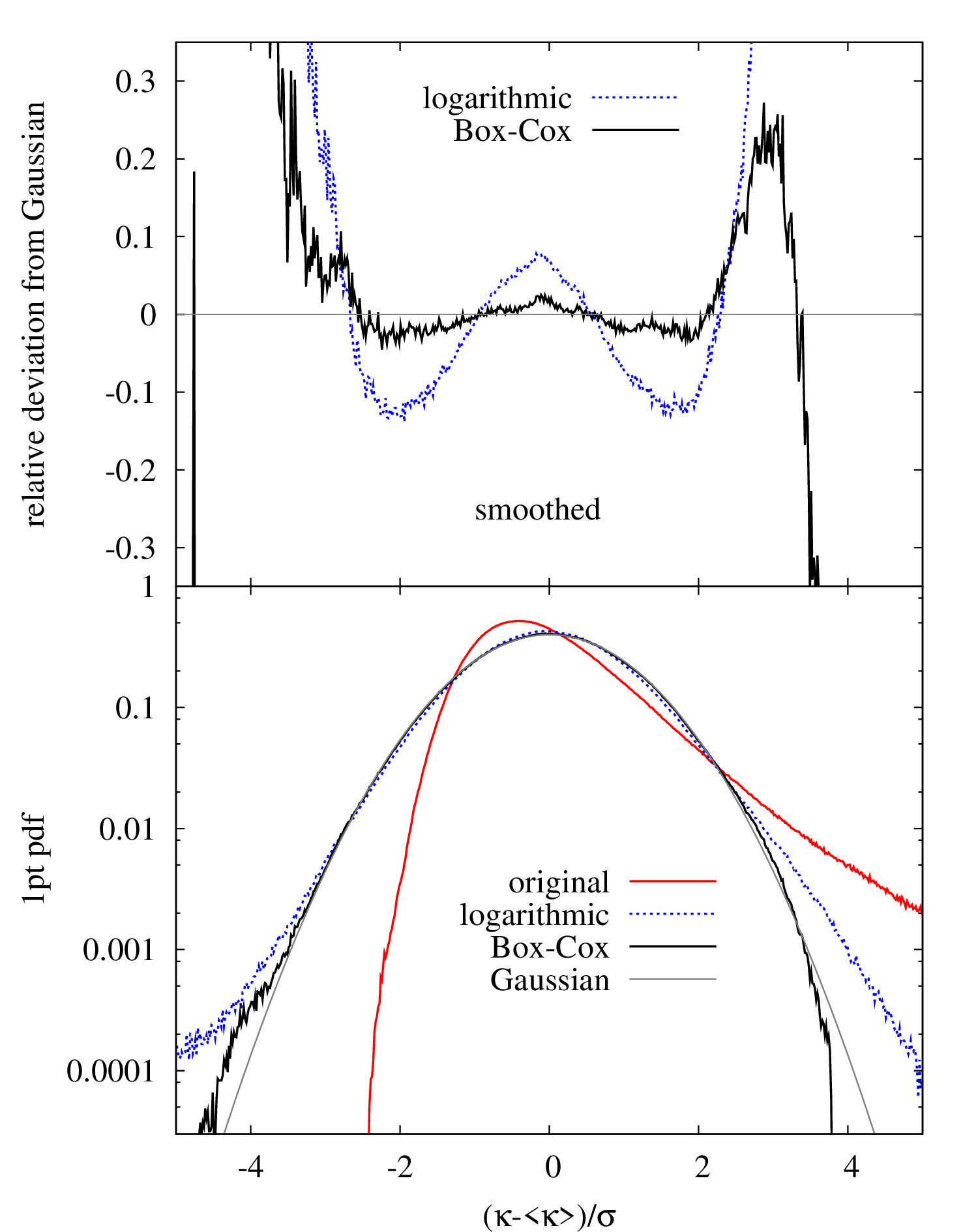}
\end{minipage}
\caption{One-point distribution of convergence values. The bottom panels show the convergence distribution, stacked for 5 randomly chosen realisations of convergence fields and corrected to zero mean and unit variance. Shown is the original distribution of the convergence as red solid line, the logarithmically transformed convergence as blue dotted line, and the Box-Cox transformed convergence as black solid line. For reference a Gaussian distribution is shown in grey. The top panels display the relative deviation from the Gaussian distribution after logarithmic and Box-Cox transformation. \textit{Left panel}: For the unsmoothed convergence. \textit{Right panel}: For the convergence smoothed with a Gaussian kernel of width 5 pixels.}
\label{fig:kappatrafo}
\end{figure*} 

In Fig.$\,$\ref{fig:bcplane} we illustrate the dependence of the properties of the one-point distribution of the transformed convergence on $\lambda$ and $a$. As diagnostics we use the skewness, excess kurtosis, and the Kullback-Leibler divergence $D_{\rm KL}$ between the convergence distribution and a Gaussian with the same mean and variance, defined as
\eq{ 
\label{eq:dkl}
D_{\rm KL} = \int \dd \kappa\; {\cal P}_{\rm 1pt}(\kappa)\; \ln \frac{{\cal P}_{\rm 1pt}(\kappa)}{{\cal P}_{\rm Gauss}(\kappa)}\;,
}
and likewise for the transformed convergence. Note that we place the Gaussian reference distribution in the denominator because it has infinite support. All considered quantities approach zero as the transformed convergence becomes more Gaussian. 

The Kullback-Leibler divergence obtains its minimum along a linear degeneracy line in the $\lambda-a$ plane. All optimal Box-Cox parameters determined from the 100 realisations come to lie close to this minimum, their distribution being excellently fit by the line
\eqa{
\label{eq:lambdaafit}
a \!&=&\! m \lambda + b ~~~\mbox{with}\\ \nn
&& \hspace*{-1.1cm} m=(-6.45 \pm 0.07) \times 10^{-2}\;; ~b=(7.02 \pm 0.08) \times 10^{-2}\;.
}
Contours of equal skewness are close to straight lines, and the region of vanishing skewness matches the valley in $D_{\rm KL}$. The kurtosis varies along this line, but apparently does not cause a significant deviation from a Gaussian distribution since $D_{\rm KL}$ as a global measure of Gaussianity remains approximately constant. The scatter of optimal Box-Cox parameter values along the degeneracy line is caused by cosmic variance and determined by the varying position of the intersection with the zero-kurtosis contour, which for the realisation used to produce Fig.$\,$\ref{fig:bcplane}, right panel, is close to the mean of $\lambda$ and $a$ taken over all realisations.

\begin{table*}
\centering
\caption{Mean and standard deviation of skewness, kurtosis, and Kullback-Leibler divergence $D_{\rm KL}$ as measured from the 100 realisations of convergence fields. The fields are either unsmoothed, smoothed by a Gaussian of width 5 pixels, or incorporate shape noise. In each case results are given for the original, the logarithmically transformed, and the Box-Cox transformed convergence fields. Values for the log-arctan transformation applied to the noisy convergence fields are listed as well.}
\begin{tabular}[t]{llrrr}
\hline\hline
convergence map & analysis & skewness & kurtosis & $D_{\rm KL}$\\
\hline
 & original    & $2.05 \pm 0.45$                  & $18.77 \pm 22.04$                 & $0.126 \pm 0.006$\\
unsmoothed & logarithmic (LOG1) & $(0.57 \pm 1.92) \times 10^{-2}$ & $(3.93 \pm 0.69) \times 10^{-1}$ & $(3.47 \pm 0.70) \times 10^{-3}$\\
 & Box-Cox (BC1)    & $(0.04 \pm 1.29) \times 10^{-2}$ & $(-2.86 \pm 3.78) \times 10^{-2}$ & $(1.33 \pm 0.18) \times 10^{-3}$\\
\hline
 & original    & $1.80 \pm 0.32$                  & $8.65 \pm 6.24$                 & $0.131 \pm 0.012$\\
smoothed & logarithmic (LOG2) & $(1.02 \pm 0.61) \times 10^{-1}$ & $(5.50 \pm 1.19) \times 10^{-1}$ & $(5.78 \pm 1.40) \times 10^{-3}$\\
 & Box-Cox (BC2)    & $(0.24 \pm 3.94) \times 10^{-2}$ & $(2.37 \pm 5.35) \times 10^{-2}$ & $(1.42 \pm 0.26) \times 10^{-3}$\\
\hline
 & original    & $0.83 \pm 0.17$                  & $3.11 \pm 2.36$                 & $0.038 \pm 0.007$\\
shape noise & logarithmic (LOGs) & $(-4.74 \pm 6.55) \times 10^{-2}$ & $(7.28 \pm 1.48) \times 10^{-1}$ & $(7.49 \pm 1.46) \times 10^{-3}$\\
 & Box-Cox (BCs)    & $(-2.32 \pm 4.31) \times 10^{-2}$ & $(4.33 \pm 0.78) \times 10^{-1}$ & $(4.22 \pm 0.84) \times 10^{-3}$\\
 & log-arctan & $(-5.04 \pm 5.70) \times 10^{-2}$ & $(-1.86 \pm 4.37) \times 10^{-2}$ & $(1.74 \pm 0.29) \times 10^{-3}$\\
\hline
\end{tabular}
\label{tab:skewcurt}
\end{table*}

In Table \ref{tab:trafos} an overview on the different transformations and their parameters is provided. The Box-Cox transformations generally prefer negative values for $\lambda$, e.g. if applied to the unsmoothed convergence fields, the optimum is close to an inverse transformation. Values of $\lambda<0$ imply that high-density peaks in the convergence fields are downweighted even stronger than for a logarithmic transformation (see Fig.$\,$\ref{fig:boxcoxillustrate}). Note that deriving the value of the shift $a$ for the logarithmic transformation from the minimum value of the convergence produces a pair of Box-Cox parameters that is also located in the valley of minimum skewness and $D_{\rm KL}$, as indicated by the blue triangle in the left panel of Fig.$\,$\ref{fig:bcplane}.

The plots of the one-point distribution of $\kappa$ shown in Fig.$\,$\ref{fig:kappatrafo} confirm that both Box-Cox and logarithmic transformation effectively remove the pronounced skewness of the original convergence distribution. The logarithmic transform features a significant deviation from the Gaussian case for values $3\sigma$ and more above the mean which is avoided in the optimal Box-Cox transform by the negative value of $\lambda$. When applied to the unsmoothed fields, the Box-Cox transformed distribution deviates less than $\pm 10\,\%$ in the range $\pm 3\sigma$ around the mean while the logarithmic transformation features slightly larger deviations for $\kappa$ values close to the mean and also differs from the Gaussian more significantly for extreme values of $\kappa$.

Smoothing the convergence field flattens high-density peaks and makes voids more shallow, so that the distribution of original convergence values is modified to look slightly more Gaussian. Nonetheless the transformations we consider perform somewhat worse in rendering the one-point distribution Gaussian, which applies in particular to the logarithmic transformation for $\kappa$ values far from the mean, see the right panel of Fig.$\,$\ref{fig:kappatrafo}. Note that the optimal Box-Cox parameters determined from the smoothed convergence fields follow a similarly well defined linear relation as the one shown in Fig.$\,$\ref{fig:bcplane}, only shifted to more negative values of $\lambda$.

Table \ref{tab:skewcurt} lists the mean and standard deviation, computed from 100 realisations, of skewness, kurtosis, and $D_{\rm KL}$ for the original and transformed convergence fields. For both unsmoothed and smoothed convergence the logarithmic transformation improves all three diagnostics by at least an order of magnitude while the Box-Cox transformation adds another factor of 10 reduction in skewness and kurtosis. The Kullback-Leibler divergence decreases less when switching from logarithmic to Box-Cox transformation, by factors of 2.6 and 4.1, respectively. To illustrate the absolute values of $D_{\rm KL}$, one can compare them to $D_{\rm KL}$ for two Gaussian distributions with identical variance but shifted means. We find that $D_{\rm KL}=0.1$, as found for the original convergence distribution, corresponds to a shift in the mean of half a standard deviation. Similarly, one obtains shifts of $0.1 \sigma$ ($0.04 \sigma$) for $D_{\rm KL}=5 \times 10^{-3}$ ($D_{\rm KL}= 10^{-3}$), which is of the same order as the results for the logarithmic and Box-Cox transformations, respectively.

\subsection{Modelling accuracy}
\label{sec:meanps}

\begin{figure}
\centering
\includegraphics[scale=.33,angle=270]{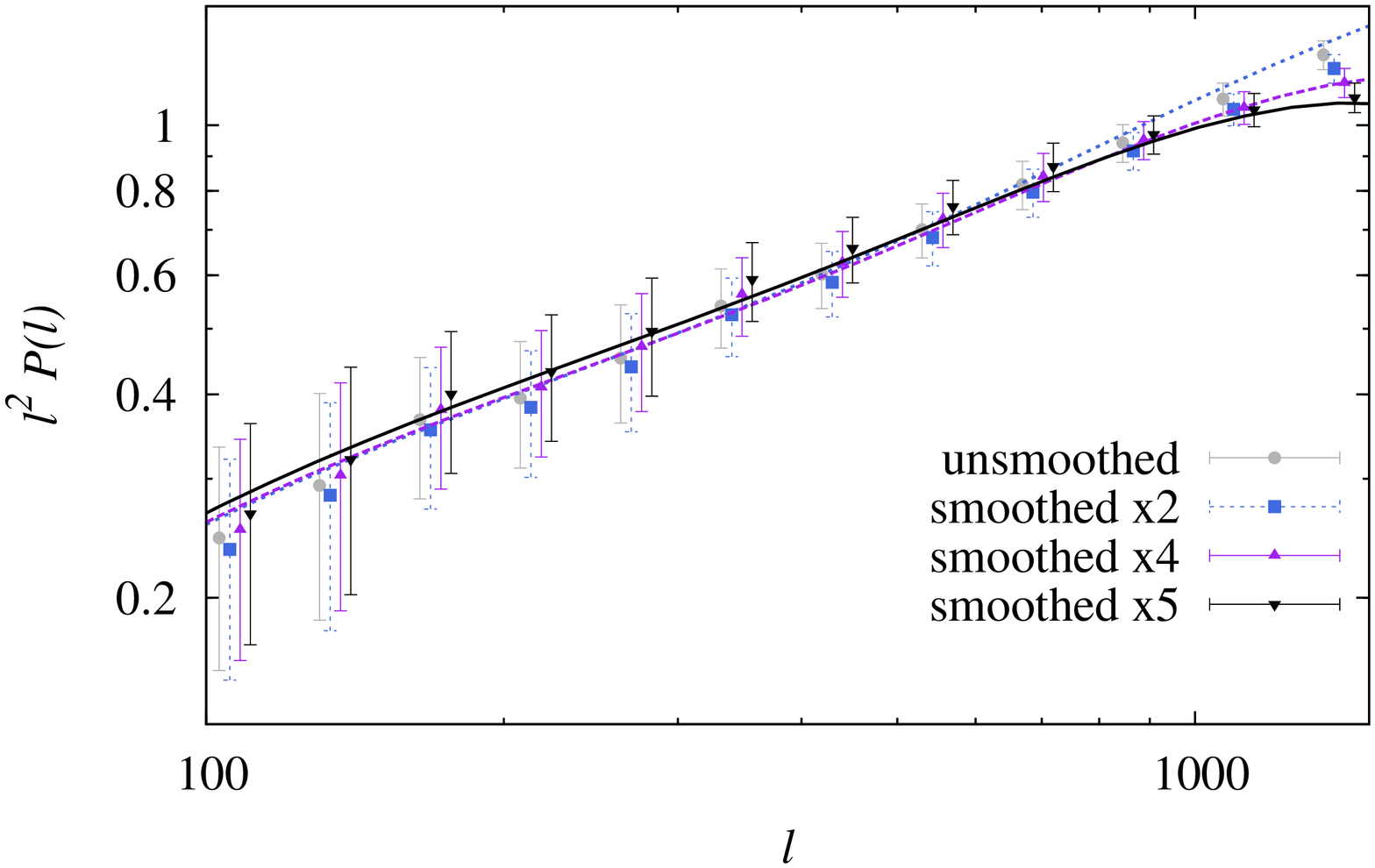}
\includegraphics[scale=.33,angle=270]{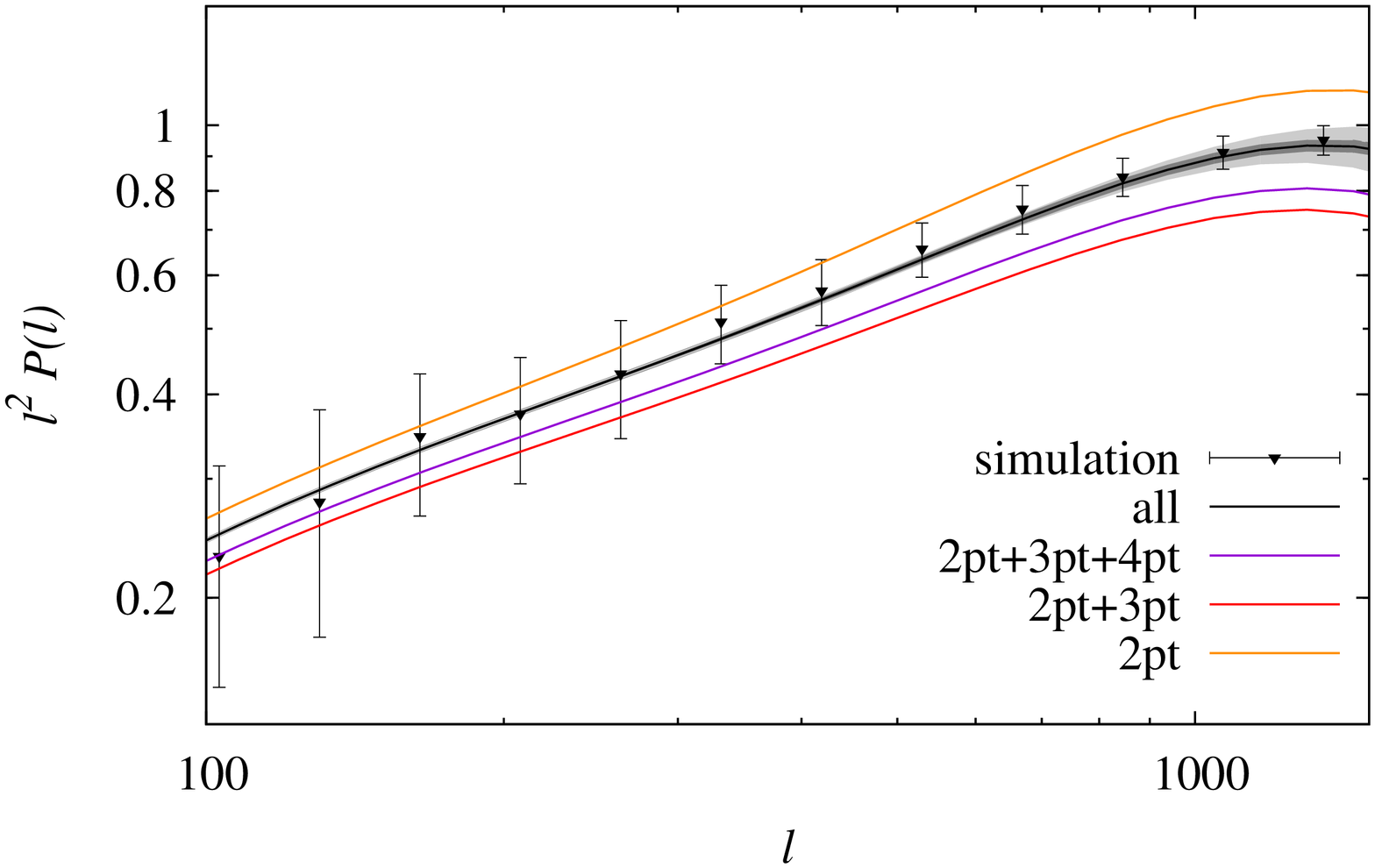}
\includegraphics[scale=.33,angle=270]{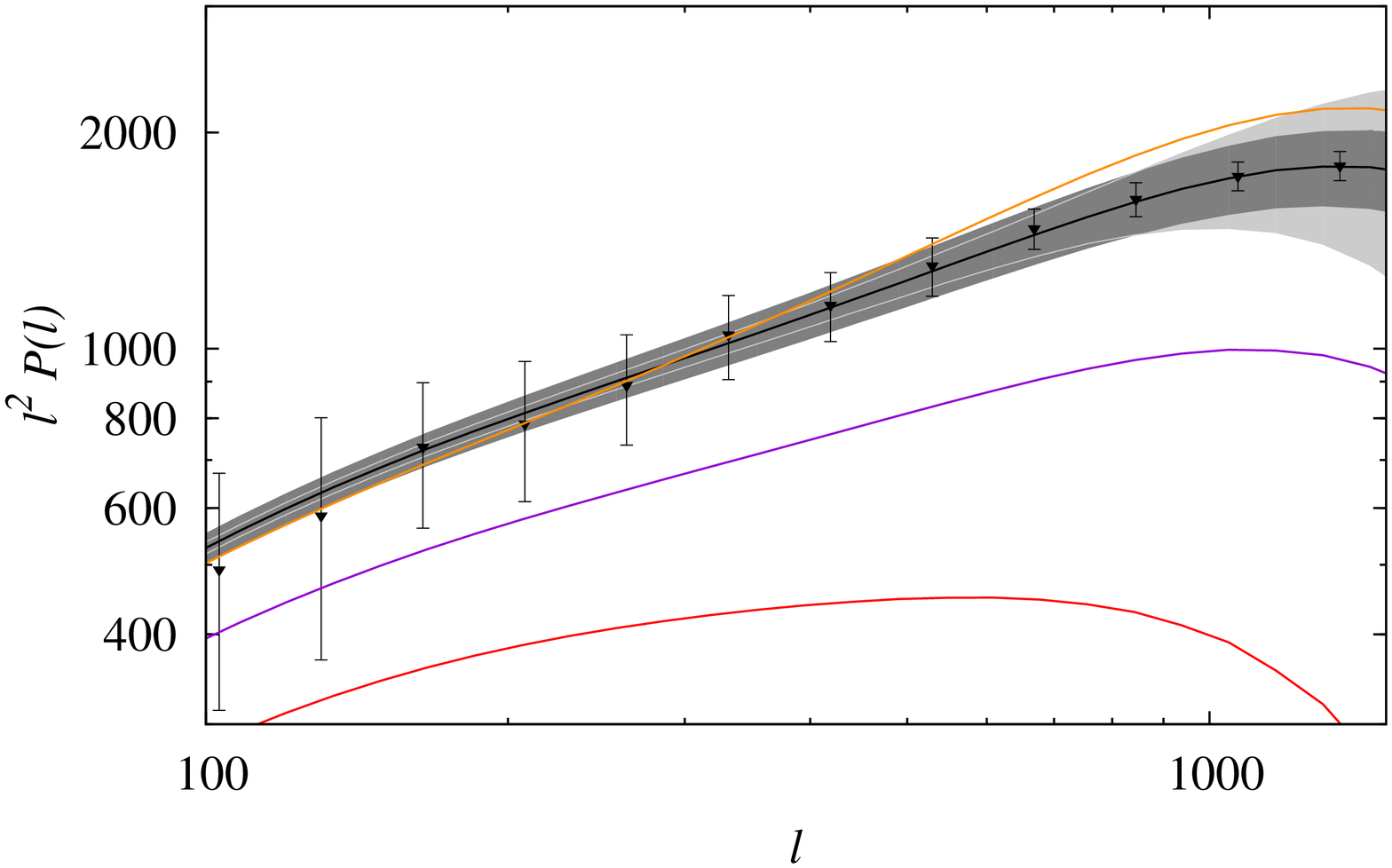}
\caption{\textit{Top panel}: Box-Cox transformed convergence power spectrum obtained from the mean of the 100 realisations. Grey circles correspond to the power spectrum computed from the unsmoothed convergence, blue squares (purple upward triangles; black downward triangles) to the power spectrum obtained from the convergence smoothed with a Gaussian kernel of width 2 (4; 5) pixels. The corresponding models for the smoothed power spectra are shown as blue dotted lines (width 2 pixels), purple dashed lines (width 4 pixels), and black solid lines (width 5 pixels). Note that points and curves have been slightly offset horizontally for clarity. \textit{Centre panel}: Contributions to the smoothed (5 pixels) Box-Cox transformed power spectrum (BC1). The model including up to two-point (three-point; Gaussian four-point; all) terms is shown as orange (red; violet; black) curve. Triangles indicate the simulation results, cf. the top panel. The light grey area indicates the uncertainty due to the modelling of the bispectrum in the non-linear regime, computed from equation (\ref{eq:modbispec}). The dark grey area shows the variation in the model resulting from the normalisation of the trispectrum contribution $r_4$; see equation (\ref{eq:higherordersummary}). \textit{Bottom panel}: Same as in the centre panel, but based on the BC2 transformation.}
\label{fig:meanps}
\end{figure} 

To model the power spectra calculated from the Box-Cox and logarithmically transformed convergence fields according to equation (\ref{eq:pstrafo_full_approx}), convergence power spectra and bispectra are required. We compute the matter power spectrum $P_\delta(k)$ for the simulation cosmology, employing the transfer function by \citet{eisenstein98} and the correction for the non-linear regime by \citet{smith03}. As was demonstrated in \citet{kiessling11}, our model power spectra match the simulation results well in the relevant angular frequency regime. 

The convergence power spectrum is then given by the Limber equation \citep{kaiser92}
\eq{
\label{eq:pslimber}
P_\kappa(\ell) = \frac{9H_0^4 \Omega_{\rm m}^2}{4 c^4} \int^{\chi_{\rm hor}}_0 \dd \chi\; \frac{g^2(\chi)}{a^2(\chi)}\; P_\delta \br{\frac{\ell}{\chi},\chi}\;,
}
where we used the lensing efficiency $g(\chi)=1-\chi/\chi(z_{\rm s})$ with $z_{\rm s}=1$. The analogous equation for the convergence bispectrum reads \citep[e.g.][]{takada04}
\eqa{
\label{eq:bispectrumlimber}
B_\kappa(\ell_1,\ell_2,\ell_3) &=& \frac{27H_0^6 \Omega_{\rm m}^3}{8 c^6} \int^{\chi_{\rm hor}}_0 \dd \chi\; \frac{g^3(\chi)}{\chi\; a^3(\chi)}\\ \nn
&& \times\; B_\delta \br{\frac{\ell_1}{\chi},\frac{\ell_2}{\chi},\frac{\ell_3}{\chi},\chi}\;.
}
The matter bispectrum $B_\delta(k_1,k_2,k_3)$ is computed via perturbation theory \citep{fry84} from the matter power spectrum, applying the corrections due to non-linear structure evolution given in \citet{scoccimarro01}. To allow for efficient interpolation, we calculate $B_\kappa$ in practice as a function of two triangle side lengths $\ell_1, \ell_2$ and their internal angle $\varphi$, with dense binning between $\ell=1$ and $\ell \sim 10,000$ (where the smoothing has safely suppressed all contributions to zero), and for $\varphi \in \bb{0;\pi}$.

\citet{kiessling11} found that in the simulations under consideration angular frequencies larger than $\ell \sim 1500$ are significantly affected by particle shot noise and thus discarded these scales in their cosmological analysis, so that we can safely choose a smoothing scale that downweights scales $\ell > 1500$. Besides we have to make sure that we limit our study to sufficiently large scales on which higher-order correlations which we are not able to model have not yet become important.

In Fig.$\,$\ref{fig:meanps}, top panel, the mean simulation power spectrum using transformation BC1 (for the definition of identifiers see Table \ref{tab:trafos}), averaged over 100 realisations, is shown without any smoothing as well as for smoothing with kernels of width 2, 4, and 5 times the pixel size of the convergence map of $0.3\,{\rm arcmin}$. Note that in this and all similar figures we use the error bars corresponding to a single realisation, i.e. a $100\,{\rm deg}^2$ patch; errors on the mean from all realisations are smaller by a factor of 10. In addition we plot the power spectrum models obtained by using the corresponding smoothing window. Increasing the smoothing scale boosts the simulation signal on large scales and significantly reduces it at $\ell \sim 1000$ and above. The model is biased high at high angular frequencies for the 2-pixel kernel, but provides a good fit to the simulation data on all scales for the 4- and 5-pixel kernels.

Using a narrow smoothing kernel, one includes more information from highly non-linear scales into the transformed power spectrum model for which the prescription of the bispectrum and the included higher-order contributions becomes insecure and the neglected higher-order statistics more important, hence the bias. Henceforth we will adopt a kernel width of 5 pixels (corresponding to $1.5\,{\rm arcmin}$) which balances the systematic offset due to inaccurate modelling and the suppression of cosmological information at high angular frequencies, visible in the decrease in the amplitude of the transformed power spectrum setting in at increasingly smaller $\ell$.

Particle shot noise is incorporated throughout in our models, computed via the analytical formula given in \citet{kiessling11}, equation (15), which yields $P_{\rm noise} \approx 2.4 \times 10^{-12}$. The 5-pixel window effectively downweights the regime where shot noise becomes important, so that the models are affected by less than $3\,\%$ in the range $100 < \ell < 1500$. 

The centre panel of Fig.$\,$\ref{fig:meanps} details the contributions of terms with different orders of $\kappa$ to the model of the transformed power spectrum, again for transformation BC1. The two-point term is simply a rescaled version of the original convergence power spectrum while the three-point contribution is negative due to $\lambda<0$, see equation (\ref{eq:pstrafo_full_approx}), and has a stronger effect at high angular frequencies. The Gaussian four-point term adds to the model almost constantly over the range of $\ell$ considered.

The combined higher-order contribution is also positive and surpasses the Gaussian four-point term in amplitude (although the latter is second order in $P_\kappa$), in particular on small scales. The full model yields a good fit to the simulation, being marginally low at high angular frequencies, but note that in this regime error bars are significantly correlated.

We construct a toy model to estimate the influence of the limited accuracy of modelling the non-linear matter bispectrum by the \citet{scoccimarro01} fitting formula. A multiplicative term $f$ is introduced which modifies the convergence bispectrum to $B(\ell_1,\ell_2,\ell_3) \rightarrow B(\ell_1,\ell_2,\ell_3)\, f(\ell_1,\ell_2)$. \citet{scoccimarro01} found little dependence of the accuracy of their fit on the internal angle of a triangle of angular frequencies, so that, without loss of generality, we assume $f$ to only depend on $\ell_1$ and $\ell_2$. Due to lack of information about any further dependence on triangle shapes, we furthermore assume that any deviation of the fit can be phrased in terms of the mean side length $\bar{\ell}=1/2(\ell_1+\ell_2)$. Judging from the plots in \citet{scoccimarro01}, the formula fits their $\Lambda$CDM simulations well up to $k \sim 0.5h^{-1}\,{\rm Mpc}$. We translate this into a scale $\bar{\ell} \sim 700$ as the sensitivity of lensing peaks at about half the source distance at $z_{\rm s}=1$. 

\begin{figure*} 
\begin{minipage}[c]{0.5\textwidth}
\centering
\includegraphics[scale=.35,angle=270]{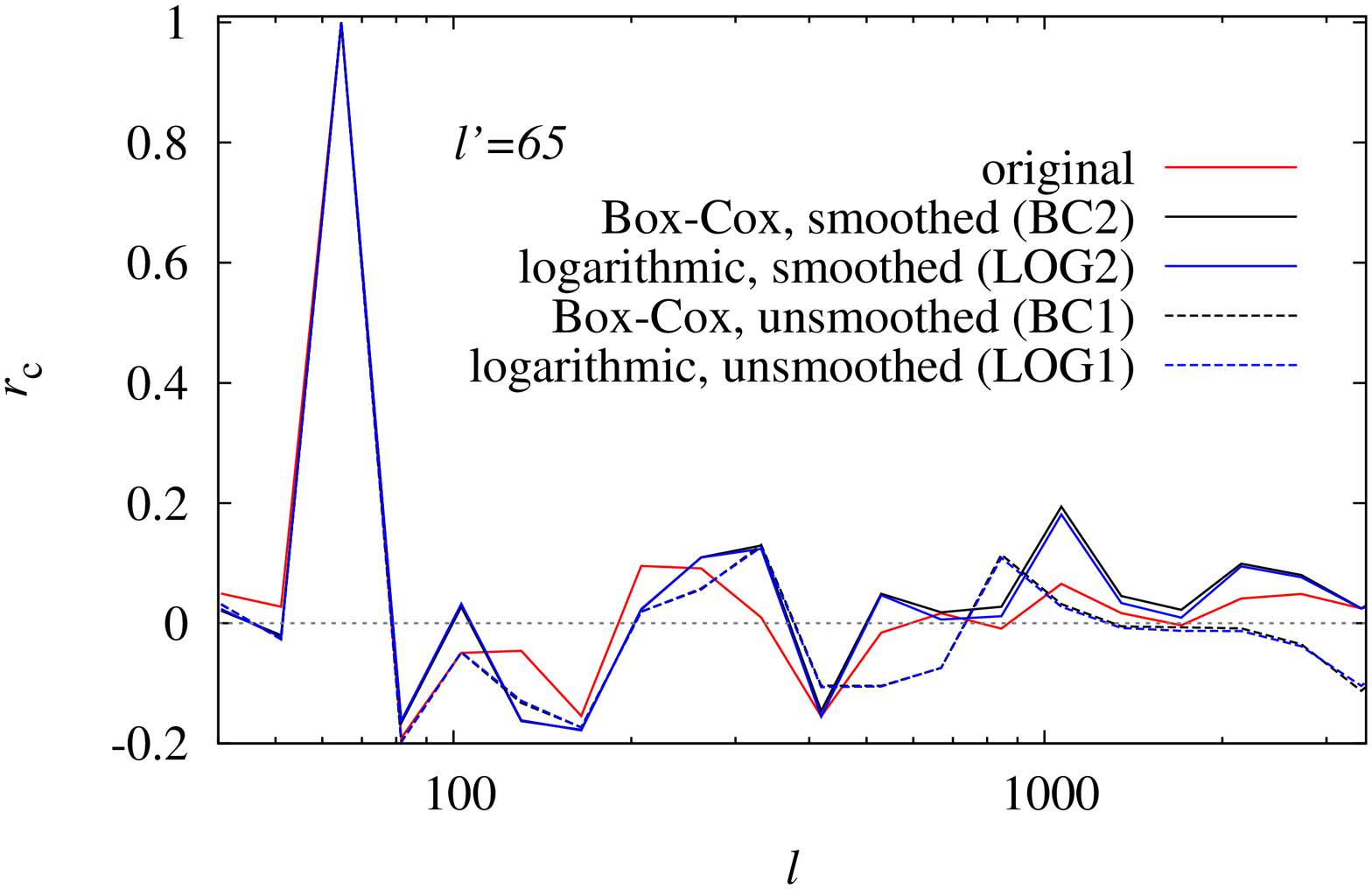}
\end{minipage}%
\begin{minipage}[c]{0.5\textwidth}
\centering
\includegraphics[scale=.35,angle=270]{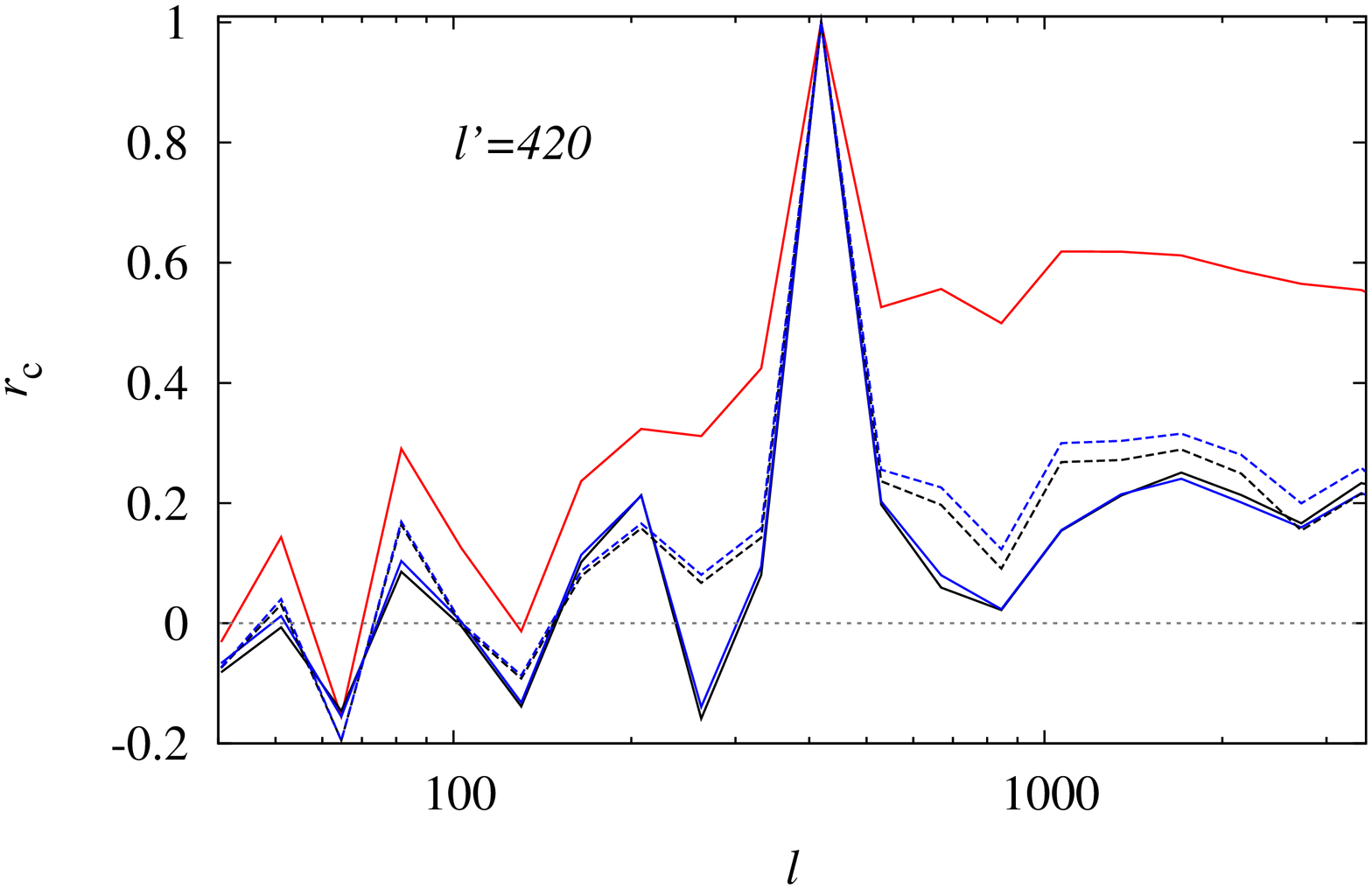}
\end{minipage}\\
\begin{minipage}[c]{0.5\textwidth}
\centering
\includegraphics[scale=.35,angle=270]{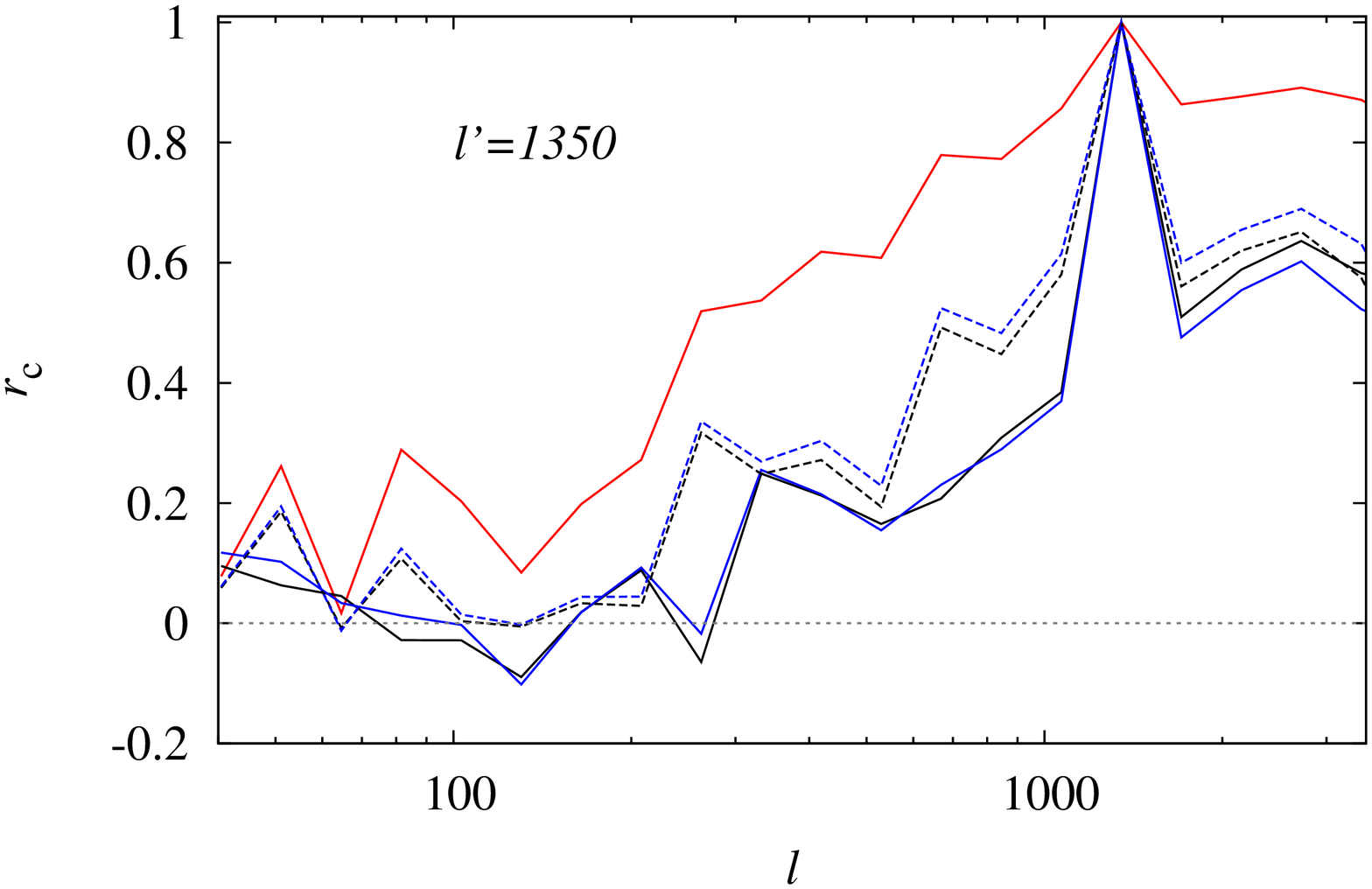}
\end{minipage}%
\begin{minipage}[c]{0.5\textwidth}
\centering
\includegraphics[scale=.35,angle=270]{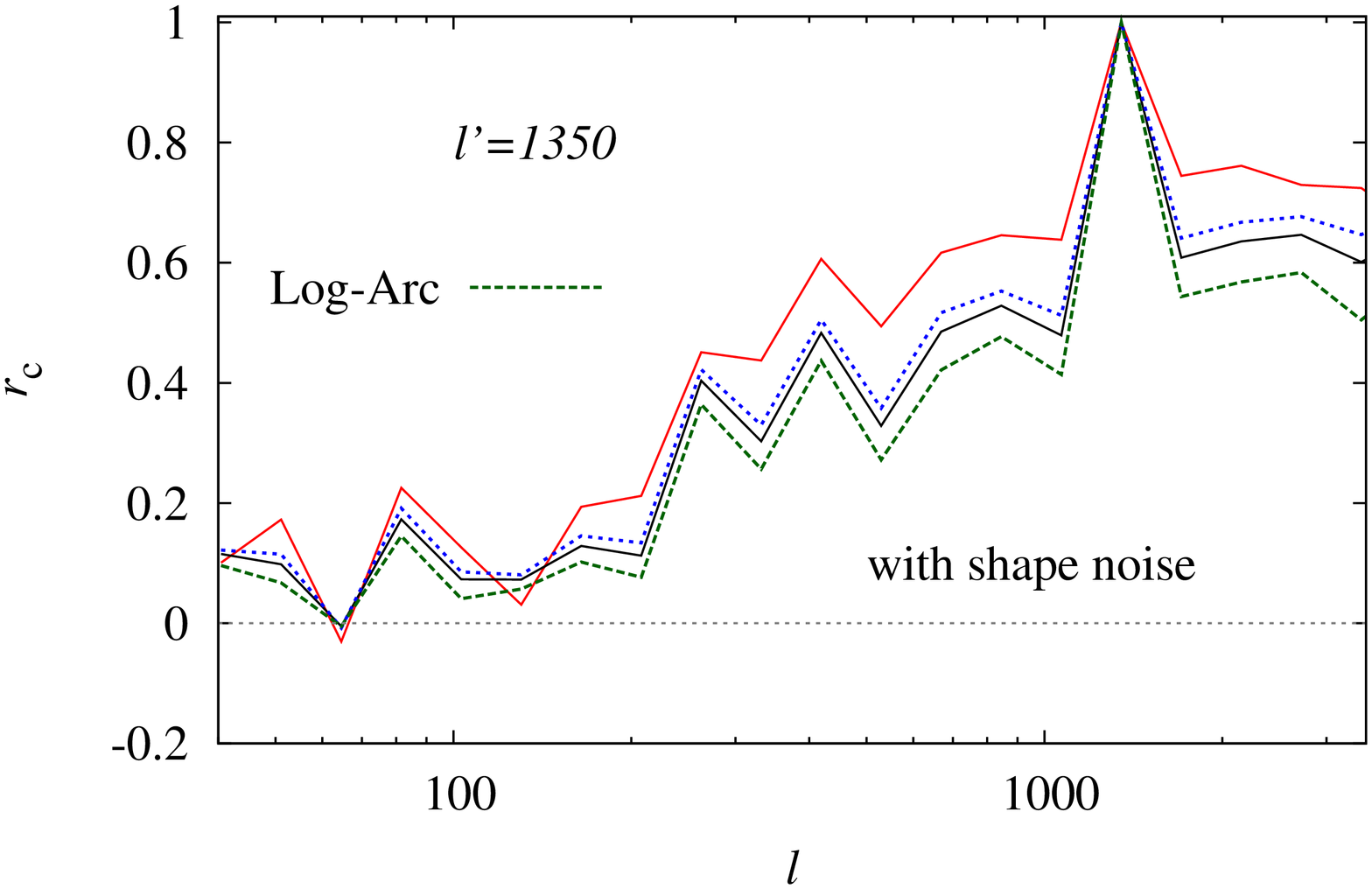}
\end{minipage}
\caption{\textit{Top panels}: Correlation coefficients $r_{\rm c}$ for the power spectrum covariance, computed from the original convergence fields (red solid lines), the Box-Cox transformed fields (black lines), and the logarithmically transformed fields (blue lines). In the two latter cases solid curves were obtained for the transformations determined from the smoothed convergence fields (BC2,LOG2), and dotted curves for the transformations determined from the unsmoothed fields (BC1, LOG1). In the left panel the correlation at low angular frequencies (around $\ell \approx 65$) is shown; in the right panel the correlation for medium high angular frequencies (around $\ell \approx 420$). \textit{Bottom panels}: Same as above, but for angular frequencies around $\ell \approx 1350$, close to the maximum used for the likelihood analysis. The left panel, like the top panels, shows results obtained without shape noise, while in the right panel shape noise is included, using the corresponding transformation parameters. We additionally show the results for the log-arctan transformation discussed in Section \ref{sec:shapenoise} as green dashed curves.}
\label{fig:pscorrelation}
\end{figure*} 

The discrepancy between fit formula and simulation seems to increase linearly at high wavenumbers, with an average accuracy of $15\,\%$ \citep{scoccimarro01}, so that we define
\eq{
\label{eq:modbispec}
f(\ell_1,\ell_2) = \left\{ \begin{array}{ll} 1 & \bar{\ell} < 700\\ 1 \pm 0.15 \br{\frac{\bar{\ell}}{700}-1} & \bar{\ell} \geq 700 \end{array} \right.\;.
}
Although the fit formula persistently underestimates the simulations used in that work, we understand the model in equation (\ref{eq:modbispec}) as a rough estimate for the general accuracy of the fit and consider both positive and negative deviations from the formula, which leads to the light grey regime of uncertainty in the transformed convergence power spectrum shown in Fig.$\,$\ref{fig:meanps}. At high $\ell$ this uncertainty amounts to about $\pm 5\,\%$ for the BC1 transformation and is thus of the same order as the Gaussian four-point contribution.

The dark grey regions shown in Fig.$\,$\ref{fig:meanps} correspond to the uncertainty induced by the measurement error of the connected fourth moment, entering $r_4$, which we determine from the simulations to normalise the lognormal trispectrum contribution in ${\cal H}(\ell)$; see equations (\ref{eq:pstrafo_full_approx}) and (\ref{eq:higherordersummary}). For the BC1 transformation the higher-order terms are small and so is the uncertainty due to $r_4$.

The bottom panel of Fig.$\,$\ref{fig:meanps} displays the model details for the Box-Cox transformation BC2 (determined from the smoothed convergence fields). Since compared to BC1 $\lambda$ is more negative and $a$ closer to zero, the higher-order contributions are boosted much stronger, which entails a larger impact of the uncertainty in the bispectrum and $r_4$, the latter having a weaker dependence on $\ell$ and thus dominating on large and intermediate scales. Despite these substantial sources of uncertainty and the high amplitudes of each of the three-point, four-point, and ${\cal H}$ terms, the full model provides an excellent fit to the simulation power spectrum also in the BC2 case.

\subsection{Noise properties}
\label{sec:sn}

If Gaussianising the one-point distribution of the convergence succeeds in turning convergence maps into approximative realisations of a Gaussian random field, one expects that the covariance of the convergence power spectrum is diagonal. Conversely, any significant cross-correlation between angular frequencies is a clear sign for a non-zero trispectrum \citep[e.g.][]{pielorz10}. We determine the power spectrum covariance
\eq{
\label{eq:covP}
{\rm Cov}_P (\ell,\ell') = \ba{\hat{P}_{\kappa}(\ell)\; \hat{P}_{\kappa}(\ell')} -\ba{\hat{P}_{\kappa}(\ell)}\;\ba{\hat{P}_{\kappa}(\ell')}\;
}
from the simulations, where angular brackets denote the average over the 100 realisations, and subsequently correlation coefficients
\eq{
\label{eq:rcorr}
r_{\rm c}(\ell,\ell') = \frac{{\rm Cov}_P (\ell,\ell')}{\sqrt{{\rm Cov}_P (\ell,\ell)\; {\rm Cov}_P (\ell',\ell')}}\;.
}
These and the following equations all hold likewise for the transformed power spectra.

In Fig.$\,$\ref{fig:pscorrelation} we show $r_{\rm c}$ for the original as well as for all transformed convergence fields. The power spectra transformed according to BC1 and LOG1 have been computed from the unsmoothed fields and those transformed according to BC2 and LOG2 from the smoothed convergence, i.e. the transformations have been applied to the cases where they should work optimally. The original convergence power spectrum features significant positive cross-correlations for $\ell,\ell' > 200$ which rise up to $r_{\rm c}=0.9$ for $\ell,\ell' > 1000$. Box-Cox and logarithmic transformations perform almost identically and reduce these correlations substantially, yet in neither case to a negligible level.

These findings are in disagreement with the results of \citet{seo11} who obtained a level of cross-correlations that is consistent with zero after a logarithmic transformation of the convergence. However, $r_{\rm c}$ does not exceed 0.4 even for their original power spectra, but direct comparison is hindered by the different angular frequency binning which affects the Gaussian contribution to the diagonal of the covariance and thereby the normalisation of $r_{\rm c}$. The parameters of the underlying simulations are similar to ours, except for considerably lower cosmological parameter values $\Omega_{\rm m}=0.24$ and $\sigma_8=0.76$ \citep{sato09}. Thus non-linear clustering might be less pronounced in these simulations and hence their mode-coupling effects easier to remove.

\begin{figure}
\centering
\includegraphics[scale=.34,angle=270]{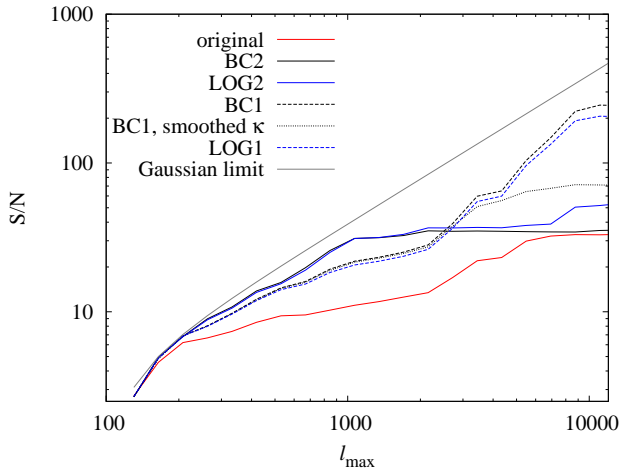}
\caption{Cumulative S/N as a function of the maximum angular frequency included. Results for the original (Box-Cox transformed; logarithmically transformed) convergence fields are shown as red (black; blue) curves. In the case of the transformed fields solid lines correspond to using parameters determined from the smoothed fields (BC2, LOG2), and dashed lines to parameters obtained from the unsmoothed fields (BC1, LOG1). The effect on the S/N of smoothing with a Gaussian kernel of width 5 pixels is indicated for the Box-Cox (BC1) transformed case by the black dotted line. For reference the Gaussian limit resulting from uncorrelated angular frequencies is shown as grey curve.}
\label{fig:signaltonoise}
\end{figure} 

A first insight into the information content is given by the cumulative signal-to-noise (S/N), defined by
\eq{
\label{eq:sn}
{\rm S/N}(\ell_{\rm max}) = \sqrt{\sum_{\ell,\ell'}^{\ell_{\rm max}} P_\kappa(\ell)\; {\rm Cov}_P^{-1} (\ell,\ell')\; P_\kappa(\ell')}\;.
}
Note that we employ the correction factor given in \citet{hartlap06} to get an unbiased estimate of the inverse covariance in the presence of simulation noise. This step removes the bias introduced when using the inverse of the sample covariance as an estimator for the inverse; see \citet{anderson03} for details.

We follow \citet{seo11} in using the maximum S/N, achieved in the limit of a Gaussian random field and given by the total number of independent modes, as a reference. As is demonstrated in Fig.$\,$\ref{fig:signaltonoise}, the cumulative S/N for the original convergence power spectra departs from this ideal already at $\ell \sim 200$, reaching a value of about $12.1$ at $\ell=1500$, the maximum angular frequency we use for the likelihood analysis.

In agreement with \citet{seo11} the cumulative S/N for the transformed power spectra remains close to the Gaussian limit up to $\ell \sim 1000$, yielding an increase in S/N by a factor of 2.6 (2.0) for transformations based on the (un-) smoothed convergence fields. Again choosing optimal Box-Cox parameters or the respective logarithmic transformation makes little difference in the performance. The BC2/LOG2 transformations concentrate the S/N into angular frequencies up to $\ell=1000$ and level off in the regime where the smoothing washes out information (the increase in S/N for the LOG2 at very high $\ell$ is probably a noise artifact in the inverted covariance). 

The cumulative S/N for the BC1/LOG1 transformations has a shallower slope for $\ell \leq 1000$, but surpasses the S/N for the BC2/LOG2 case beyond $\ell \sim 2500$, even if the convergence fields are also smoothed with the same kernel. This behaviour is reflected also in $r_{\rm c}$ where the BC2/LOG2 transformations suppress cross-correlations better in the range $200 < \ell < 1000$. The strong rise in S/N for the original convergence power spectra and the BC1/LOG1-transformed power spectra for $\ell>3000$ could be due to either noise or cosmological information from the highly non-linear regime, but is in any case inaccessible to us because of the limitations in modelling.

\subsection{Likelihood analysis}
\label{sec:like}

Due to the computational costs of calculating the transformed power spectrum models according to equation (\ref{eq:pstrafo_full_approx}) we restrict ourselves to the cosmological parameters $\Omega_{\rm m}$ and $\sigma_8$ in the likelihood analysis. This should allow us to study the effects of Gaussianising transformations on the joint constraints on cosmology as well as the potential to break the characteristic degeneracy in the $\Omega_{\rm m}-\sigma_8$ plane appearing in standard analyses of weak lensing two-point statistics. The signals from all 100 realisations are combined, so that we reach an effective survey size of $10,000\,{\rm deg}^2$.

\begin{figure*}
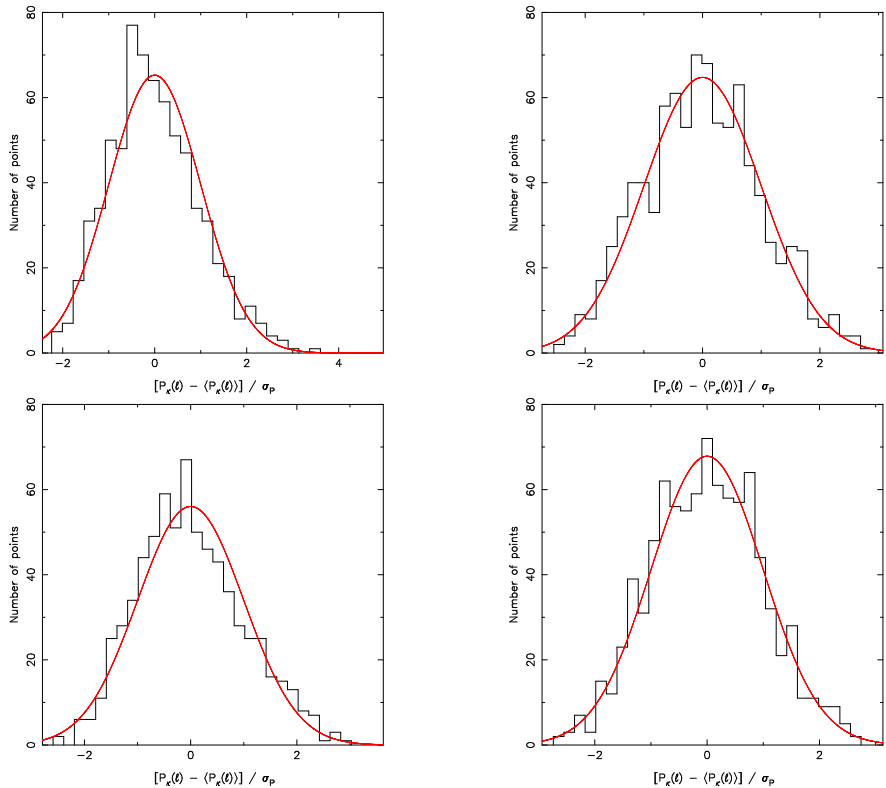

\begin{minipage}[c]{0.75\textwidth}
\begin{minipage}[c]{0.5\textwidth}
\centering
\includegraphics[scale=.34,angle=270]{Hist_low_orig.ps}
\end{minipage}%
\begin{minipage}[c]{0.5\textwidth}
\centering
\includegraphics[scale=.34,angle=270]{Hist_high_orig.ps}
\end{minipage}
\begin{minipage}[c]{0.5\textwidth}
\centering
\includegraphics[scale=.34,angle=270]{Hist_low_meanbc.ps}
\end{minipage}%
\begin{minipage}[c]{0.5\textwidth}
\centering
\includegraphics[scale=.34,angle=270]{Hist_high_meanbc.ps}
\end{minipage}
\end{minipage}%
\begin{minipage}[c]{0.25\textwidth}
\caption{Histograms of the distribution of convergence power spectra, transformed to zero mean and unit variance, for $100<\ell<500$ (left panels), and $500<\ell<3500$ (right panels). The top row corresponds to power spectra computed from the original convergence fields, and the bottom row to those obtained from the Box-Cox transformed fields (BC1). The red curves are unit Gaussians with a normalisation adapted to the number of data in the histograms.}
\label{fig:psdistribution}
\end{minipage}
\end{figure*} 

We make the assumption of a Gaussian likelihood for both the original and transformed convergence power spectra,
\eqa{
\label{eq:likepkappa}
L(\bc{P_\kappa},\vek{p}) \!\!&\propto&\!\! \exp\; \biggl\{- \frac{1}{2} \sum_{i,j=1}^{N_\ell} \bb{\hat{P}_\kappa(\ell_i) - P_\kappa(\ell_i,\vek{p})}\\ \nn
&& \times\; {\rm Cov}_P^{-1} (\ell_i,\ell_j) \bb{\hat{P}_\kappa(\ell_j) - P_\kappa(\ell_j,\vek{p})} \biggr\}\;,
}
where $N_\ell$ is the number of angular frequency bins. The measured power spectrum $\hat{P}_\kappa$ and the covariance are extracted from the simulations, while the cosmology-dependent models $P_\kappa(\ell,\vek{p})$ are calculated from equation (\ref{eq:pstrafo_full_approx}).

In total $N_\ell=10$ bins in the range $150 < \ell < 1500$ are included in the likelihood. We exclude lower angular frequencies to avoid systematic effects in the simulation power spectra due to discreteness error caused by the limited number of Fourier modes per angular frequency bin at low $\ell$; see \citet{kiessling11} for details. At $\ell > 1500$ shot noise becomes relevant, and the power spectra are largely suppressed by the smoothing.

If the Gaussian approximation were of different accuracy for the original and the transformed power spectra, a fair comparison of the resulting parameter constraints would be hampered. Therefore we inspect the distribution of power spectrum values from all realisations in the linear regime $100 < \ell < 500$ and the non-linear regime $500 < \ell < 3500$ where scales are not yet dominated by shot noise. The results for the original and the BC1-transformed power spectra are presented in Fig.$\,$\ref{fig:psdistribution}. 

At high $\ell$ the distribution of the original convergence power spectra is marginally left-skewed, which is removed after Box-Cox transformation. In the linear regime $\kappa$ should be Gaussian distributed anyway, so that the power spectrum histogram should be well described by a $\chi^2$ distribution, which is expected to be very close to Gaussian due to the large number of modes included. However, both original and transformed distributions are mildly skewed, an effect which was also evident in the results of \citet{kiessling11}. We suspect that this is caused by the apodisation in the power spectrum estimation and will investigate this effect elsewhere. Since the deviation from Gaussianity is small and similar for all power spectra considered, and since the bulk of the cosmological information stems from angular frequencies above 500 (see the error bars in Fig.$\,$\ref{fig:meanps}), we conclude that the assumption of a Gaussian likelihood is justified.

\begin{figure*}
\begin{minipage}[c]{0.5\textwidth}
\centering
\includegraphics[scale=.34,angle=270]{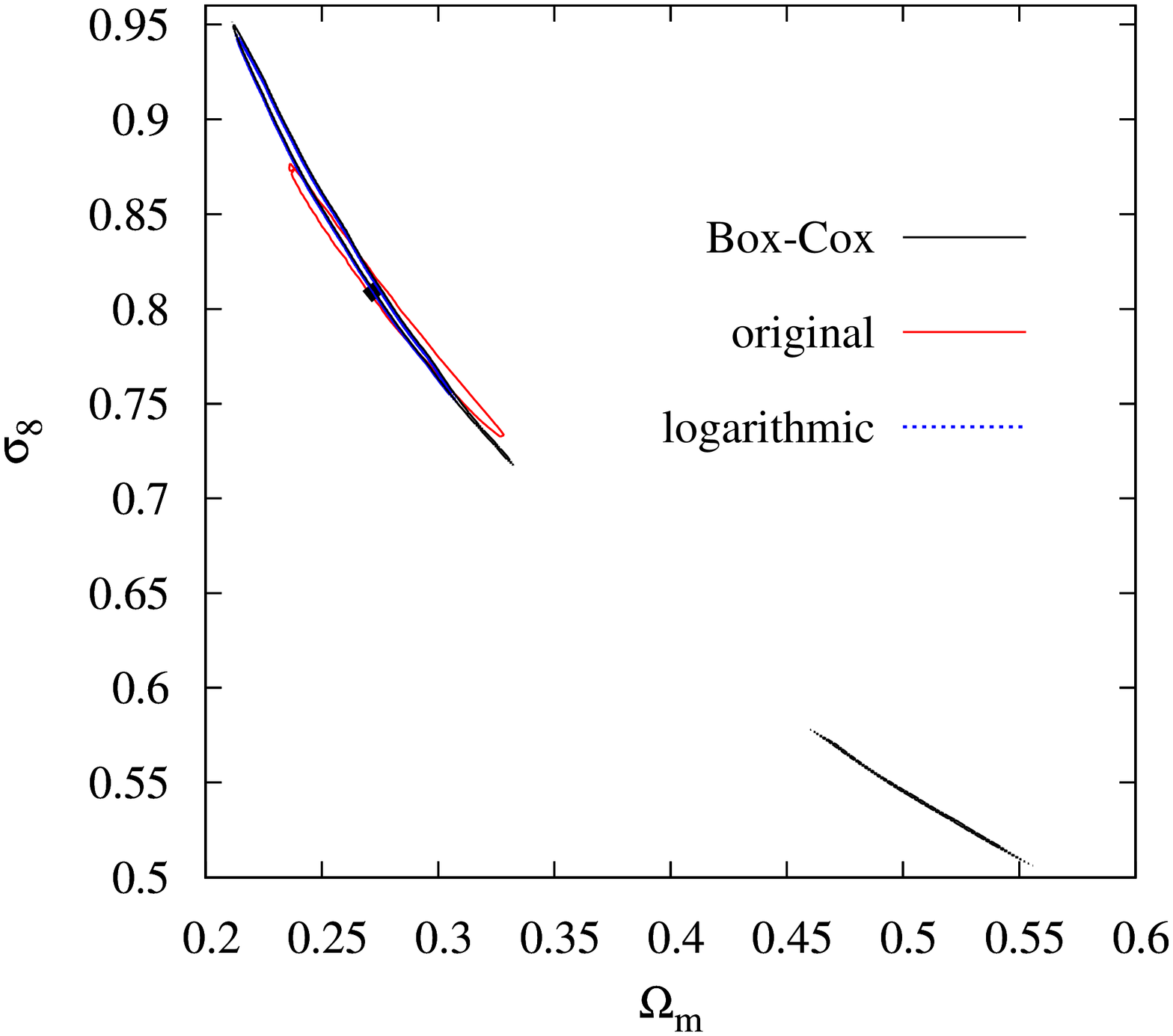}
\end{minipage}%
\begin{minipage}[c]{0.5\textwidth}
\centering
\includegraphics[scale=.34,angle=270]{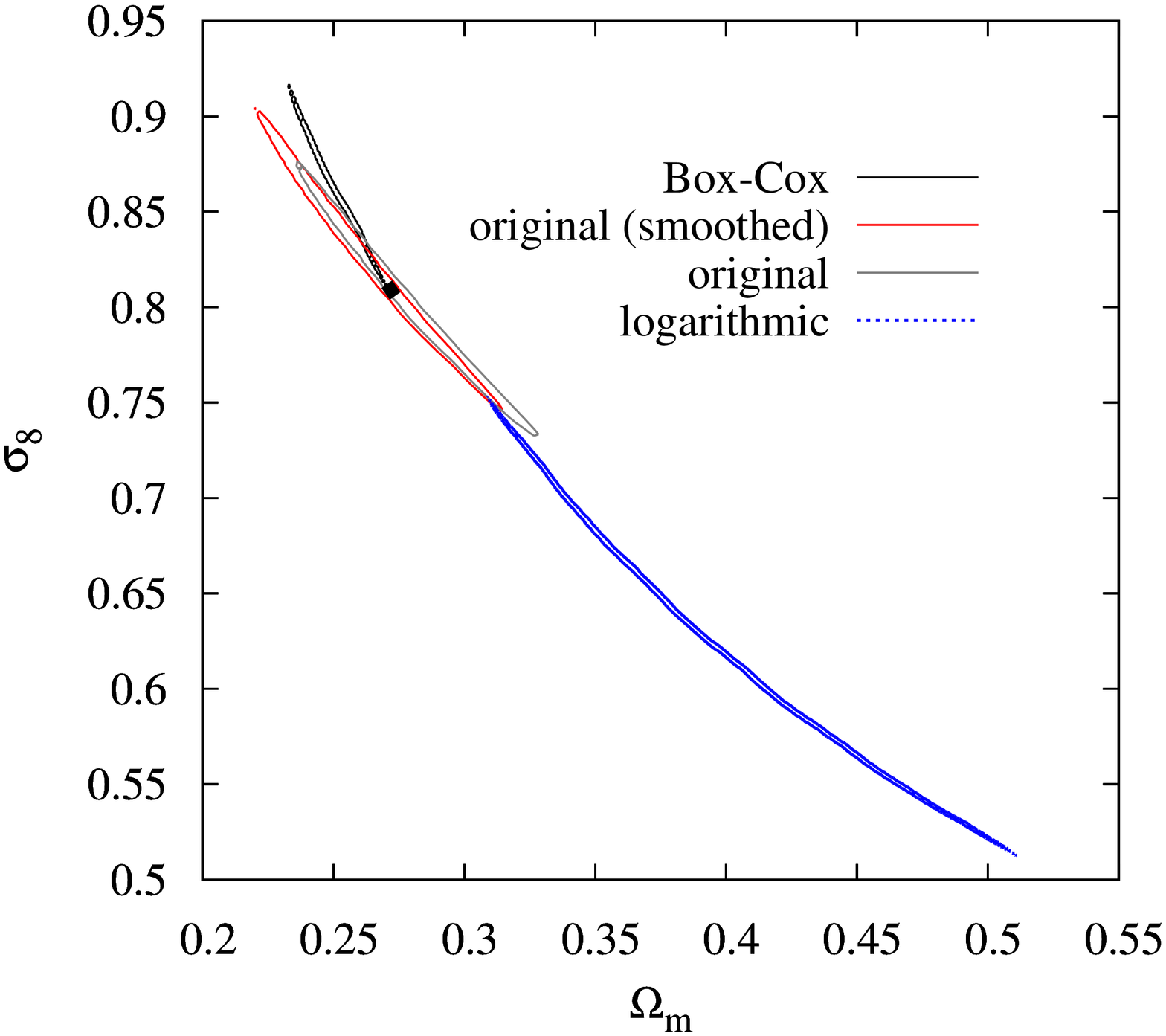}
\end{minipage}
\caption{\textit{Left panel}: $2\sigma$ confidence levels for the likelihood analysis based on the original convergence power spectra (red solid lines), the Box-Cox transformed power spectra (black solid lines), and the logarithmically transformed power spectra (blue dotted lines). No smoothing is used in the analysis of the original power spectra; the convergence transformations are based on parameters determined from the unsmoothed convergence fields (BC1, LOG1). The fiducial cosmology of the simulation is marked by the black point. Note the secondary likelihood peak at high $\Omega_{\rm m}$ and low $\sigma_8$ in the Box-Cox transformed case. \textit{Right panel}: Same as above, but including smoothing in the case of the original power spectra and using parameters determined from the smoothed convergence fields (BC2, LOG2). For reference contours for the unsmoothed original likelihood analysis are shown in grey. Note the strong bias in the LOG2 case.}
\label{fig:boxcoxkappa_likelihood}
\end{figure*} 

We compute power spectrum models from equation (\ref{eq:pstrafo_full_approx}) on a grid in the $\Omega_{\rm m}-\sigma_8$ plane with boundaries $\Omega_{\rm m} \in \bb{0.15;0.70}$ and $\sigma_8 \in \bb{0.45;1.20}$ which we treat as conservative top-hat priors. The resulting $2\sigma$ confidence levels from the subsequent likelihood evaluation are shown in Fig.$\,$\ref{fig:boxcoxkappa_likelihood}. The constraints from the original convergence power spectrum feature the typical banana-shaped degeneracy. The fiducial cosmology at $\Omega_{\rm m}=0.27$ and $\sigma_8=0.81$ is enclosed in the contours, nearly coinciding with the maximum likelihood point. 

The Box-Cox and logarithmic transformations BC1/LOG1 produce similar constraints which are very narrow transverse to the degeneracy line, but the extent of the contours alongside the degeneracy is increased, in particular in the case of the BC1 transformation for which a secondary, very elongated peak along the degeneracy line can be found at high $\Omega_{\rm m}$ and low $\sigma_8$. This indicates a nearly perfect degeneracy between $\Omega_{\rm m}$ and $\sigma_8$ which is even more pronounced for the BC1 transformation although it results in a higher cumulative S/N. The confidence region of the transformed power spectra is slightly tilted with respect to the original one, but the $2\sigma$ contours of both the LOG1 and BC1 results enclose the fiducial cosmology. The maximum likelihood is located at slightly higher values of $\sigma_8$ than the fiducial one which is in agreement with the model for the fiducial cosmology being marginally low in overall amplitude; see the centre panel of Fig.$\,$\ref{fig:meanps}).

\begin{table}
\centering
\caption{Figure of merit in terms of $q$-values in the $\Omega_{\rm m}-\sigma_8$ plane. The likelihood analysis has been performed for the original convergence fields, the Box-Cox transformed field, and the logarithmically transformed fields. The second column contains results based on transformations determined from the unsmoothed convergence (BC1, LOG1), the third column those based on transformations determined from the smoothed convergence (BC2, LOG2), and the fourth column those for the analysis including shape noise. Values of $q$ are given in units of $10^{-5}$. Note that smaller $q$-values correspond to tighter parameter constraints.}
\begin{tabular}[t]{lrrr}
\hline\hline
analysis & unsmoothed & smoothed & shape noise\\
\hline
original       &  7.2 &  7.5 &  7.6\\  
logarithmic    & 51.6 & 37.2 & 27.0\\
Box-Cox        & 83.8 &  1.7 & 12.9\\
\hline
\end{tabular}
\label{tab:qvalues}
\end{table}

Given the persistent degeneracy between $\Omega_{\rm m}$ and $\sigma_8$, marginal errors on theses parameters are of little value. Instead we employ $q$-values, defined as $q=\sqrt{\det Q_{\mu\nu}}$ \citep{kilbinger04}, as a figure of merit, being a measure of the area enclosed by the confidence contours. They are based on the quadrupole of the posterior distribution
\eq{
\label{eq:qvalue}
Q_{\mu\nu} = \sum_{\vek{p}} L(\bc{P_\kappa},\vek{p})\; (p_\mu - p_{{\rm max},\mu}) (p_\nu - p_{{\rm max},\nu})\;,
}
where $\vek{p}_{\rm max}$ marks the point of maximum likelihood. The $q$-values which correspond to the parameter constraints shown in Fig.$\,$\ref{fig:boxcoxkappa_likelihood} are summarised in Table \ref{tab:qvalues}. Note that smaller $q$-values correspond to tighter parameter constraints and hence a better performance of the transformations. Transforming the convergence according to the parameter sets BC1/LOG1 yields an increase in this figure of merit, i.e. a degradation of constraints, by roughly an order of magnitude.

The S/N discussed in the foregoing section is equivalent to the Fisher matrix with the amplitude of the power spectrum as the single inferred parameter. Hence the S/N can also be used as a measure for the change in constraints when only $\sigma_8$ is varied. Contrasting the doubling in S/N with the pronounced increase in $q$, it is clearly the failure of breaking the $\Omega_{\rm m}-\sigma_8$ degeneracy that hinders a stronger improvement in the figure of merit.

We repeat the likelihood analysis for the original convergence power spectra smoothed with the same kernel as the transformed convergence fields; see Fig.$\,$\ref{fig:boxcoxkappa_likelihood}, right panel. The smoothing affects the area of the confidence region only marginally, but the suppression of the signal from high angular frequencies shifts the contours upwards along the degeneracy line. The corresponding $q$-value increases slightly.

The likelihood analysis for the transformation BC2, which boosts the cumulative S/N stronger than BC1/LOG1, results in $q$-values that are a factor of 4.4 smaller than for the original convergence (Table \ref{tab:qvalues}). The confidence region still features a degeneracy between $\Omega_{\rm m}$ and $\sigma_8$, but has shrunk considerably. We observe a similar tilt of the degeneracy line as for the BC1/LOG1 case and a mild bias, the $2\sigma$ confidence level touching the point of the fiducial cosmology.

In stark contrast to this, the LOG2 transformed models fail to fit the simulation power spectra, leading to a strong bias in cosmological parameters and a very strong degeneracy between $\Omega_{\rm m}$ and $\sigma_8$. The LOG2 transformation features by far the smallest value of the shift parameter $a$ and therefore the strongest boost of higher-order contribution. As we detail in Appendix \ref{app:higherorder}, terms of the order $P_\kappa^4$ and higher, which we are unable to model, are likely to become relevant in this case, particularly on small scales where the amplitude of the model is correspondingly low (Fig.$\,$\ref{fig:meanps_logsmooth}).

\subsection{Effect of shape noise}
\label{sec:shapenoise}

So far we have considered an idealistic experiment with noise levels that cannot be achieved even by future weak lensing experiments. For instance, the deep, space-based COSMOS survey contains $n_{\rm gal}=76\,{\rm arcmin}^{-2}$ galaxies usable for shape measurement \citep{schrabback09}, which would still produce noise power more than an order of magnitude larger than the shot noise level. We assume $n_{\rm gal}=30\,{\rm arcmin}^{-2}$, which the planned Euclid mission aims for, resulting in a noise power spectrum $P_{\rm noise} = \sigma_\epsilon^2 / (2 n_{\rm gal}) \approx 1.7 \times 10^{-10}$. Transformations are only determined from the smoothed convergence fields as otherwise the (Gaussian) shape noise would dominate the one-point distribution of convergence values and hence obscure any cosmological effects.

\begin{figure}
\centering
\includegraphics[scale=.33,angle=270]{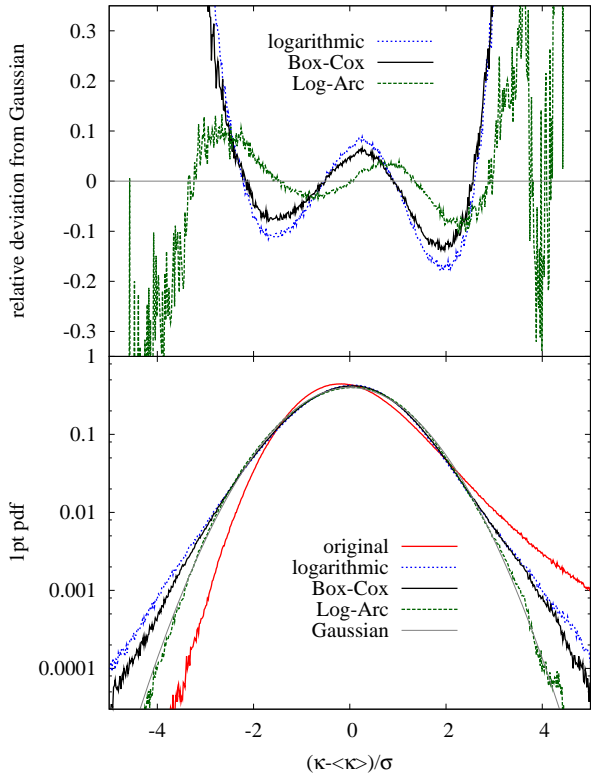}
\caption{Same as Fig.$\,$\ref{fig:kappatrafo}, but for the case with shape noise added to the convergence fields. Additionally, the resulting convergence distribution after a log-arctan transformation is shown as green dashed line.}
\label{fig:kappatrafo_shapenoise}
\end{figure} 

Since the overall minimum of the convergence is very similar to the case of the unsmoothed noise-free maps, we set again $a=0.07$. The optimisation procedure for the Box-Cox parameters prefers strongly negative values of $\lambda \ll -10$ mainly to reduce the residual kurtosis, which causes numerical issues, e.g. due to a very small variance of the transformed convergence. Hence we restrict $\lambda$ to a moderately negative value of approximately $-7.5$ and choose $a$ such that $(\lambda,a)$ lies on the degeneracy line of close to optimal Gaussianity, analogous to the one observed in Fig.$\,$\ref{fig:bcplane}; see Table \ref{tab:trafos} for an overview on the transformation parameters.

Figure \ref{fig:kappatrafo_shapenoise} shows that both logarithmic and Box-Cox transformations (designated LOGs and BCs, respectively) struggle to render the one-point distribution of the transformed convergence Gaussian. The original distribution still features a long positive tail caused by clustering, but values of $\kappa$ below the mean now have a shallower slope closer to a Gaussian due to shape noise. The transformations are capable of reducing the skewness of this hybrid distribution to negligible values, but the Mexican-hat shaped residuals in Fig.$\,$\ref{fig:kappatrafo_shapenoise}, top panel, indicate that a significant positive excess kurtosis remains; see Table \ref{tab:skewcurt} for the statistics. Consequently, $D_{\rm KL}$ for the transformed fields is larger than in the cases without shape noise while the opposite holds for the original convergence, so that one expects overall less improvement through the Gaussianising transformations.

\begin{figure}
\centering
\includegraphics[scale=.34,angle=270]{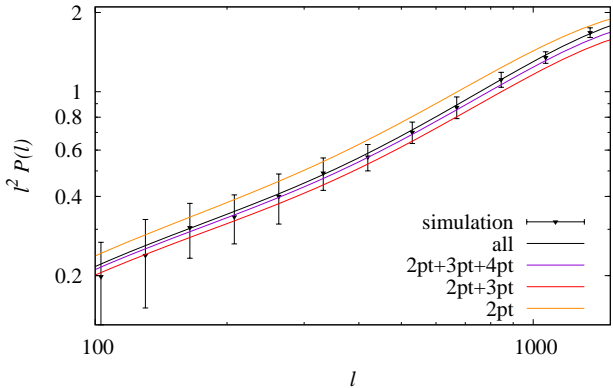}
\caption{Same as Fig.$\,$\ref{fig:meanps}, bottom panel, but showing the Box-Cox transformed power spectrum in the presence of shape noise, again smoothed with a Gaussian kernel of width 5 pixels. Note that the shape noise is included both in the simulation power spectrum and the model.}
\label{fig:meanps_shapenoise}
\end{figure} 

To ensure that the limited ability of Box-Cox-type transformations to arrive at a Gaussian one-point distribution (which could in principle be overcome by a rank-order Gaussianisation procedure) does not mislead our conclusions on the information content of Gaussianised fields, we introduce yet another type of transformation which fares better in the case of noisy convergence fields. Noting that the main flaw of the BCs/LOGs transformations is a significantly leptocurtic result, we define
\eq{
\label{eq:logarc}
\bar{\kappa}_i(s,a) = \arctan \bb{s\, \ln (\kappa_i + a)}\;,
}
i.e. after the LOGs transformation we apply in addition the arc-tangent, using a free scaling $s$ as a second free parameter. We illustrate the mapping by equation (\ref{eq:logarc}) in Fig.$\,$\ref{fig:boxcoxillustrate}. Via a straightforward generalisation of the Box-Cox formalism one can derive an optimisation for the transformation parameters in analogy to equation (\ref{eq:lmaxboxcox}), as well as models for the transformed power spectrum by means of the procedure presented in Section \ref{sec:transformedps} and Appendix \ref{app:higherorder}.

In our models we incorporated terms up to third order in $\kappa$, where the $\kappa^3$ contribution only entered the first four-point term in equation (\ref{eq:pstrafo_basic_approx}); see also Appendix \ref{app:transformedps}. Apart from an irrelevant overall rescaling with $s$, only this term is modified, as is readily seen by consulting the Taylor expansion $\arctan x = x - 1/3 x^3 + {\cal O}(x^5)$. To leading order, the arc-tangent is the identity transform, and the next-to-leading order can contribute only to terms that are third order in $\kappa$ or higher. As Fig.$\,$\ref{fig:kappatrafo_shapenoise} demonstrates, this log-arctan transformation indeed results in a Gaussianised one-point distribution for the convergence with an accuracy compatible to the noise-free case (see also Table \ref{tab:skewcurt}).

\begin{figure}
\centering
\includegraphics[scale=.34,angle=270]{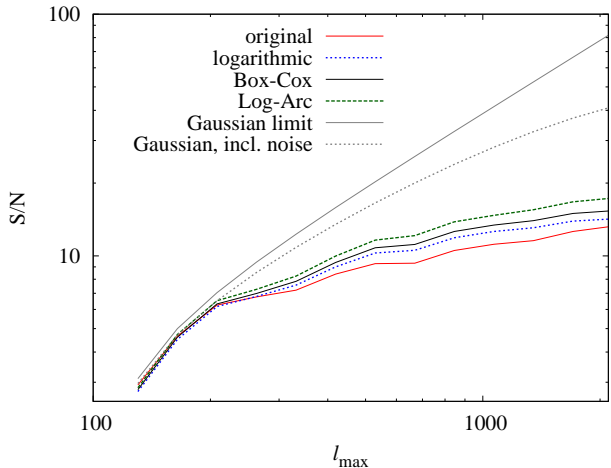}
\caption{Same as Fig.$\,$\ref{fig:signaltonoise}, but based on convergence fields with shape noise. Results for the original (Box-Cox transformed; logarithmically transformed) convergence fields are again shown as red solid (black solid; blue dotted) curves. In addition we plot the S/N obtained with the log-arctan transformation as green dashed line. All fields have undergone smoothing with a Gaussian kernel of width 5 pixels. The Gaussian limit (attainable without shape noise) is shown as grey curve; the limit including shape noise in the covariance is shown as grey dotted curve}. Note the different scaling of the abscissa compared to Fig.$\,$\ref{fig:signaltonoise}.
\label{fig:signaltonoise_shapenoise}
\end{figure} 

Figure \ref{fig:meanps_shapenoise} shows the contributions to the model of the BCs-transformed power spectrum, again obtained by using equation (\ref{eq:pstrafo_full_approx}), which provides a good fit to the mean from the simulation. Although $\lambda \ll 0$, the three-point, four-point, and higher-order contributions are small and sequentially decline in amplitude because $a$ is almost an order of magnitude larger than for the noise-free Box-Cox transformations. Note that both model and simulation include shape noise which is visible as the bump at $\ell > 500$.

Gaussian shape noise adds only to the diagonal of the power spectrum covariance and thus reduces the importance of off-diagonal terms. This implies a decrease in $r_{\rm c}$ for the original covariance power spectrum, as is evident in the bottom right panel of Fig.$\,$\ref{fig:pscorrelation}. The gain in decorrelation due to the transformation of the convergence is largely reduced as $r_{\rm c}$ changes little compared to the noise-free case and even marginally increases for $\ell \sim 500$. The different transformations perform similarly, where as a trend we find that the closer the transformed one-point distribution for $\kappa$ is to a Gaussian, the smaller the cross-correlations between angular frequencies.

The same conclusion holds for the cumulative S/N displayed in Fig.$\,$\ref{fig:signaltonoise_shapenoise}, yielding improvements of $34\,\%$ ($20\,\%$; $13\,\%$) by the log-arctan (BCs; LOGs) transformation over the S/N of the original convergence power spectrum at $\ell=1500$. However, all curves deviate largely from the Gaussian limit (which can only be reached if noise contributions are negligible) from $\ell \approx 200$ onwards. This remains true even if we consider the S/N using a Gaussian covariance with shape noise included, so that the comparatively low S/N is mainly caused by the cross-correlation of angular frequencies, and not by the higher noise levels. Again these results differ from the findings by \citet{seo11} who assume the same number density of galaxies, but whose cumulative S/N degrades less in the presence of shape noise. We can only speculate at this point that this discrepancy might, like in the noise-free case, be related to the different levels of non-linear structure evolution in the underlying N-body simulations.

Note that shape noise of course adds to the covariance, but should not be included in the signal, i.e. not enter the power spectra used in equation (\ref{eq:sn}). The usual approach of subtracting the shape noise power spectrum from the observed signal does not work after non-linear transformations of the convergence which spread noise contributions to all terms of even order in $\kappa$, see equation (\ref{eq:pstrafo_full_approx}). Since in this case our analytic models fit the simulation well, we recompute the model without shape noise and use this result in the S/N computation.

\begin{figure}
\centering
\includegraphics[scale=.34,angle=270]{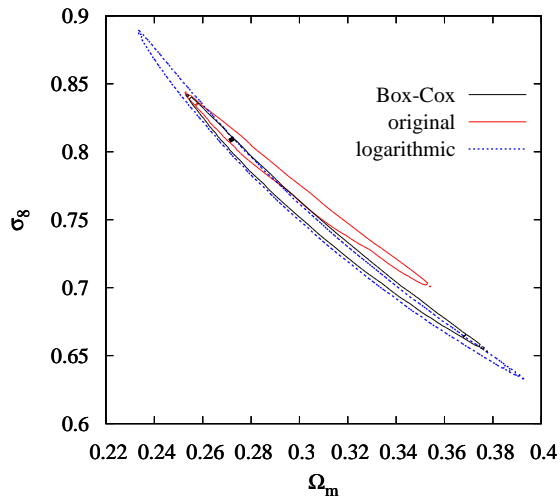}
\caption{Same as Fig.$\,$\ref{fig:boxcoxkappa_likelihood}, but for the case with shape noise included in the convergence fields. Again the fiducial cosmology is indicated by the black point.}
\label{fig:boxcoxkappa_likelihood_shapenoise}
\end{figure} 

Both $q$-values and contours change less compared to the noise-free transformations after Gaussianising the convergence, see Fig.$\,$\ref{fig:boxcoxkappa_likelihood_shapenoise} and Table \ref{tab:qvalues}. Despite the increase in S/N, and although the transformations have been optimised for the smoothing and noise level present in the convergence fields, parameter constraints mildly degrade. The $2\sigma$ contours for the Box-Cox and logarithmic transformation have a similar form, being slightly more concentrated in the former case (hence we expect analogous results, with possibly marginally tighter constraints still, for the log-arctan transformation). The degeneracy between the cosmological parameters is once again more pronounced than for the original likelihood analysis, the degeneracy line being tilted in the same way as in the noise-free cases. The confidence regions comfortably enclose the fiducial cosmology, so that in the most realistic situation of a convergence with shape noise our modelling is reliable and thus our conclusions robust.

\section{Interpretation and Discussion}
\label{sec:discussion}

\subsection{Performance of Gaussianising transformations}
\label{sec:trafodiscussion}

Generally we can confirm earlier results that a logarithmic transformation of the weak lensing convergence renders its one-point distribution close to Gaussian, mainly via removing the skewness induced by structure evolution. Optimised Box-Cox transformations perform in all considered cases significantly better in Gaussianising the convergence distribution, but do not necessarily produce better constraints on cosmology than the logarithmic transformation. This suggests that any fine-tuning on the shape of the transformed one-point distribution has only a modest effect on the amount of cosmological information in the transformed two-point statistics, implying that little could be gained by using a perfect rank-order Gaussianisation as in \citet{yu11}.

For all three situations we study, both logarithmic and Box-Cox transformations fail to reduce correlations between power spectra at different angular frequencies to a negligible level, so that a non-vanishing connected trispectrum must be present in the transformed convergence fields. Together with the measurement of a non-zero bispectrum from a perfectly (one-point) Gaussianised field by \citet{yu11}, this provides firm evidence that manipulating the one-point distribution is insufficient in turning the convergence into a Gaussian random field. As discussed in \citet{yu11}, this also implies that the assumption of a Gaussian copula \citep{scherrer10,sato11} to describe the convergence field is of limited accuracy.

In addition to concentrating cosmological information into two-point statistics, a Gaussianised convergence would allow one to use an exact functional form for the likelihood. Instead of assuming a Gaussian likelihood for weak lensing two-point statistics, which cannot be accurate due to the effects of non-linear structure formation \citep{hartlap09} and because of theoretical arguments \citep{Schneider09}, one treats $\kappa$ itself as the data for which the Gaussian assumption then holds. In Appendix \ref{app:likekappa} we outline the likelihood formalism for $\kappa$ and show that the Fisher information in the likelihood for $\kappa$ is equivalent to that in the likelihood for $P_\kappa$ if the latter is Gaussian and contains a Gaussian covariance.

\begin{figure}
\centering
\includegraphics[scale=.30,angle=270]{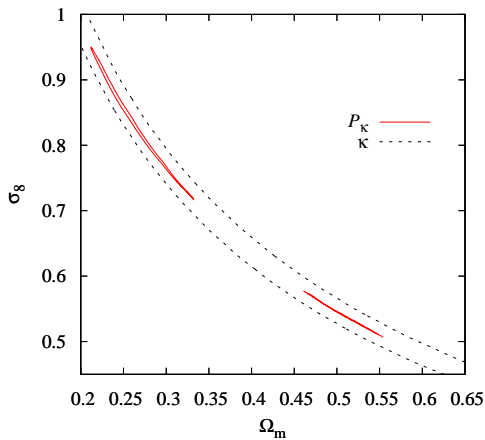}
\caption{Comparison between the $2\sigma$ confidence levels obtained from the power spectrum likelihood analysis (red solid line; cf. Fig.$\,$\ref{fig:boxcoxkappa_likelihood}) and the convergence likelihood analysis (black dashed line) after Box-Cox transformation BC1.}
\label{fig:likelihood_pskappa}
\end{figure}

We compare the constraints from the two likelihood formalisms for a Box-Cox transformed (BC1) convergence without shape noise in Fig.$\,$\ref{fig:likelihood_pskappa}. The resulting confidence levels are largely different, the likelihood based on $\kappa$ as the data-vector having considerably less constraining power. The difference can be ascribed to the residual connected four-point correlations in the convergence fields which can be incorporated into the power spectrum likelihood via the simulation covariance matrix with its off-diagonal terms, but not into the likelihood for $\kappa$ which includes at most terms that are second order in $\kappa$. This result suggests that the residual non-Gaussianity of the convergence after transformation is not a small effect, and that the information in the transformed connected trispectrum helps considerably constraining cosmological parameters.

To ameliorate the performance, it is therefore necessary to go beyond transformations of the one-point distribution. Box-Cox transformations are readily applied to multi-dimensional data \citep{velilla93}, so that one could in principle compose a large data-vector of all convergence values on the gridded $\kappa$ map and assign an individual pair of Box-Cox parameters $(\lambda,a)$ to each entry. As a consequence the transformation becomes scale-dependent, which violates the statistical translational invariance of the convergence fields and is thus undesirable\footnote{This is readily shown by introducing a dependence of $\lambda$ and $a$ on $\vek{\theta}$ in equation (\ref{eq:kappatrafo}) and then repeating the computation of the correlator $\ba{\bar{\kappa}(\vek{\ell})\; \bar{\kappa}^*(\vek{\ell'})}$.}. The same holds for a global transformation of the Fourier-transformed convergence values $\kappa_{\svek{\ell}}$ which couples spherical harmonics, i.e. angular frequencies $\ell'$ with $2\ell,\,\ell/2,\,3\ell,\,\ell/3$ etc.

Hence, the only practical option seems retaining a global transformation of the real-space convergence, but using a multi-dimensional $\kappa$ data-vector, thereby taking spatial correlations within the convergence map into account. Then the variance in equation (\ref{eq:lmaxboxcox}) needs to be replaced by the full covariance of the $\kappa$ values in the data-vector, readily obtained by measuring the correlation function $\xi_\kappa(\theta)=\xi_+(\theta)$ from the fields. As an aside, note that it is not obvious how to generalise rank-order Gaussianisation procedures to more than one-dimensional data as they rely on the concept of a cumulative probability density function. The necessary statistics to optimise the transformation parameters in this multivariate case could either be obtained from a large number of simulation realisations or by exploiting translational and rotational invariance of a single simulation or observational data.

Although fairly comprehensive, the flexibility of Box-Cox transformations encapsulated in the parameters $\lambda$ and $a$ might not suffice to Gaussianise the multivariate distribution of convergence values to the desired accuracy. The formalism used in this work to find optimal transformation parameters via equation (\ref{eq:lmaxboxcox}) and develop models of the transformed power spectrum (see Section \ref{sec:transformedps}) is applicable to any parametrised, analytical set of transformations, so it could e.g. be used to explore general parametrisations with a larger number of free parameters.

It would be desirable to derive a more physically motivated set of transformations which ideally describe a bijective mapping from the present-day convergence field to a nearly Gaussian convergence that one would have observed at an early stage of structure formation. Since matter in high-density regions is virialised and thus can by definition not remember its original trajectory, such a mapping can principally only exist down to a certain spatial scale. We defer the investigation of advanced transformations of the convergence as outlined above to future work.

\subsection{Extraction of cosmological information}
\label{sec:constraintsdiscussion}

Throughout this work we use invertible transformations, so that the mapping should preserve the cosmological information contained in the convergence fields (we ignore the smoothing for the moment). This information is distributed over the n-point statistics of the field, and in the dependence of these statistics on the amplitude and phases of the angular frequencies (or equivalently angular scales) involved. The power spectrum only depends on the absolute value of $\vek{\ell}$ and not its phase while higher-order statistics also vary e.g. as a function of the internal angles of the triangle, quadrangle, etc. they are evaluated at. If the convergence was transformed into a perfect Gaussian random field, all information in the amplitude and phase dependence of all n-point statistics would be transferred into the amplitude dependence of the transformed power spectrum.

With this in mind we will attempt to elucidate why none of the transformations could efficiently break the degeneracy between $\Omega_{\rm m}$ and $\sigma_8$, thereby limiting the improvement in, or even degrading, the figure of merit. Our models include both the power spectrum and bispectrum as the dominant contributions, and their combined analysis has been proven to break this parameter degeneracy \citep{takada04,berge10}. However, these works simplistically assumed that there is no cross-variance between two- and three-points statistics. Since working with Box-Cox transformed power spectra does not suffer from this simplification, our findings could indicate that the five-point cross-variance between power spectrum and bispectrum partly eliminates the complementarity of these statistics.

Alternatively, the transformations we considered might have failed to incorporate information which is capable of breaking the $\Omega_{\rm m}-\sigma_8$ from e.g. the bispectrum into the transformed power spectrum. Note that only integrals over the higher-order spectra contribute to $P_{\bar{\kappa}}$; see equation (\ref{eq:pstrafo_basic_approx}). For instance, the triangles of angular frequencies at which the bispectrum is evaluated have one fixed side length $\ell$, and all possible positions of the third point of the triangle are averaged over in the integration, thereby diluting the independent phase information in $B_\kappa$.

\begin{figure}
\centering
\includegraphics[scale=.34,angle=270]{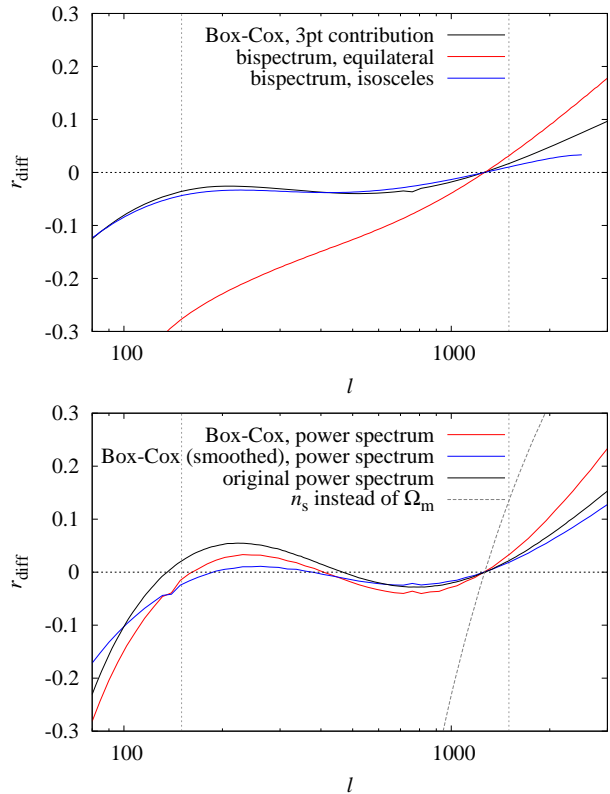}
\caption{Relative difference $r_{\rm diff}$ between derivatives with respect to $\Omega_{\rm m}$ and $\sigma_8$ as a function of angular frequency. \textit{Top panel}: Normalised $r_{\rm diff}$ for the three-point contribution to the Box-Cox transformed power spectrum (black solid line), the bispectrum of the original convergence using equilateral triangles of side length $\ell$ (red solid line), and the bispectrum of the original convergence using isosceles triangles with two side lengths fixed at $\ell_1=\ell_2 \approx 1265$ and third side length $\ell$ (blue solid line). \textit{Bottom panel}: Normalised $r_{\rm diff}$ for the full Box-Cox transformed power spectrum with parameters BC1 (red solid line) and BC2 (blue solid line), and the power spectrum of the original convergence (black solid line). For comparison we have also plotted $r_{\rm diff}$ for the convergence power spectrum with the derivative with respect to $\Omega_{\rm m}$ replaced by the derivative with respect to $n_{\rm s}$ as grey dashed line. In both panels the range used for the likelihood analysis is marked by vertical lines. All curves have been normalised to zero at $\ell \approx 1300$, i.e. within the angular frequency range with highest S/N entering the likelihood analysis.}
\label{fig:derivatives}
\end{figure} 

We demonstrate the effect on the sensitivity to cosmological parameters by comparing the derivatives of the quantities involved with respect to $\Omega_{\rm m}$ and $\sigma_8$. A perfect degeneracy between the two parameters is expected if their derivatives have exactly the same dependence on angular frequency over the range considered for the likelihood analysis. Thus we use the relative difference in the derivatives,
\eqa{
\label{eq:rdiff}
r_{\rm diff}(\ell) &=& 1 - \left. \frac{\partial S}{\partial \Omega_{\rm m}} \right|_\ell / \left. \frac{\partial S}{\partial \sigma_8} \right|_\ell\\ \nn
&& \;\; \times\; \br{\left. \frac{\partial S}{\partial \Omega_{\rm m}}\right|_{\ell=1300} / \left. \frac{\partial S}{\partial \sigma_8}\right|_{\ell=1300}}^{-1}\;,
}
as a measure for degeneracy-breaking capabilities. Here, $S$ stands for the quantity whose properties are tested, i.e. the convergence power spectrum or bispectrum, the transformed power spectrum, as well as its three-point contribution. To simplify the visual inspection, we normalise the ratio of derivatives to unity at $\ell=1300$, i.e. in the regime where the S/N is highest. Hence, a flat $r_{\rm diff}$ indicates a strong degeneracy between parameters whereas a strongly varying $r_{\rm diff}$ and in particular a steep slope at the pivot $\ell$ signify non-degenerate constraints.

As is shown in Fig.$\,$\ref{fig:derivatives}, $r_{\rm diff}$ for the convergence power spectrum varies only moderately with a rather shallow slope at $\ell=1300$, remaining relatively close to zero within the angular frequency range entering the likelihood analysis. For reference we also show $r_{\rm diff}$ with the derivative with respect to $\Omega_{\rm m}$ replaced by the one with respect to the slope of the initial matter power spectrum $n_{\rm s}$. This parameter predominantly affects the slope of the convergence power spectrum while $\sigma_8$ only changes its amplitude, so that these parameters are close to orthogonal. Correspondingly, $r_{\rm diff}$ has a steep slope at the pivot point and attains values more than an order of magnitude larger than the original $r_{\rm diff}$ with $\Omega_{\rm m}$.

The curve for the Box-Cox transformed (BC1) power spectrum has a similar form, so that no significant improvement in the parameter degeneracy can be expected. In fact, the degeneracy proves to be much more pronounced in this case (see Fig.$\,$\ref{fig:boxcoxkappa_likelihood}), which might also be related to the size and correlation of the errors on the power spectrum. In the case of the BC2 transform for which we found strong constraints $r_{\rm diff}$ is even slightly closer to zero over a large portion of the angular frequency range, i.e. the degeneracy is still present. However, we find that the relative change of the BC2-transformed power spectrum with $\Omega_{\rm m}$ and $\sigma_8$ individually is much stronger than for the original $P_{\kappa}$, hence the substantial shrinkage of the confidence region.

Furthermore the three-point contribution to the transformed power spectrum, i.e. the integrated bispectrum in equation (\ref{eq:pstrafo_basic_approx}), produces again a slowly varying $r_{\rm diff}$, albeit with a differing dependence on angular frequency. Contrasting this with $r_{\rm diff}$ for the convergence bispectrum evaluated at two exemplary sets of triangle shapes, one finds a very similar functional form as for the three-point contribution to the Box-Cox transformed power spectrum for isosceles, as well as a more strongly varying curve with steeper slope for equilateral triangles (which has little effect in practice as the S/N for equilateral triangles is small; see e.g. \citealp{berge10}). Indeed constraints from the bispectrum alone are also degenerate, but with a different degeneracy line from the two-point case, so that a joint likelihood analysis yields much tightened constraints. Contrary to this, in the likelihood analysis of Box-Cox or logarithmically transformed power spectra only a linear combination of these two- and three-point statistics enters, cancelling the degeneracy-breaking capabilities to a large degree.

To summarise, the particular way in which the convergence statistics are combined to arrive at the Box-Cox transformed power spectrum implies a dilution and partial cancellation of cosmological parameter dependencies, thereby yielding much less improvement in the breaking of the degeneracy between $\Omega_{\rm m}$ and $\sigma_8$ in the angular frequency range we consider than if the statistics were analysed separately and their constraints combined afterwards. See also the perfect cancellation of terms at different order in the case of a lognormal distributed convergence (Appendix \ref{app:higherorder}). Since the transformed power spectrum is rather featureless (see Fig.$\,$\ref{fig:meanps}), and thus parameter dependencies generally difficult to disentangle, we expect similar conclusions to hold if a larger set of cosmological parameters is considered.

All conclusions made in this paper are restricted to the limited range of angular frequencies available for analysis. One has little control over which angular frequencies cosmological information is transferred to by the different transformations\footnote{Note however the results of Fig.$\,$\ref{fig:signaltonoise}: when optimising the transformations on the smoothed convergence fields, the main increase in S/N is on scales which are not affected by smoothing whereas no independent information is added in the angular frequency range where smoothing is important. Note further that in practice not only cosmological information would be re-distributed, but also remaining systematic effects, complicating the analysis.}, so it might well be possible that the gain from Gaussianising the convergence is much higher when extending the analysis farther into the non-linear regime.

In our case the restrictions in $\ell$ were given by the resolution of the simulation, but were more stringently determined by the limitations of analytical modelling of weak lensing statistics on non-linear scales. While no significant progress on (semi-) analytical models is to be expected in the near future, a potential remedy could be provided by the path integral marginalisation technique developed by \citet{kitching10}. This way all model uncertainties could be accurately accounted for, allowing e.g. two-point statistics at higher angular frequencies or a tree-level perturbation trispectrum to add to the constraints without risking parameter bias.

Modelling issues can be circumvented by resorting to a massive suite of simulations to sample parameter space \citep[e.g.][]{neyrinck11b}. The immense computational costs are not necessarily a downside of the Gaussianisation approach as also standard weak lensing measurements will eventually require a large simulation effort to obtain precise models for two- and higher-order statistics and their covariances. Even when fully simulating Gaussianised signals, however, one needs to carefully account for noise and resolution effects, as our analytical models demonstrate.

\subsection{Prospects for an application to real data}
\label{sec:prospects}

To be a viable alternative to the standard statistical analysis of weak lensing data, Gaussianisation methods have to work in the presence of a realistic level of shape noise. We find a rather poor performance, even after introducing a transformation that accurately Gaussianises the one-point convergence distribution, with an only modest increase in S/N and a small degradation in constraints on cosmology. The main reason for this is that shape noise partly takes over the Gaussianisation of the data by turning the one-point distribution more Gaussian and decorrelating angular frequencies, so that there is less room for information gain. In addition, Gaussianising the one-point distribution does worse in bringing the convergence close to a Gaussian random field, as can be concluded from the increased correlation coefficient at intermediate scales $300 < \ell < 1000$; compare the bottom panels of Fig.$\,$\ref{fig:pscorrelation}.

Furthermore this and foregoing work are based on convergence fields, which however are not directly observable. Convergence maps can be constructed from the gravitational shear via grid-based techniques (see \citealp{seitz97} and references therein) or pseudo-$C_\ell$ methods (\citealp{wandelt01}; \citealp{brown05}; see \citealp{hikage11} for an application to weak lensing). In practice one needs to take into account the complex masks applied to weak lensing surveys, which will modify the distribution of convergence values\footnote{Note that in the case of the CMB analysis the pseudo-harmonic coefficients are a linear combination of the original Gaussian distributed coefficients, and thus also follow a Gaussian distribution. However, a linear combination of non-normal random variables generally results in another non-normal but differently distributed quantity, as applies to the weak lensing convergence.}. Hence a more flexible convergence transformation than a fixed logarithm, as provided by the Box-Cox formalism, could prove fruitful in this case.

As already pointed out above, the results on more realistic data might be improved by going beyond transforming only the one-point distribution. Even if that were successful, and e.g. the bi- and trispectrum in the transformed maps removed, one still could not rule out the transfer of information into n-point correlations of the transformed field with $n \geq 5$. Thus one would not be spared the usage of large sets of simulations to verify that the noisy $n$th moment or n-point statistics are negligible, similar to the need for covariances of higher-order statistics in the standard analysis of weak lensing data. Yet, one may be able to build up an alternative way to inference on cosmology from weak lensing data, based on Gaussianised convergence fields, which relies on different assumptions than the standard approach and therefore provides valuable complementarity.

\section{Conclusions}
\label{sec:conclusions}

In this work we investigated the information on cosmology contained in Gaussianised weak lensing convergence fields, using Box-Cox transformations of the one-point distribution of the convergence $\kappa$. We derived an expression for the power spectrum of the transformed convergence in terms of the statistics of the original fields and computed models including contributions up to sixth order in $\kappa$ and taking the dependence on noise and smoothing into account.

From a set of 100 convergence maps obtained via N-body simulations we measured the correlation properties and the cumulative S/N of transformed power spectra for a number of different transformations. Using our analytical models, we performed a likelihood analysis jointly on all simulated maps, deriving constraints on the parameters $\Omega_{\rm m}$ and $\sigma_8$.

Our main findings can be summarised as follows:
\begin{enumerate}
\item Optimal Box-Cox transformations prefer in all cases considered an even stronger downweighting of high-density regions in the convergence map than a logarithmic transformation and yield excellent results on the Gaussianisation of the one-point convergence distribution. The logarithmic transformation has results close to this optimum, performing slightly worse in Gaussianising the convergence, but in some cases producing similar constraints on cosmological parameters. The best results are obtained when extracting the transformation parameters from a convergence field that has already undergone the same smoothing as the fields used for the power spectrum estimation and likelihood analysis.

However, none of the transformations were capable of rendering the transformed convergence fields close to a Gaussian random field, despite the one-point distribution being very close to Gaussian. We found significant residual correlations between power spectra at different angular frequencies, indicative of a non-vanishing connected trispectrum of the transformed convergence, and demonstrated that these have a strong effect on cosmological constraints if ignored. We discussed possible remedies by going beyond transformations of the one-point distribution and advertised the Box-Cox formalism outlined in this work to be readily applicable to the multivariate case and alternative parametrised forms of transformations.
\item The accuracy of analytical models for the transformed power spectra is limited by the uncertainty in the modelling of higher-order convergence statistics as well as by systematic deviations of the fit formulae for the convergence power spectrum and particularly the bispectrum in the non-linear regime. Due to the non-linearity of the transformations this uncertainty affects all scales of the transformed statistics. 

Suppressing contributions from small scales substantially via smoothing, our models yields good fits to the simulations, enclosing the true combination of cosmological parameters within the $2\sigma$ confidence limits in five of six cases. The modelling fails for a logarithmic transformation with very small shift parameter $a$, which we demonstrate to be caused by important contributions from higher-order terms in the convergence beyond those that we can include. As stronger smoothing modifies the convergence distribution such that optimal Box-Cox parameters cause contributions by higher-order correlations to be even more important (more negative $\lambda$; $a$ closer to zero), the resulting bias in this one case can only be removed by further increasing the overall modelling accuracy.
\item The cumulative S/N of the convergence power spectrum in the range $100 \leq \ell \leq 1500$ increases by a factor of up to 2.6 after applying Box-Cox or logarithmic transformations, in qualitative agreement with the results by \citet{seo11}. We find that the S/N is only a rough indicator of the strength of cosmological constraints, primarily because it does not account for degeneracies between parameters. 

Measuring the size of the confidence region in the $\Omega_{\rm m}-\sigma_8$ plane in terms of $q$-values, we obtain a significant degradation due to a near-perfect degeneracy between $\Omega_{\rm m}$ and $\sigma_8$ if the transformations are determined from the unsmoothed convergence fields, and a decrease in $q$ by up to a factor of 4.4 if transformations are optimised for the smoothing. Although contributions from e.g. the convergence bispectrum enter the transformed models, the degeneracy between $\Omega_{\rm m}$ and $\sigma_8$ is broken in neither case, which we ascribe to the cancellation of information through the integration over the phase dependence of the bispectrum (and higher-order correlations) as well as the summation over the two-point, three-point, and higher-order terms.
\item If a realistic level of galaxy shape noise is added to the convergence fields, transformations achieve an increase in the cumulative S/N by up to $34\,\%$, but leave the statistical errors of and correlations between cosmological parameters practically unchanged, if not mildly degraded. The failure to boost the information contained in the transformed power spectrum is firstly caused by the fact that shape noise already renders the distribution of convergence values more Gaussian, so that there is less to gain by a subsequent Gaussianisation, and secondly, the decorrelation of angular frequencies by the transformations performs worse. The latter result means that the approximation that Gaussianising the one-point distribution renders the full convergence field Gaussian is worse in the more realistic case with noise.
\end{enumerate}

All of the conclusions above depend on the angular frequency range included in the analysis since cosmological information might be re-distributed to scales outside this regime by the transformations and hence not recovered in our study. The low maximum $\ell=1500$ in our likelihood analysis (plus a suppression of signal by smoothing relevant for $\ell > 1000$) was governed by the limitations of analytical modelling. One way to extend the analysis deeper into the non-linear regime is the inclusion of marginalisation over free functional forms \citep{kitching10} to account for uncertainty in the modelling, which of course would degrade cosmological parameter constraints. Otherwise one has to resort to simulations to explore parameter space for the likelihood of the transformed power spectrum, as already proposed by \citet{yu11}; see also \citet{neyrinck11b}.

To prove that Gaussianising transformations of convergence fields are a viable and worthwhile approach to the analysis of upcoming weak lensing surveys, it is foremost necessary to demonstrate that cosmological information can be gained and the statistical properties of the transformed two-point statistics substantially improved under realistic conditions which include shape noise, a distribution of source galaxies in redshift, and the effects of masks on the convergence fields. It will be the subject of follow-up work to investigate whether this goal can be achieved by improved modelling, an extended angular frequency range, or by going beyond transformations of the one-point distribution.

\section*{Acknowledgments}

The authors would like to thank Alan Heavens for a careful reading of the manuscript and valuable comments. We are grateful to Gary Bernstein for useful suggestions and to our referee for an insightful report. BJ acknowledges support by the European DUEL network, project MRTN-CT-2006-036133, and a UK Space Agency Euclid grant. AK is supported by DUEL and a University of Edinburgh studentship.

\bibliographystyle{mn2e}

\onecolumn
\appendix

\section{Calculation of the Box-Cox transformed power spectrum}
\label{app:transformedps}

In this appendix we detail the calculation of the Box-Cox transformed convergence power spectrum from the statistics of the original convergence field. We take into account that the convergence has an additive noise component originating from the random intrinsic ellipticities of galaxies, indicated by a subscript n to $\kappa$. Furthermore we consider the smoothed convergence $\kappa'(\vek{\theta}) = \bb{\kappa_{\rm n}*W}(\vek{\theta})$, where $W$ is the Gaussian smoothing kernel. Via Taylor expansion we find
\eqa{
\label{eq:kappatrafo_detail}
\bar{\kappa}(\vek{\theta}) &=& \frac{1}{\lambda} \bc{\br{\kappa'(\vek{\theta})+a}^\lambda-1} = \frac{1}{\lambda} \br{a^\lambda-1} + a^{\lambda-1} \kappa'(\vek{\theta}) + \frac{\lambda-1}{2}\; a^{\lambda-2} {\kappa'}^2(\vek{\theta}) + \frac{(\lambda-1) (\lambda-2)}{6}\; a^{\lambda-3} {\kappa'}^3(\vek{\theta})\\ \nn
&& +\; \frac{(\lambda-1) (\lambda-2) (\lambda-3)}{24}\; a^{\lambda-4} {\kappa'}^4(\vek{\theta}) + \frac{(\lambda-1) (\lambda-2) (\lambda-3) (\lambda-4)}{120}\; a^{\lambda-5} {\kappa'}^5(\vek{\theta})\; +\; {\cal O}(\kappa^6)\;.
}
Except for the irrelevant zeroth-order term the expansion for the case $\lambda=0$ is identical. In analogy to equation (\ref{eq:kappatrafo}) we change to Fourier space and apply the convolution theorem on the powers of $\kappa'(\vek{\theta})$. Writing $\kappa'(\vek{\ell}) = \bb{\kappa_{\rm n} \times W}(\vek{\ell})$, one obtains for the two-point correlator of the transformed convergence ($\vek{\ell},\vek{\ell'} \neq 0$)
\eqa{\nn
\ba{\bar{\kappa}(\vek{\ell})\; \bar{\kappa}^*(\vek{\ell'})} &=& a^{2\lambda-2} \ba{\kappa'(\vek{\ell})\; \kappa'(\vek{-\ell'})} + \frac{\lambda-1}{2}\; a^{2\lambda-3} \int \frac{\dd^2 \ell_1}{(2 \pi)^2}\; \bigl\{ \ba{\kappa'(\vek{\ell})\; \kappa'(\vek{-\ell}_1)\; \kappa'(\vek{\ell}_1-\vek{\ell'})} + \ba{\kappa'(-\vek{\ell'})\; \kappa'(\vek{\ell}_1)\; \kappa'(\vek{\ell}-\vek{\ell}_1)} \bigr\}\\ \nn
&& \hspace*{-2.7cm} + \frac{(\lambda-1) (\lambda-2)}{6}\; a^{2\lambda-4}\!\! \int \frac{\dd^2 \ell_1}{(2 \pi)^2} \int \frac{\dd^2 \ell_2}{(2 \pi)^2}\; \bigl\{ \ba{\kappa'(\vek{\ell})\; \kappa'(-\vek{\ell}_1)\; \kappa'(-\vek{\ell}_2)\; \kappa'(\vek{\ell}_1+\vek{\ell}_2-\vek{\ell'})}  + \ba{\kappa'(-\vek{\ell'})\; \kappa'(\vek{\ell}_1)\; \kappa'(\vek{\ell}_2)\; \kappa'(\vek{\ell}-\vek{\ell}_1-\vek{\ell}_2)} \bigr\}\\ 
\label{eq:kappacorrelator1}
&& \hspace*{-2.7cm} +\; \frac{(\lambda-1)^2}{4}\; a^{2\lambda-4} \int \frac{\dd^2 \ell_1}{(2 \pi)^2} \int \frac{\dd^2 \ell_2}{(2 \pi)^2}\; \ba{\kappa'(\vek{\ell}_1)\; \kappa'(\vek{\ell}-\vek{\ell}_1)\; \kappa'(-\vek{\ell}_2)\; \kappa'(\vek{\ell}_2-\vek{\ell'})}\;  +\; {\cal O}(\kappa^5)\;.
}
Since the convergence is real, we could replace $\kappa^*(\vek{\ell})=\kappa(-\vek{\ell})$. The first four-point term is produced by correlating the first- and third-order contributions in equation (\ref{eq:kappatrafo_detail}); the second four-point term by correlating the second-order term of the original convergence. Correlations of higher order than $\kappa^4$ will be considered in Appendix \ref{app:higherorder}.

The noise contribution to the convergence results in an additional scale-independent power spectrum, so that $\ba{\kappa_{\rm n}(\vek{\ell})\; \kappa_{\rm n}(\vek{\ell'})} = (2\pi)^2\; \delta^{(2)}(\vek{\ell}+\vek{\ell'})\; \bb{P_\kappa(\ell) + P_{\rm noise}}$. Higher-order statistics are not affected by noise, so that equation (\ref{eq:defspectra}) can be used. We apply Wick's theorem to split up the four-point correlators of $\kappa$, arriving at
\eqa{
\nn
\ba{\kappa(\vek{\ell})\, \kappa(-\vek{\ell}_1)\, \kappa(-\vek{\ell}_2)\, \kappa(\vek{\ell}_1+\vek{\ell}_2-\vek{\ell'})} &=& \ba{\kappa(\vek{\ell})\, \kappa(-\vek{\ell}_1)\, \kappa(-\vek{\ell}_2)\, \kappa(\vek{\ell}_1+\vek{\ell}_2-\vek{\ell'})}_{\rm c} + \ba{\kappa(\vek{\ell})\, \kappa(-\vek{\ell}_1)} \ba{\kappa(-\vek{\ell}_2)\, \kappa(\vek{\ell}_1+\vek{\ell}_2-\vek{\ell'})}\\ \nn
&& \hspace*{-5.5cm}  + \ba{\kappa(\vek{\ell})\, \kappa(-\vek{\ell}_2)} \ba{\kappa(-\vek{\ell}_1)\, \kappa(\vek{\ell}_1+\vek{\ell}_2-\vek{\ell'})} + \ba{\kappa(\vek{\ell})\, \kappa(\vek{\ell}_1+\vek{\ell}_2-\vek{\ell'})} \ba{\kappa(-\vek{\ell}_2)\, \kappa(-\vek{\ell}_1)}\\ \nn
&& \hspace*{-5.7cm} = (2\pi)^2\; \delta^{(2)}(\vek{\ell}-\vek{\ell'})\; T_\kappa(\vek{\ell},-\vek{\ell}_1,-\vek{\ell}_2,\vek{\ell}_1+\vek{\ell}_2-\vek{\ell}) + (2\pi)^4\; \bigl\{ \delta^{(2)}(\vek{\ell}-\vek{\ell}_1)\; \delta^{(2)}(\vek{\ell'}-\vek{\ell}_1)\; P_\kappa(\ell)\; P_\kappa(\ell_2)\\ 
\label{eq:4ptkappacorrelator}
&& \hspace*{-5.5cm}  +\, \delta^{(2)}(\vek{\ell}-\vek{\ell}_2)\; \delta^{(2)}(\vek{\ell'}-\vek{\ell}_2)\; P_\kappa(\ell)\; P_\kappa(\ell_1) + \delta^{(2)}(\vek{\ell}-\vek{\ell'}+\vek{\ell}_1+\vek{\ell}_2)\; \delta^{(2)}(\vek{\ell}_1+\vek{\ell}_2)\; P_\kappa(\ell)\; P_\kappa(\ell_1) \bigr\}\;,
}
and likewise for $\kappa'$ and the other four-point correlators. In each of the power spectrum terms one delta function disappears after performing one of the angular frequency integrals, turning the other one into $\delta^{(2)}(\vek{\ell}-\vek{\ell'})$. Moreover, after renaming the remaining integration variable to $\vek{\ell}_1$, all products of power spectra in equation (\ref{eq:4ptkappacorrelator}) can be written as $P_\kappa(\ell)\, P_\kappa(\ell_1)$. Treating the other correlators analogously, one hence obtains
\eqa{
\label{eq:kappacorrelator2}
\ba{\bar{\kappa}(\vek{\ell})\; \bar{\kappa}^*(\vek{\ell'})} &=& (2\pi)^2\; \delta^{(2)}(\vek{\ell}-\vek{\ell'})\; a^{2\lambda-2}\; \Biggl\{ \bb{ P_\kappa(\ell) + P_{\rm noise}}\; W^2(\ell)\\ \nn
&& \hspace*{-2.7cm} +\; (\lambda-1)\; a^{-1}\; W(\ell) \int \frac{\dd^2 \ell_1}{(2 \pi)^2}\; B_\kappa(\ell,\ell_1,|\vek{\ell}-\vek{\ell}_1|)\; W(\ell_1)\; W(|\vek{\ell}-\vek{\ell}_1|)\\ \nn
&& \hspace*{-2.7cm} +\; \frac{(\lambda-1) (\lambda-2)}{3}\; a^{-2} \biggl\{ 3 \bb{ P_\kappa(\ell) + P_{\rm noise}}\; W^2(\ell) \int \frac{\dd^2 \ell_1}{(2 \pi)^2}\; \bb{ P_\kappa(\ell_1) + P_{\rm noise}}\; W^2(\ell_1)\\ \nn
&& \hspace*{-2.0cm} +\; W(\ell) \int \frac{\dd^2 \ell_1}{(2 \pi)^2} \int \frac{\dd^2 \ell_2}{(2 \pi)^2}\; T_\kappa(\vek{\ell},-\vek{\ell}_1,-\vek{\ell}_2,\vek{\ell}_1+\vek{\ell}_2-\vek{\ell})\; W(\ell_1)\; W(\ell_2)\; W(|\vek{\ell}_1+\vek{\ell}_2-\vek{\ell}|) \biggr\}\\ \nn
&& \hspace*{-2.7cm} +\; \frac{(\lambda-1)^2}{4}\; a^{-2} \biggl\{ 2 \int \frac{\dd^2 \ell_1}{(2 \pi)^2}\; \bb{ P_\kappa(\ell_1) + P_{\rm noise}}\; W^2(\ell_1) \bb{ P_\kappa(|\vek{\ell}-\vek{\ell}_1|) + P_{\rm noise}}\; W^2(|\vek{\ell}-\vek{\ell}_1|)\\ \nn
&& \hspace*{-2.0cm} +\; \int \frac{\dd^2 \ell_1}{(2 \pi)^2} \int \frac{\dd^2 \ell_2}{(2 \pi)^2}\; T_\kappa(\vek{\ell}_1,\vek{\ell}-\vek{\ell}_1,-\vek{\ell}_2,\vek{\ell}_2-\vek{\ell})\; W(\ell_1)\; W(|\vek{\ell}-\vek{\ell}_1|)\; W(\ell_2)\; W(|\vek{\ell}_2-\vek{\ell}|) \biggr\} + {\cal O}(\kappa^5) \Biggr\}\;.
}
Note that the Fourier transform of $W$ depends only on the modulus of the angular frequency. Invoking equation (\ref{eq:defspectra}) again, an expression for the Box-Cox transformed power spectrum immediately follows, which reduces to equation (\ref{eq:pstrafo_basic_approx}) if smoothing and noise are neglected. Equation (\ref{eq:pstrafo_full_approx}) follows from equation (\ref{eq:kappacorrelator2}) if one defers the connected trispectrum terms to the higher-order contribution ${\cal H}(\ell)$, and if one re-writes the two-dimensional integration over $\vek{\ell}_1$ as a radial integral over $\ell_1$ and an integral over the angle $\varphi$ between $\vek{\ell}$ and $\vek{\ell}_1$, noting that $|\vek{\ell}-\vek{\ell}_1|^2=\ell^2 + \ell_1^2 - 2 \ell \ell_1 \cos \varphi$.

\section{Modelling higher-order contributions to the transformed power spectrum}
\label{app:higherorder}

Including contributions only up to the Gaussian four-point level into the model of the transformed power spectrum results in significantly biased parameter constraints for both Box-Cox and logarithmic transformations. Hence higher-order terms are important and need to be included. Since Gaussianising transformations of the convergence using the logarithm perform well at least at the one-point level, we make the assumption that the original convergence is lognormal distributed, which allows one to proceed with analytic means.

Under the lognormal assumption we derive in Appendix \ref{app:lognormal} a closed-form relation between the original and transformed two-point convergence statistics which, however, still yields poor fits to the simulations. Yet, by means of this relation we are able to calculate expressions for higher-order correlations of the lognormal distributed convergence in Appendix \ref{app:lognormalcorr}. By comparing the moments of the convergence obtained from the simulations and from our models in Appendix \ref{app:moments}, we normalise the different contributions to match results from the simulated original convergence fields, arriving at the final expression (\ref{eq:higherordersummary}) for the higher-order contribution to the transformed power spectrum models.

\subsection{The lognormal model}
\label{app:lognormal}

A logarithmic transformation renders the one-point distribution of the convergence, both with and without shape noise, Gaussian to good approximation; see above, and e.g. \citet{taruya02}. Hence it is reasonable to assume that the original convergence follows a lognormal distribution although the non-vanishing higher-order statistics of the logarithmically transformed fields indicate that the lognormal assumption cannot be perfect.

Analogous to equation (\ref{eq:boxcox}) for $\lambda=0$ we write for the transformed convergence $\bar{\kappa}= \ln (\kappa+a) + N$, where we have now introduced a normalisation $N$ to ensure that $\bar{\kappa}$ has vanishing expectation. Solving this equation for $\kappa$ and choosing $N$ such that $\langle \bar{\kappa} \rangle=0$, one obtains
\eq{
\label{eq:lognormal1pt}
\kappa + a = a\; \exp \bc{\bar{\kappa} - \frac{\bar{\sigma}^2}{2}}\;,
}
where $\bar{\sigma}$ denotes the variance of the transformed convergence. Note that throughout this appendix we do not include the effects of noise and smoothing on the convergence to keep the notation tractable.

If $\kappa$ is lognormal distributed, $\bar{\kappa}$ follows a Gaussian distribution which allows us to compute expectation values analytically via Gaussian integration. In the case of two-point convergence statistics, this results in
\eq{
\label{eq:derivelognormalxi}
\ba{\bb{\kappa(\vek{x})+a}\; \bb{\kappa(\vek{x}+\vek{\theta})+a}} 
= \xi_\kappa(\theta) + a^2
= a^2 \ba{\exp \bc{\bar{\kappa}(\vek{x}) - \frac{\bar{\sigma}^2}{2}}\; \exp \bc{\bar{\kappa}(\vek{x}+\vek{\theta}) - \frac{\bar{\sigma}^2}{2}}} 
= a^2\; \exp \bc{\xi_{\bar{\kappa}}(\theta)}\;,
}
where we have made use of $\langle \kappa \rangle=0$ and the definition of the correlation function $\xi$. Solving for the correlation function of $\bar{\kappa}$, one arrives at
\eq{
\label{eq:lognormal2pt}
\xi_{\bar{\kappa}}(\theta) = \ln \bb{1 + a^{-2}\, \xi_\kappa(\theta)}
= a^{-2}\, \xi_\kappa(\theta) - \frac{1}{2} a^{-4}\, \xi_\kappa^2(\theta) + \frac{1}{3} a^{-6}\, \xi_\kappa^3(\theta) + {\cal O}(\xi_\kappa^4)\;,
}
where the second equality is derived from a Taylor expansion. The first equality provides us with a closed-form relation between transformed and original correlation functions (see also \citealp{hilbert11}, as well as \citealp{coles91} for a similar result) which can readily be converted into a closed-form relation between transformed and original power spectra as $\xi_\kappa$ and $P_\kappa$ are Hankel transform pairs \citep[e.g.][]{SvWM02}.

\begin{figure}
\begin{minipage}[c]{0.5\textwidth}
\centering
\includegraphics[scale=.33,angle=270]{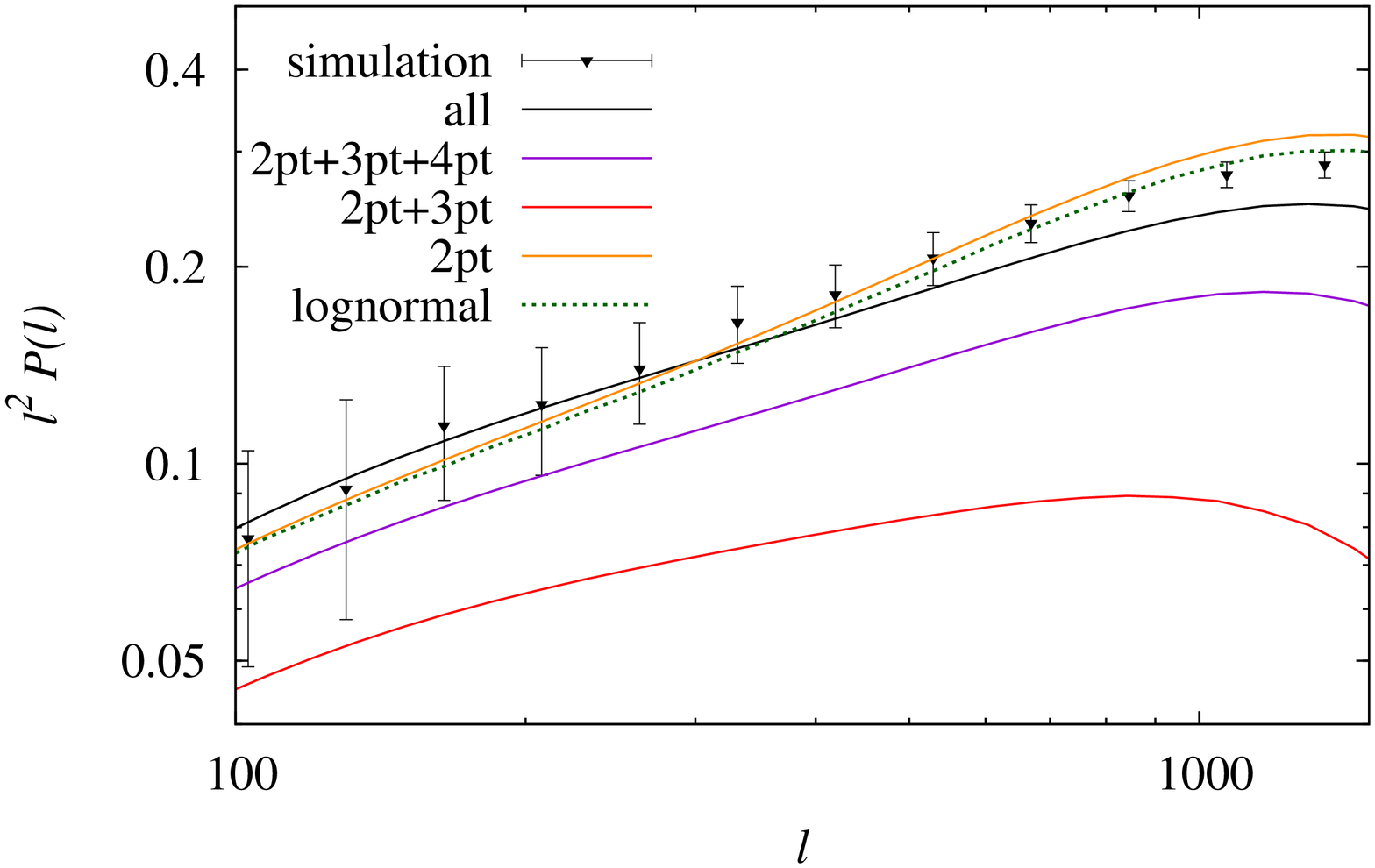}
\end{minipage}%
\begin{minipage}[c]{0.5\textwidth}
\caption{Same as Fig.$\,$\ref{fig:meanps}, bottom panel, but showing the logarithmically transformed power spectrum (LOG2). In addition the model based on the lognormal prediction is shown as green dotted curve. Note that both modelling approaches fail to match the simulation data on small scales.}
\label{fig:meanps_logsmooth}
\end{minipage}
\end{figure}

However, although not suffering from the truncation at a certain order in $\kappa$, the lognormal model for the transformed power spectrum fails to fit the simulations on small scales, as illustrated in Fig.$\,$\ref{fig:meanps_logsmooth} for the LOG2 transformation (we find similar results for LOG1). It is interesting to note that the lognormal model remains very close to the two-point contribution $a^{-2} P_\kappa(\ell)$ which implies that the higher-order terms almost cancel each other; see the alternating signs of the expansion in equation (\ref{eq:lognormal2pt}).

We ascribe the shortcomings of the lognormal model to the failure of providing a fair representation of the three-point, four-point and possibly higher-order statistics of the simulated convergence fields. Hence we seek to study the different orders of an expansion in $\kappa$ individually within the lognormal framework and match the resulting models to the simulations. A further motivation for this approach is that it will enable us to calculate expressions for arbitrary $\lambda$. 

Transforming the expansion in equation (\ref{eq:lognormal2pt}) to Fourier space yields
\eqa{
\nn
P_{\bar{\kappa}}(\ell) &=& a^{-2}\, P_\kappa(\ell) - \frac{1}{2} a^{-4} \int \frac{\dd^2 \ell_1}{(2 \pi)^2}\; P_{\kappa}(\ell_1) P_{\kappa}(|\vek{\ell}-\vek{\ell}_1|) + \frac{1}{3} a^{-6} \int \frac{\dd^2 \ell_1}{(2 \pi)^2} \int \frac{\dd^2 \ell_2}{(2 \pi)^2}\; P_{\kappa}(\ell_1) P_{\kappa}(\ell_2) P_{\kappa}(|\vek{\ell}-\vek{\ell}_1-\vek{\ell}_2|)+ {\cal O}(P_\kappa^4)\\ 
\label{eq:lognormal2ptps}
&\equiv& a^{-2}\, P_\kappa(\ell) - \frac{1}{2} a^{-4}\, {\cal B}(\ell) + \frac{1}{3} a^{-6}\, {\cal C}(\ell) + {\cal O}(P_\kappa^4)\;,
}
where we defined the shorthands ${\cal B}(\ell)$ and ${\cal C}(\ell)$ for convenience. Note that ${\cal B}(\ell)$ is the Fourier transform of $\xi_\kappa^2$, and ${\cal C}(\ell)$ of $\xi_\kappa^3$. The first term in equation (\ref{eq:lognormal2ptps}) is identical to the one in equation (\ref{eq:pstrafo_basic_approx}) for $\lambda=0$, as expected. The second term is of the same form as one of the Gaussian four-point contributions, but has a different prefactor. As we shall see below, this term receives additional contributions from the lognormal bispectrum. The third term collects contributions of order $P_\kappa^3$ which are not included in equation (\ref{eq:pstrafo_basic_approx}), but which may be important particularly for small $a$. In the next section we will identify and explicitly calculate all contributions that are third order in the two-point statistic within the lognormal framework.

\subsection{Lognormal higher-order correlations}
\label{app:lognormalcorr}

In complete analogy to the steps performed in equation (\ref{eq:derivelognormalxi}) one can derive relations between the n-point correlations of $\kappa$ and the correlation function $\xi_{\bar{\kappa}}$. Subsequently applying equation (\ref{eq:lognormal2pt}), one can express the n-point correlations in terms of the two-point statistic of the original convergence field as follows,
\eq{
\label{eq:lognormalgeneral}
\ba{\bb{\kappa(\vek{x}_1)+a}\; \dots\; \bb{\kappa(\vek{x}_n)+a}}
= a^n\; \exp \bc {\sum_{i < j}^n \xi_{\bar{\kappa}}(|\vek{x}_j-\vek{x}_i|)}
= a^n \prod_{i < j}^n \bb{1 + a^{-2}\, \xi_{\kappa}(|\vek{x}_j-\vek{x}_i|)}\;.
}
In the case of three-point statistics the correlator on the left-hand side can be expanded into
\eq{
\label{eq:derivelognormalgamma}
\ba{\bb{\kappa(\vek{x})+a}\; \bb{\kappa(\vek{x}+\vek{\theta}_1)+a}\; \bb{\kappa(\vek{x}+\vek{\theta}_2)+a}} 
= \Gamma_\kappa(\vek{\theta}_1,\vek{\theta}_2) +a \bc{\xi_{\kappa}(\theta_1) + \xi_{\kappa}(\theta_2) + \xi_{\kappa}(\theta_{12})} + a^3\;,
}
where the shorthand notation $\theta_{ij} = |\vek{\theta}_j-\vek{\theta}_i|$ was introduced. We defined $\Gamma_\kappa(\vek{\theta}_1,\vek{\theta}_2)$ as the three-point correlation function of the convergence, i.e. the Fourier transform of the convergence bispectrum. Due to the homogeneity of the convergence field, two angular vectors suffice to specify $\Gamma_\kappa$; we have not invoked rotational invariance at this stage.

Together with equation (\ref{eq:lognormalgeneral}), one can derive the lognormal three-point correlation function
\eq{
\label{eq:lognormal3pt}
\Gamma_{\kappa,{\rm LN}}(\vek{\theta}_1,\vek{\theta}_2) = a^{-1} \bc{\xi_{\kappa}(\theta_1) \xi_{\kappa}(\theta_2) + \xi_{\kappa}(\theta_1) \xi_{\kappa}(\theta_{12}) + \xi_{\kappa}(\theta_2) \xi_{\kappa}(\theta_{12})} + a^{-3}\; \xi_{\kappa}(\theta_1) \xi_{\kappa}(\theta_2) \xi_{\kappa}(\theta_{12})\;,
}
which, after Fourier transformation, yields the lognormal bispectrum
\eq{
\label{eq:lognormal3ptbs}
B_{\kappa,{\rm LN}}(\ell_1,\ell_2,\ell_3) = a^{-1} \bc{ P_\kappa(\ell_1) P_\kappa(\ell_2) + P_\kappa(\ell_1) P_\kappa(\ell_3) + P_\kappa(\ell_2) P_\kappa(\ell_3) } + a^{-3} \int \frac{\dd^2 \ell_4}{(2 \pi)^2}\; P_{\kappa}(\ell_4) P_{\kappa}(|\vek{\ell}_1+\vek{\ell}_4|) P_{\kappa}(|\vek{\ell}_2-\vek{\ell}_4|)\;.
}
Note that we use a subscript LN to label expressions that were obtained under the assumption of lognormality of $\kappa$. This result allows us to compute the third-order integral expression entering equation (\ref{eq:pstrafo_basic_approx}) in the lognormal case,
\eq{
\label{eq:lognormal3ptbsint}
\int \frac{\dd^2 \ell_1}{(2 \pi)^2}\; B_{\kappa,{\rm LN}}(\ell,\ell_1,|\vek{\ell}-\vek{\ell}_1|) = a^{-1} \bc{ 2\, \sigma_\kappa^2\; P_\kappa(\ell) + {\cal B}(\ell)} + a^{-3} \sigma_\kappa^2\; {\cal B}(\ell)\;,
}
where we defined the variance of the convergence given by
\eq{
\label{eq:defsigmakappa}
\sigma_\kappa^2 \equiv \xi_{\kappa}(0) = \int_0^\infty \frac{\dd \ell\; \ell}{2 \pi}\; P_\kappa(\ell)\;.
}
Terms second order in $P_\kappa$ contributing to equation (\ref{eq:lognormal3ptbsint}) are of exactly the same form as the Gaussian four-point contributions. Taking the prefactors into account, see equation (\ref{eq:pstrafo_basic_approx}), the terms proportional to $\sigma_\kappa^2\; P_\kappa(\ell)$ cancel, while the terms proportional to ${\cal B}(\ell)$ reduce to the second-order contribution to equation (\ref{eq:lognormal2ptps}).

The procedure spelled out in equations (\ref{eq:lognormalgeneral}) and (\ref{eq:derivelognormalgamma}) can readily be applied to four-point statistics. Note however that the four-point correlator splits into connected and unconnected parts, the latter reproducing the Gaussian four-point contribution, and the former yielding the connected four-point correlation function for which we find
\eq{
\label{eq:lognormal4pt}
\eta_{\kappa,{\rm LN}}(\vek{\theta}_1,\vek{\theta}_2,\vek{\theta}_3) = a^{-2} \bc{\xi_{\kappa}(\theta_1) \xi_{\kappa}(\theta_2) \xi_{\kappa}(\theta_3) + \xi_{\kappa}(\theta_1) \xi_{\kappa}(\theta_2) \xi_{\kappa}(\theta_{13}) + \xi_{\kappa}(\theta_1) \xi_{\kappa}(\theta_2) \xi_{\kappa}(\theta_{23}) + \mbox{13 perm.} } + {\cal O}(\xi_\kappa^4)\;.
}
Note that equations (\ref{eq:lognormal3pt}) and (\ref{eq:lognormal4pt}) are in agreement with the results found by \citet{hilbert11}.

The corresponding expression for the lognormal trispectrum is readily derived but lengthy, containing also 16 terms that are third order in $P_\kappa$. Considerable simplification is achieved by performing the two integrals entering equation (\ref{eq:pstrafo_basic_approx}), yielding
\eqa{
\label{eq:lognormal4ptts}
\int \frac{\dd^2 \ell_1}{(2 \pi)^2} \int \frac{\dd^2 \ell_2}{(2 \pi)^2}\; T_{\kappa,{\rm LN}}(\vek{\ell},-\vek{\ell}_1,-\vek{\ell}_2,\vek{\ell}_1+\vek{\ell}_2-\vek{\ell}) &=& a^{-2} \bc{9\, \sigma_\kappa^4\; P_\kappa(\ell) + 6\, \sigma_\kappa^2\; {\cal B}(\ell) + {\cal C}(\ell)} + {\cal O}(P_{\kappa}^4)\;;\\ \nn
\int \frac{\dd^2 \ell_1}{(2 \pi)^2} \int \frac{\dd^2 \ell_2}{(2 \pi)^2}\; T_{\kappa,{\rm LN}}(\vek{\ell}_1,\vek{\ell}-\vek{\ell}_1,-\vek{\ell}_2,\vek{\ell}_2-\vek{\ell}) &=& a^{-2} \bc{4\, \sigma_\kappa^4\; P_\kappa(\ell) + 8\, \sigma_\kappa^2\; {\cal B}(\ell) + 4\, {\cal C}(\ell)} + {\cal O}(P_{\kappa}^4)\;.
}
Note that the lognormal trispectrum and all lognormal n-point correlations with $n \geq 5$ do not contain terms proportional to $P_{\kappa}^2$ anymore.

The connected part of the five-point convergence correlation contains at least powers of 4 in the two-point statistic, so that we only have to consider the unconnected parts, consisting of products of two- and three-point statistics. Inserting the terms proportional to $\xi_{\kappa}^2$ in equation (\ref{eq:lognormal3pt}) into the five-point analogues of equations (\ref{eq:lognormalgeneral}) and (\ref{eq:derivelognormalgamma}), one obtains
\eq{
\label{eq:lognormal5pt}
\ba{\kappa(\vek{x})\; \kappa(\vek{x}+\vek{\theta}_1)\; \dots\; \kappa(\vek{x}+\vek{\theta}_4)}_{\rm LN} 
= a^{-1} \bc{ \xi_{\kappa}(\theta_1) \bb{\xi_{\kappa}(\theta_{23}) \xi_{\kappa}(\theta_{24}) + \xi_{\kappa}(\theta_{23}) \xi_{\kappa}(\theta_{34}) + \xi_{\kappa}(\theta_{24}) \xi_{\kappa}(\theta_{34})} + \mbox{9 perm.} } + {\cal O}(\xi_{\kappa}^4)\;.
}
Writing the real-space analogue of the computation done in equation (\ref{eq:kappacorrelator1}), one realises that only correlators of the form $\ba{\kappa^i(\vek{x})\; \kappa^j(\vek{x}+\vek{\theta})}$ are required to calculate the transformed convergence power spectrum, i.e. in the five-point case it is sufficient to determine
\eqa{
\label{eq:5ptterms}
\ba{\kappa(\vek{x})\; \kappa^4(\vek{x}+\vek{\theta})}_{\rm LN}
= \ba{\kappa^4(\vek{x})\; \kappa(\vek{x}+\vek{\theta})}_{\rm LN}
&=& a^{-1} \bc{6\, \sigma_\kappa^2\; \xi_{\kappa}^2(\theta) + 24\, \sigma_\kappa^4\; \xi_{\kappa}(\theta)} + {\cal O}(\xi_{\kappa}^4)\;;\\ \nn
\ba{\kappa^2(\vek{x})\; \kappa^3(\vek{x}+\vek{\theta})}_{\rm LN} 
= \ba{\kappa^3(\vek{x})\; \kappa^2(\vek{x}+\vek{\theta})}_{\rm LN} 
&=& a^{-1} \bc{6\, \xi_{\kappa}^3(\theta) + 15\, \sigma_\kappa^2\; \xi_{\kappa}^2(\theta) + 6\, \sigma_\kappa^4\; \xi_{\kappa}(\theta) + 3\, \sigma_\kappa^6} + {\cal O}(\xi_{\kappa}^4)\;.
}
Likewise, at the six-point level only the following correlators are needed,
\eqa{
\label{eq:6ptterms}
\ba{\kappa(\vek{x})\; \kappa^5(\vek{x}+\vek{\theta})} 
= \ba{\kappa^5(\vek{x})\; \kappa(\vek{x}+\vek{\theta})} 
&=& 15\, \sigma_\kappa^4\; \xi_{\kappa}(\theta) + {\cal O}(\xi_{\kappa}^4)\;;\\ \nn
\ba{\kappa^2(\vek{x})\; \kappa^4(\vek{x}+\vek{\theta})} 
= \ba{\kappa^4(\vek{x})\; \kappa^2(\vek{x}+\vek{\theta})} 
&=& 12\, \sigma_\kappa^2\; \xi_{\kappa}^2(\theta) + 3\, \sigma_\kappa^6 + {\cal O}(\xi_{\kappa}^4)\;;\\ \nn
\ba{\kappa^3(\vek{x})\; \kappa^3(\vek{x}+\vek{\theta})} 
&=& 6\, \xi_{\kappa}^3(\theta) + 9\, \sigma_\kappa^4\; \xi_{\kappa}(\theta) + {\cal O}(\xi_{\kappa}^4)\;.
}
Note that to third order in $\xi_{\kappa}$ only the Gaussian terms, i.e. triple products of two-point correlators, contribute to equation (\ref{eq:6ptterms}), so that the results do not rely on the lognormal assumption (hence the omission of the subscript LN).

Noting again the correspondence $\xi_{\kappa}^2(\theta) \leftrightarrow {\cal B}(\ell)$ and $\xi_{\kappa}^3(\theta) \leftrightarrow {\cal C}(\ell)$, equations (\ref{eq:5ptterms}) and (\ref{eq:6ptterms}) are readily transformed to Fourier space. Extending the calculation of equation (\ref{eq:kappacorrelator1}), with all necessary terms in the expansion of $\bar{\kappa}$ given by equation (\ref{eq:kappatrafo_detail}), we obtain the five- and six-point contribution to $P_{\bar{\kappa}}$
\eqa{
\label{eq:transformedps5plus6}
P_{\bar{\kappa}}^{(5+6)}(\ell) \!\!\! &=& \!\!\! a^{2\lambda-6} \Biggl\{ \frac{(\lambda-1)(\lambda-2)(\lambda-3)}{2} \bb{4\, \sigma_\kappa^4\; P_{\kappa}(\ell) + \sigma_\kappa^2\; {\cal B}(\ell)} + \frac{(\lambda-1)^2(\lambda-2)}{2} \bb{2\, \sigma_\kappa^4\; P_{\kappa}(\ell) + 5\, \sigma_\kappa^2\; {\cal B}(\ell) + 2\, {\cal C}(\ell)}\\ \nn
&& \hspace*{-2.2cm}+ \frac{(\lambda-1)(\lambda-2)(\lambda-3)(\lambda-4)}{4} \sigma_\kappa^4\; P_{\kappa}(\ell) + \frac{(\lambda-1)^2(\lambda-2)(\lambda-3)}{2} \sigma_\kappa^2\; {\cal B}(\ell) + \frac{(\lambda-1)^2(\lambda-2)^2}{12} \bb{3\, \sigma_\kappa^4\; P_{\kappa}(\ell) + 2\, {\cal C}(\ell)} \Biggr\}\! +\! {\cal O}(P_{\kappa}^4)\;.
}
Inserting equation (\ref{eq:lognormal4ptts}) and the $P_\kappa^3$ contribution to equation (\ref{eq:lognormal3ptbsint}) into the general expansion given by equation (\ref{eq:pstrafo_basic_approx}), plus adding the equation above, one can show for $\lambda=0$ that the terms proportional to $\sigma_\kappa^4\; P_{\kappa}(\ell)$ and $\sigma_\kappa^2\; {\cal B}(\ell)$ cancel, whereas the terms containing ${\cal C}(\ell)$ add up to reproduce the third term in equation (\ref{eq:lognormal2ptps}), as desired. This implies that indeed terms of the same power in $\xi_\kappa$ from different orders in $\kappa$ nearly or fully cancel each other under the lognormal assumption, resulting in only small corrections to the lognormal model. However, if the amplitudes of the higher-order statistics of the simulations are not well represented by the lognormal model, these cancellations will not occur anymore and cause substantially different signals.

\subsection{Moment normalisation}
\label{app:moments}

We compare the connected third and fourth moments of the original convergence fields as determined from the simulations and the models we employ. If finding a discrepancy, we normalise the corresponding contributions to the models to match the amplitude of the simulations. This implicitly assumes that the angular dependence of the various contributions is modelled correctly. We choose the moment as the quantity used for this comparison because it is given by the integration over all angular dependencies of the corresponding polyspectrum, thereby testing the models on all relevant scales.

\begin{table}
\begin{minipage}[c]{0.65\textwidth}
\centering
\caption{Normalisations $r_i$ of contributions to the transformed power spectrum of order $i=3,\,..\,,6$ in $\kappa$, as determined from the $i$th moment measured from the 100 simulation realisations. The second (third) column displays results for the case without (with) shape noise in the convergence fields. The coefficients $r_4$ and $r_5$ are different because they are determined from the lognormal prediction which depends on $a_{\rm LN}$, the free parameter in the lognormal distribution.}
\label{tab:momentnorm}
\end{minipage}%
\begin{minipage}[c]{0.35\textwidth}
\centering
\begin{tabular}[t]{lrr}
\hline\hline
order & noise-free        & shape noise       \\
\hline
$r_3$ & $(1.30 \pm 0.04)$ & $(1.29 \pm 0.04)$ \\
$r_4$ & $(4.06 \pm 0.42)$ & $(5.19 \pm 0.50)$ \\
$r_5$ & $(1.65 \pm 0.05)$ & $(1.42 \pm 0.04)$ \\
$r_6$ & $1$               & $1$               \\
\hline
$a_{\rm LN}$  & 0.03       & 0.07              \\
\hline
\end{tabular}
\end{minipage}
\end{table}

At the three-point level we continue to use the bispectrum model based on perturbation theory and the fitting formula by \citet{scoccimarro01}. We calculate the third-order moment from the bispectrum via
\eq{
\label{eq:moment3rd}
\ba{\kappa^3} = \int_0^\infty \frac{\dd \ell_1\; \ell_1}{2\pi} \int_0^\infty \frac{\dd \ell_2\; \ell_2}{2\pi} \int_0^\pi \frac{\dd \varphi}{\pi}\; B_\kappa\bb{\ell_1,\ell_2,\ell_\Delta(\ell_1,\ell_2,\varphi)}\;,
}
where $\ell_\Delta$ is given by equation (\ref{eq:thirdell}). The normalisation is then defined as $r_i \equiv \langle \kappa^i \rangle_{\rm sim.}/\langle \kappa^i \rangle_{\rm model}$, the results summarised in Table \ref{tab:momentnorm}. The uncertainty quoted in the table originates from the error on the mean simulation moment measured from the 100 realisations. For both the convergence fields with and without shape noise we find that the moment as obtained from the simulations is about $30\,\%$ higher than predicted by the perturbation theory bispectrum model. The sign of this deviation is in agreement with the underestimation of the bispectrum in the $\Lambda$CDM case by the fit formula, but is larger than the quoted $15\,\%$ average discrepancy. We have also estimated that the uncertainty due to the fit formula should increase with $\ell$, see Section \ref{sec:meanps}, but nonetheless find that a simple rescaling with $r_3$ yields satisfactory fits of the models for $P_{\bar{\kappa}}$ to the simulations.

At the five-point level we intend to include the unconnected parts, consisting of products of two- and three-point correlation functions, into the modelling. We refrain from using the perturbation theory bispectrum in this case as this would necessitate computationally expensive convolutions of power spectra and bispectra. Instead, we work under the lognormal assumption which allows us to use the simple expressions contained in equation (\ref{eq:transformedps5plus6}). While the second moments of simulated and modelled convergence should agree well, and indeed do, we need to match the third moment of the simulation with the lognormal one, given by
\eq{
\label{eq:moment3rd2}
\ba{\kappa^3}_{\rm LN} \equiv \Gamma_{\kappa,{\rm LN}}(0,0) = \frac{3}{a}\; \sigma_\kappa^4 + \frac{1}{a^3}\; \sigma_\kappa^6 \;,
}
cf. equation (\ref{eq:lognormal3pt}). The lognormal model underestimates the third moment even stronger than the one based on \citet{scoccimarro01}, producing $r_5 \sim 1.5$. The normalisation $r_5$ is different for the noise-free convergence fields and those with shape noise because in the latter case the variance $\sigma_\kappa^2$ is larger due to the noise, changing the result of equation (\ref{eq:moment3rd2}). Moreover the parameter $a$ which enters the lognormal models, chosen to render $\bar{\kappa}$ as close to Gaussian as possible, is not the same, see Table \ref{tab:trafos}. 

Repeating the steps that lead to the expressions for the lognormal polyspectra, now setting $\theta_1= \dots =\theta_n=0$, it is straightforward, though tedious, to calculate higher moments,
\eqa{
\label{eq:momenthigher}
\ba{\kappa^4}_{\rm LN} &=& 3\, \sigma_\kappa^4 + 16\, a^{-2}\; \sigma_\kappa^6 + 15\, a^{-4}\; \sigma_\kappa^8 + 6\, a^{-6}\; \sigma_\kappa^{10} + a^{-8}\; \sigma_\kappa^{12}\;;\\ \nn
\ba{\kappa^5}_{\rm LN} &=& 30\, a^{-1}\; \sigma_\kappa^6 + 135\, a^{-3}\; \sigma_\kappa^8 + 222\, a^{-5}\; \sigma_\kappa^{10} + 205\, a^{-7}\; \sigma_\kappa^{12} + {\cal O}(\kappa^{14})\;;\\ \nn
\ba{\kappa^6}_{\rm LN} &=& 15\, \sigma_\kappa^6 + 330\, a^{-2}\; \sigma_\kappa^8 + 1581\, a^{-4}\; \sigma_\kappa^{10} + 3760\, a^{-6}\; \sigma_\kappa^{12} + {\cal O}(\kappa^{14})\;.
}
The first term contributing to $\langle \kappa^4 \rangle_{\rm LN}$ stems from the unconnected part and corresponds to the Gaussian four-point term. The remaining terms originate from the trispectrum, where the second one can be identified with equation (\ref{eq:lognormal4pt}). Comparing the connected fourth moments, we obtain $r_4 \approx 4$ for the noise-free convergence and $r_4 \approx 5$ in the case with shape noise, as also shown in Table \ref{tab:momentnorm}. This means that the rescaling of the trispectrum contributions to the model is quite substantial and is in addition associated with a $10\,\%$ uncertainty; see Fig.$\,$\ref{fig:meanps} for an illustration of the effect on the transformed power spectrum.

We set $r_6=1$ since only Gaussian terms contribute to our models at the six-point level which should be modelled accurately. Connected fifth and sixth moments can only be determined with large error bars from the simulation, but as a tendency we find that they are considerably above the lognormal prediction given by equation (\ref{eq:momenthigher}), the ratios surpassing $r \sim 10$. While the simulations thus seem to favour even stronger higher-order contributions, we can still use the lognormal expressions in equation (\ref{eq:momenthigher}) for a conservative estimate on how much an error is introduced when truncating the series of contributions to the transformed power spectrum after terms containing $P_\kappa^3$.

For the smallest value of $a$ we consider, $a=0.03$ for LOG2, the term proportional to $\sigma_\kappa^8$ contributes $12\,\%$ to the leading term of the connected fourth moment; for $a=0.07$ (LOG1) this reduces to a $2\,\%$ contribution. Similarly, at the five-point level we find the ratio of next-to-leading over leading term to be $57\,\%$ ($a=0.03$, LOG2) and $10\,\%$ ($a=0.07$, LOG1), respectively. For the sixth moment the higher-order contributions can even dominate, yielding ratios over the first, Gaussian term of 2.76 ($\sigma_\kappa^8$ term), 1.66 ($\sigma_\kappa^{10}$ term), and 0.5 ($\sigma_\kappa^{12}$ term) for $a=0.03$. We understand these findings as the most likely explanation for the clear failure of our models in the LOG2 case (see Fig.$\,$\ref{fig:meanps_logsmooth}). Due to the small value of $a$ higher-order terms are boosted, with particularly strong contributions from positive six-point correlations, which could be the cause of the simulation signal being high at large $\ell$ compared to the model. All other transformations considered in this work have $a \geq 0.07$ for which the terms that are not included into the models may not be completely negligible, but are clearly subdominant.

If we want the higher-order modelling to be of practical use for all Box-Cox transformations, we have to differentiate between powers of $a$ originating from the expansion of $\bar{\kappa}$, e.g. those appearing in equation (\ref{eq:pstrafo_basic_approx}), from those entering via the lognormal models as in equations (\ref{eq:lognormal3ptbsint}) and (\ref{eq:lognormal4ptts}). Note that we have not yet made this distinction in equation (\ref{eq:transformedps5plus6}), so that all contributions of order $P_\kappa^3$ had a prefactor $a^{2\lambda-6}$. We add a subscript LN to $a$ from the lognormal models, keeping these parameters fixed at $a_{\rm LN}=0.03$ (noise-free) and $a_{\rm LN}=0.07$ (shape noise). Collecting all higher-order contributions, and incorporating the normalisations to the simulation moments, we finally obtain the model term
\eqa{
\label{eq:higherordersummary}
{\cal H}(\ell) &=& C_A\; \sigma_\kappa^4\; P_{\kappa}(\ell) + C_B\; \sigma_\kappa^2\; {\cal B}(\ell) + C_C\; {\cal C}(\ell)  \hspace*{0.6cm} \mbox{with}\\ \nn
C_A &=& r_4\; a^{2\lambda-4} a_{\rm LN}^{-2}\; (\lambda-1) (4\lambda-7) + r_5\; a^{2\lambda-5} a_{\rm LN}^{-1}\; (\lambda-1) (\lambda-2) (3\lambda-7) + r_6\; a^{2\lambda-6}\; (\lambda-1) (\lambda-2) (\lambda^2-5\lambda+7)/2\;;\\ \nn
C_B &=& r_4\; a^{2\lambda-4} a_{\rm LN}^{-2}\; 2 (\lambda-1) (2\lambda-3) + r_5\; a^{2\lambda-5} a_{\rm LN}^{-1}\; (\lambda-1) (\lambda-2) (3\lambda-4) + r_6\; a^{2\lambda-6}\; (\lambda-1)^2 (\lambda-2) (\lambda-3)/2\;;\\ \nn
C_C &=& r_4\; a^{2\lambda-4} a_{\rm LN}^{-2}\; (\lambda-1) (4\lambda-5)/3 + r_5\; a^{2\lambda-5} a_{\rm LN}^{-1}\; (\lambda-1)^2 (\lambda-2) + r_6\; a^{2\lambda-6}\; (\lambda-1)^2 (\lambda-2)^2/6\;.
}
To summarise, ${\cal H}(\ell)$ contains contributions from the leading term of the lognormal trispectrum, the unconnected five-point correlations also in the lognormal framework, and the Gaussian six-point term. Note that the effects of noise and smoothing still need to be incorporated into equation (\ref{eq:higherordersummary}).

\section{Likelihood estimation based on the convergence as the data-vector}
\label{app:likekappa}

If one succeeds in transforming convergence maps such that they are close to a Gaussian random field, it becomes advantageous to compute a likelihood for $\kappa$ instead of the widespread likelihood analysis for two-point statistics. The latter are generally chosen because two-point statistics provide a certain amount of data compression and are expected to be closer to Gaussian distributed than the underlying field. However, weak lensing two-point statistics are not accurately modelled by a Gaussian distribution \citep[see e.g.][]{hartlap09,Schneider09} and require the computation of four-point statistics to obtain covariances \citep[e.g.][]{takada09,pielorz10}. If one treats the convergence itself as the data-vector, the covariance is a two-point statistic which contains the cosmological information \citep[e.g.][]{heavens03}.

In the following we will outline our formalism for the likelihood for $\kappa$ and compare its Fisher information to the one of the power spectrum likelihood. We continue to work in Fourier space and compose the data-vector of the values $\kappa_{\svek{\ell}}$ of the convergence on the grid of a discrete Fourier transformation, defined via
\eq{
\label{eq:kappadft}
\kappa(\vek{\theta}) = \int \frac{\dd^2 \ell}{(2\pi)^2}\; \kappa(\vek{\ell})\; \expo{\ic \svek{\theta} \cdot \svek{\ell}} \approx \sum_i \br{ \frac{\delta \ell}{2\pi} }^2 \kappa(\vek{\ell}_i)\; \expo{\ic \svek{\theta} \cdot \svek{\ell}_i} \equiv \sum_i \kappa_{\svek{\ell}_i}\; \expo{\ic \svek{\theta} \cdot \svek{\ell}_i}\;,
}
where $\delta \ell$ denotes the grid spacing of the discrete Fourier transformation such that $A_{\rm field}=(2\pi)^2/\delta \ell^2$ for a square survey region. The $\kappa_{\svek{\ell}}$ are Gaussian distributed if the convergence is a Gaussian random field, unless real-world effects such as masking become important. The covariance of the $\kappa_{\svek{\ell}}$, averaged over annuli of radius $\ell$ and $\ell'$, with width $\Delta \ell$ each, is given by
\eqa{
\label{eq:covkappa}
{\rm Cov}_\kappa(\ell,\ell') &=& \sum_{\svek{\ell}_i \in\, {\rm shell}(\ell)} \sum_{\svek{\ell}_j \in\, {\rm shell}(\ell')} \ba{\kappa_{\svek{\ell}_i} \kappa_{\svek{\ell}_j}} = \sum_{\svek{\ell}_i \in\, {\rm shell}(\ell)} \sum_{\svek{\ell}_j \in\, {\rm shell}(\ell')} \br{\frac{\delta \ell}{2\pi}}^4\; \ba{\kappa(\vek{\ell}_i) \kappa(\vek{\ell}_j)}\\ \nn
&=& \sum_{\svek{\ell}_i \in\, {\rm shell}(\ell)} \sum_{\svek{\ell}_j \in\, {\rm shell}(\ell')} \br{\frac{\delta \ell}{2\pi}}^2\; \delta_{ij}\; P_\kappa(\ell_i) = \delta_{\ell \ell'}\; \frac{\ell \Delta \ell}{2\pi}\; P_\kappa(\ell)\;,
}
where we used the equality
\eq{
\label{eq:nlinshell}
\sum_{\svek{\ell}_i \in\, {\rm shell}(\ell)} 1 = \frac{A_{\rm shell(\ell)}}{A_{\rm cell}} = \frac{2\pi \ell \Delta \ell}{\delta \ell^2} = \frac{A_{\rm field} \ell \Delta \ell}{2\pi}\;.
}
Here, $A_{\rm shell(\ell)}$ denotes the area covered by an annulus and $A_{\rm cell}=\delta \ell^2$ the area of a pixel of the Fourier transformed convergence map. If the shells over which the average is performed do not overlap, the $\kappa_{\svek{\ell}}$ are uncorrelated, i.e. the covariance in equation (\ref{eq:covkappa}) is diagonal.

The likelihood for the Gaussian covariance is then given by \citep[e.g.][]{bond00}
\eqa{
\label{eq:likekappa}
L(\bc{\kappa},\vek{p}) &=& \bb{(2\pi)^{N_\ell}\; \det \br{{\rm Cov}_\kappa}}^{-1/2}\; \exp \bc{-\frac{1}{2}\; \sum_{i=1}^{N_\ell} \sum_{\svek{\ell}_j \in\, {\rm shell}(\ell_i)} \frac{|\kappa_{\svek{\ell}_j}|^2}{{\rm Cov}_\kappa(\ell_i)} }\\ \nn
&=& \bb{\prod_{i=1}^{N_\ell} \ell_i\; \Delta \ell_i\; P_\kappa(\ell_i,\vek{p})}^{-1/2}\; \exp \bc{-\frac{1}{2}\; \sum_{i=1}^{N_\ell} \frac{\hat{P}_\kappa(\ell_i)}{P_\kappa(\ell_i,\vek{p})}}\;,
}
where $N_\ell$ denotes the number of angular frequency bins. To obtain the second equality, we inserted equation (\ref{eq:covkappa}) and the estimator of equation (\ref{eq:psestimator}). Hence one can continue to measure two-point statistics, in this case power spectra, where the data determines $\hat{P}_\kappa$ and the dependence on the set of cosmological parameters $\vek{p}$ enters via the model $P_\kappa(\ell,\vek{p})$. Note that shape noise is not yet included in equation (\ref{eq:likekappa}). 

To assess the information content in $L(\bc{\kappa},\vek{p})$, we compute the Fisher matrix. If the convergence is Gaussian distributed and the dependence on cosmology in its covariance, the Fisher matrix reads \citep{tegmark97}
\eqa{
\label{eq:fisherkappa}
F_{\mu\nu} &=& \frac{1}{2}\; {\rm Tr} \br{{\rm Cov}_\kappa^{-1}\; \frac{\partial {\rm Cov}_\kappa}{\partial p_\mu}\; {\rm Cov}_\kappa^{-1}\; \frac{\partial {\rm Cov}_\kappa}{\partial p_\nu}} = \frac{1}{2}\; \sum_{i=1}^{N_\ell} \sum_{\svek{\ell}_j \in\, {\rm shell}(\ell_i)} \frac{\partial P_\kappa(\ell_i)}{\partial p_\mu} P_\kappa^{-2}(\ell_i) \frac{\partial P_\kappa(\ell_i)}{\partial p_\nu}\\ \nn
&=& \sum_{i=1}^{N_\ell} \frac{\partial P_\kappa(\ell_i)}{\partial p_\mu} \bc{\frac{4\pi}{A_{\rm field}\, \ell_i\, \Delta \ell_i} P_\kappa^2(\ell_i)}^{-1}\frac{\partial P_\kappa(\ell_i)}{\partial p_\nu}\;,
}
where in the last step we again made use of equation (\ref{eq:nlinshell}). The term in curly brackets is identical to the Gaussian power spectrum covariance \citep{joachimi08}, and thus the information content of $L(\bc{\kappa},\vek{p})$ and $L(\bc{P_\kappa},\vek{p})$ is the same for a Gaussian distributed convergence. Illustratively, a Gaussian random field is fully specified by its power spectrum, and therefore the power spectrum covariance (a four-point statistic) cannot yield additional information. Conversely, if the convergence is not perfectly Gaussian, e.g. manifested via a non-vanishing connected trispectrum, $L(\bc{P_\kappa},\vek{p})$ can still yield accurate constraints when incorporating the now more complex power spectrum covariance whereas $L(\bc{\kappa},\vek{p})$ fails to account for such changes.

\label{lastpage}
\end{document}